\documentclass[preprints,article,accept,pdftex,moreauthors]{Definitions/mdpi} 

\firstpage{1} 
\pubvolume{1}
\issuenum{1}
\articlenumber{0}
\pubyear{2025}
\copyrightyear{2025}
\datereceived{ } 
\daterevised{ } 
\dateaccepted{ } 
\datepublished{ } 
\hreflink{https://doi.org/} 

\usepackage{verbatim}
\usepackage{isotope}
\usepackage{braket}
\usepackage{siunitx}
\DeclareSIUnit\torr{Torr} 
\DeclareSIUnit\atom{atoms} 
\DeclareSIUnit\gauss{G}
\usetikzlibrary {decorations.pathmorphing,arrows.meta,bending,positioning}

\Title{Trapping, manipulating and probing ultracold atoms: a quantum technologies tutorial}

\TitleCitation{Trapping, manipulating and probing ultracold atoms: a quantum technologies tutorial}

\Author{Louise Wolswijk $^{1,2,\dagger}$\orcidA{}, Luca Cavicchioli $^{1,2,\dagger}$\orcidB{}, Giuseppe Vinelli $^{3,\dagger}$\orcidD{}, Mauro Chiarotti $^{3,2,1}$\orcidP{}, Ludovica Donati $^{1,2}$\orcidO{}, Marcia Frometa Fernandez $^{1,2,4}$\orcidN{}, Diego Hern\'andez Rajkov $^{1,2,4}$\orcidM{}, Christian Mancini $^{5,6}$\orcidJ{}, Paolo Vezio $^{3,2}$\orcidF{}, \\Tianwei Zhou $^{3}$\orcidK{}, Giulia Del Pace $^{3,2,1,4}$\orcidC{}, Chiara Mazzinghi $^{1,2}$\orcidG{}, Nicolò Antolini $^{7,2}$\orcidH{}, Leonardo Salvi $^{3,2,4}$\orcidI{} and Vladislav Gavryusev $^{3,2}$*\orcidL{}}

\AuthorNames{Louise Wolswijk, Luca Cavicchioli, Giuseppe Vinelli, Mauro Chiarotti, Ludovica Donati, Marcia Frometa Fernandez, Diego Hern\'andez Rajkov, Christian Mancini, Paolo Vezio, Tianwei Zhou, Giulia Del Pace, Chiara Mazzinghi, Nicolò Antolini, Leonardo Salvi and Vladislav Gavryusev}

\isAPAStyle{%
       \AuthorCitation{Lastname, F., Lastname, F., \& Lastname, F.}
         }{%
        \isChicagoStyle{%
        \AuthorCitation{Lastname, Firstname, Firstname Lastname, and Firstname Lastname.}
        }{
        \AuthorCitation{Wolswijk, L., Cavicchioli, L., Vinelli, G., Chiarotti, M., Donati, L., Frometa Fernandez, M., Rajkov, D.H., Mancini, C., Vezio, P., Zhou, T., Del Pace, G., Mazzinghi, C., Antolini, N., Salvi, L., and Gavryusev, V.}
        }
}

\address{%
$^{1}$ \quad National Institute of Optics (CNR-INO), National Research Council, Via N. Carrara 1, 50019, Sesto Fiorentino, Italy\\
$^{2}$ \quad European Laboratory for Non-Linear Spectroscopy (LENS), University of Florence, Via N. Carrara 1, 50019, Sesto Fiorentino, Italy\\
$^{3}$ \quad Department of Physics and Astronomy, University of Florence, Via G. Sansone 1, 50019, Sesto Fiorentino, Italy\\
$^{4}$ \quad Istituto Nazionale di Fisica Nucleare (INFN), Via Sansone 1, 50019 Sesto Fiorentino, Italy \\
$^{5}$ \quad Scuola Superiore Meridionale, Largo San Marcellino 10, I-80138, Napoli, Italy\\
$^{6}$ \quad Istituto Nazionale di Fisica Nucleare, Sezione di Napoli, Complesso Universitario di Monte S. Angelo, Via Cinthia Edificio 6, I-80126, Napoli, Italy\\
$^{7}$ \quad National Institute of Optics (CNR-INO), National Research Council, Via Moruzzi 1, 56124 Pisa, Italy

}

\corres{vladislav.gavryusev@unifi.it}

\firstnote{These authors contributed equally to this work.}

\abstract{Engineered ultracold atomic systems are a valuable platform for fundamental quantum mechanics studies and the development of quantum technologies. At near zero absolute temperature, atoms exhibit macroscopic phase coherence and collective quantum behavior, enabling their use in precision metrology, quantum simulation, and even information processing. This review provides an introductory overview of the key techniques used to trap, manipulate, and detect ultracold atoms, while highlighting the main applications of each method. We outline the principles of laser cooling, magnetic and optical trapping, and the most widely used techniques, including optical lattices and tweezers. Next, we discuss the manipulation methods of atomic internal and external degrees of freedom, and we present atom interferometry techniques and how to leverage and control interatomic interactions. Next, we review common ensemble detection strategies, including absorption and fluorescence imaging, state-selective readout, correlation and quantum non-demolition measurements and conclude with high-resolution approaches. This review aims to provide newcomers to the field with a broad understanding of the experimental toolkit that underpins research in ultracold atom physics and its applications across quantum science and technology.}


\begin{document}


\section{Introduction}
\label{sec:intro}

The research field of ultracold atomic physics has a central role in modern quantum science and technology, thanks to the reliable capabilities of cooling, trapping, manipulating and probing neutral atoms that have been accrued over the last few decades. At ultralow temperatures near absolute zero, atomic ensembles are mostly devoid of thermal effects and exhibit macroscopic phase coherence and quantum correlations, providing a versatile platform for both fundamental studies and technological applications~\citep{Bloch2008, Daley2023, YagoMalo2024, Onofrio2025}.

The breakthroughs in laser cooling~\citep{Chu1998, Phillips1998} and the invention of the magneto-optical trap (MOT)~\citep{Raab1987} have enabled the preparation of atomic samples at microkelvin temperatures. The subsequent development of evaporative cooling techniques~\citep{Ketterle1996} pushed temperatures down to the nanokelvin scale, while the concurrent engineering of tightly confining magnetic and optical traps~\citep{Grimm2000} brought the samples to high densities, both in real and phase space. This progress allowed to obtain in 1995 the first Bose-Einstein condensate (BEC) in dilute alkali vapors by Cornell and Wieman, and independently by Ketterle~\citep{Anderson1995, Davis1995}. Few years later a quantum degenerate Fermi gas was demonstrated by DeMarco and Jin~\citep{DeMarco1999}. These achievements started the studies of degenerate Bose and Fermi gases~\citep{Ketterle1999, Ketterle2008}, focusing first on their inherent properties and later, with increased control, leveraging them as well engineered models for performing quantum simulations of more complex or hardly accessible physical systems~\citep{Bloch2008, Gross2017, Schaefer2020}.

In table~\ref{tab:timeline} we list a selection of historical milestones in ultracold atomic physics over the past three decades. The introduction of optical lattices and the control of short-range interactions via Feshbach resonances allowed the observation of the superfluid-Mott insulator transition in a Bose gas~\citep{Greiner2002} and of the BEC-Bardeen–Cooper–Schrieffer (BCS) crossover in Fermi gases~\citep{Regal2004, Zwierlein2004}, backing the concept of quantum simulation with ultracold quantum gases. Next, the production of dipolar quantum gases~\citep{Santos2000, Griesmaier2007} and the acquisition of control over highly excited atomics states, called \emph{Rydberg} states~\citep{Gallagher1994, Gallagher2008}, have enabled studies with long-range interactions. Then, the development of quantum gas microscopes with single-site resolution in optical lattices~\citep{Bakr2009, Sherson2010, Haller2015, Cheuk2015, Ott2016} has greatly expanded the range of accessible observables, from the many-body to the single-particle ones~\citep{Gross2017}. Precise control over the multi-level internal structure of atoms trapped in optical lattices has made it possible to engineer synthetic gauge fields and spin-orbit coupling in ultracold gases~\citep{Goldman2014, Mancini2015, Goldman2016}, opening new avenues to simulate topological phases and gauge theories with cold atoms. Moreover, the controlled interplay between long-range interactions and superfluid order in dipolar condensates has led to the observation of novel states of matter such as supersolidity~\citep{Tanzi2019, Boettcher2019, Chomaz2019}, which combines crystalline density modulation with global phase coherence. In parallel, the deterministic assembly of defect-free arrays of hundreds of neutral atoms in optical tweezer arrays~\citep{Barredo2016, Endres2016} has paved the way for large-scale, programmable Rydberg-based quantum simulators with nearest neighbor and long-range couplings~\citep{Browaeys2020, Kaufman2021, Ebadi2021, Maskara2025}. These systems now represent a credible and quickly improving platform for noisy intermediate-scale quantum (NISQ) computation~\citep{Bluvstein2022, Bluvstein2023, Bluvstein2025, Zhou2025}. Lastly, these methods have also advanced metrology: ultracold atoms in optical lattices now form the basis of optical atomic clocks with fractional frequency uncertainties below $\qty{e-18}{}$~\citep{Hinkley2013,Ludlow2015,Liu2025c}, enabling next-generation time standards and precision tests of fundamental physics.

\begin{table}[t]
\centering
\caption{Selected historical milestones in ultracold atomic physics over the last 30 years.}
\label{tab:timeline}
\begin{tabular}{p{2.5cm} p{11cm}}
\hline
\textbf{Year(s)} & \textbf{Milestone} \\
\hline
1995--2001 & Realization of Bose--Einstein condensation~\citep{Anderson1995,Davis1995} and Fermi degeneracy~\citep{DeMarco1999} in ultracold gases.\\
2002 & Observation of the superfluid--Mott insulator transition in a Bose gas~\citep{Greiner2002}.\\
2002--2004 & Control of interactions via Feshbach resonances; observation of the BEC--BCS crossover in Fermi gases~\citep{Regal2004,Zwierlein2004}.\\
2005--2009 & Production of dipolar quantum gases, enabling long-range interaction studies~\citep{Santos2000,Griesmaier2007}, also using Rydberg states~\citep{Gallagher1994,Gallagher2008}.\\
2009--2011 & Development of quantum gas microscopes with single-site resolution in optical lattices~\citep{Bakr2009,Sherson2010}.\\
2011--2015 & Realization of synthetic gauge fields and spin--orbit coupling in ultracold atoms~\citep{Goldman2014, Mancini2015, Goldman2016}.\\
2016--present & Optical tweezer arrays~\citep{Barredo2016,Endres2016} and large-scale programmable Rydberg quantum simulators~\citep{Browaeys2020,Kaufman2021,Ebadi2021,Maskara2025} and NISQ computers~\citep{Bluvstein2022,Bluvstein2023,Bluvstein2025,Zhou2025}.\\
2019 & Observation of supersolidity in dipolar condensates~\citep{Tanzi2019,Boettcher2019, Chomaz2019}.\\
2000s-present & Optical lattice clocks achieving $\qty{e-18}{}$ precision~\citep{Hinkley2013,Ludlow2015,Liu2025c}, redefining standards in metrology.\\
\hline
\end{tabular}
\end{table}

In this review, we provide an introductory survey of the key methods used to \emph{trap}, \emph{manipulate}, and \emph{detect} ultracold atoms. Our aim is to familiarize newcomers with the experimental foundations of the field, while concurrently highlighting seminal and recent advances that make ultracold atoms a leading platform for quantum technologies. We begin by describing the cooling and trapping methods in Sec.~\ref{sec:trap}, then in Sec.~\ref{sec:manip} we discuss techniques for manipulating external and internal degrees of freedom (DOF) and for interaction control, and we conclude in Sec.~\ref{sec:det} with an overview of detection strategies for bulk ensembles and with single particle and state-selective resolution, including quantum non-demolition measurements. The level of detail for each topic is meant to be introductory, providing numerous references for a more comprehensive discussion, and relates to the authors' experience in the field, therefore it is not meant to fully cover all the developments and results in each area.

\section{Trapping and cooling}
\label{sec:trap}

Techniques for cooling atoms from room temperature to near absolute zero are essential because they suppress their thermal motion, allowing their quantum-mechanical behavior to dominate. Concurrently, trapping methods are required to collect and hold these cold atoms and engineer precise potential landscapes for subsequent experiments.

In the following sections, we begin by reviewing laser cooling, covering Doppler cooling (Sec.~\ref{sec:trap:cool:doppler}), including applications to antimatter, and Sub-Doppler cooling (Sec.~\ref{sec:trap:cool:subdoppler}). Magneto-optical traps (MOTs) are then introduced in Sec.~\ref{sec:trap:cool:MOT}, where atoms are simultaneously cooled and confined. 
Next, trapping methods are detailed: magnetic traps (Sec.~\ref{sec:trap:magnetic_trap}), optical dipole traps (Sec.~\ref{sec:trap:odt:fort}), and tailored optical potentials (Sec.~\ref{sec:trap:odt:tailored_odt}), where spatial and temporal beam shaping techniques are discussed, while optical tweezers (Sec.~\ref{sec:trap:odt:tweezers}) allow single-atom trapping in reconfigurable arrays. Sec.~\ref{sec:trap:optical_lattices} introduces optical lattices, where atoms confined in a periodic potential become a tool for implementing quantum simulations.
To reach even lower temperatures, further cooling methods are needed: in Sec.~\ref{sec:trap:evaporative_cooling} we describe evaporative cooling, where the highest energy particles are selectively removed from a conservative trap (either magnetic or optical).
Finally, miniaturization of the setup by using atom chips is discussed in Sec.~\ref{sec:trap:atom_chips}. An overview table~\ref{tab:CoolTrap} summarizes the key information on these topics.

\subsection{Laser cooling}
\label{sec:trap:cool}

\subsubsection{Doppler cooling}
\label{sec:trap:cool:doppler}

\begin{figure}[htb!]
  \centering
  \includegraphics[width=0.78\textwidth]{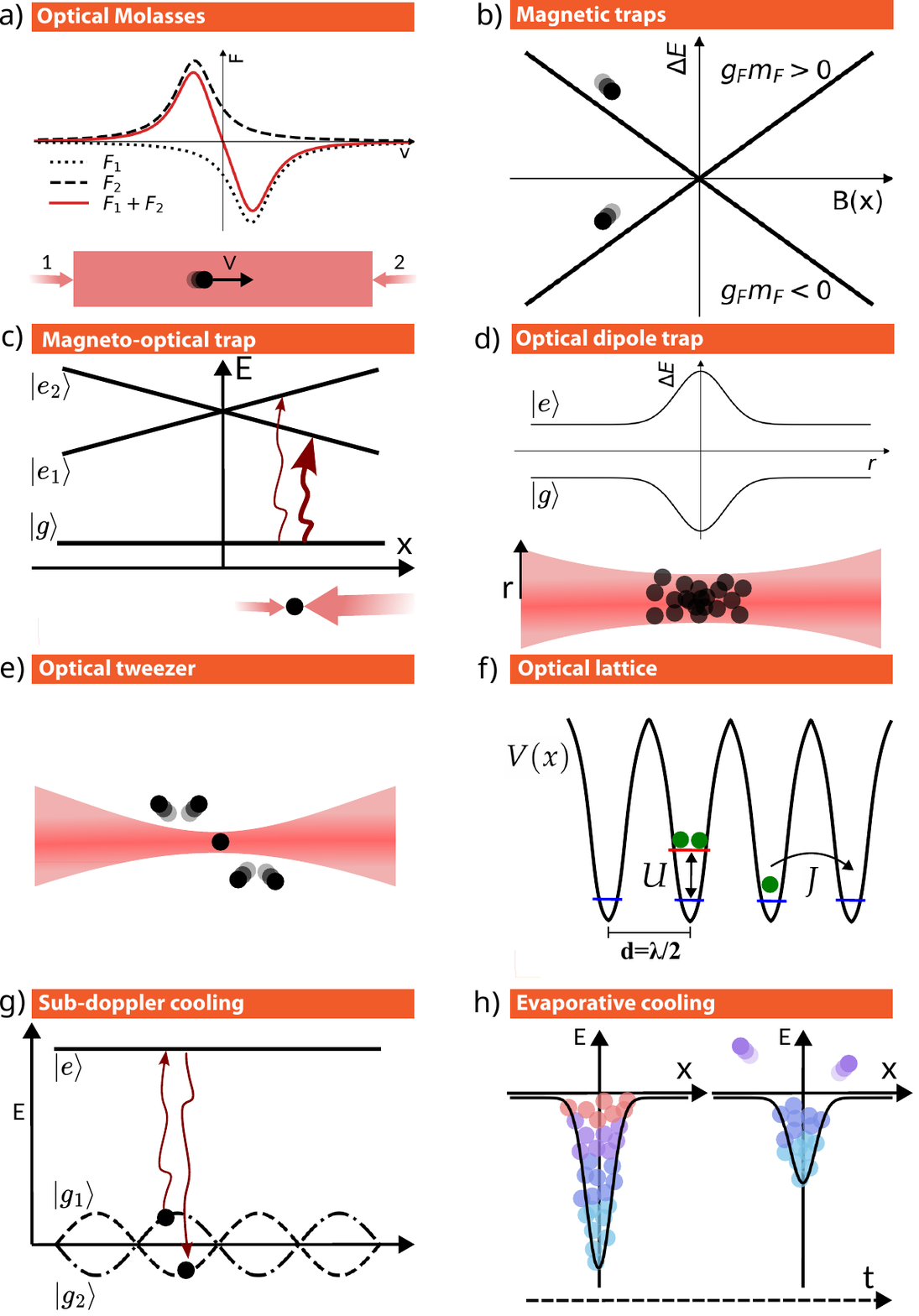}
  \caption{Outline of cooling and trapping methods. 
  (\textbf{a}) Optical Molasses: the Doppler shift of the counterpropagating beam creates a velocity dependent force on the atom. 
  (\textbf{b}) Magnetic traps: atoms in \emph{low-field seeking} states ($g_F m_F > 0$) are confined in regions of weak magnetic field, while \emph{high-field seeking} states are expelled. 
  (\textbf{c}) Magneto-optical trap: an atom that is displaced from the centre absorbs preferentially from one of the two beams, thus creating a trapping force, in addition to regular Doppler cooling. 
  (\textbf{d}) In an optical dipole trap atoms experience a potential proportional to the light intensity, allowing confinement at the focus of a far-detuned laser beam.
  (\textbf{e}) Optical tweezers are tightly focused dipole traps, where individual atoms can be trapped exploiting pair-wise collisions that remove pairs of additional atoms, until at most one particle is left inside the trap.
  (\textbf{f}) 1D optical lattice scheme. The lattice is formed by the interference between two counter-propagating laser beams, generating a potential with spatial periodicity $\lambda/2$. The schematic also illustrates the on-site interaction $U$ and the tunneling amplitude $J$ relevant to the Bose–Hubbard model.
  (\textbf{g}) Sub-Doppler cooling. An atom in a lin $\perp$ lin beam configuration is optically pumped from the hyperfine level which is upwardly shifted, in that position, to that which is lower; it must therefore spend additional kinetic energy to climb again the potential.
  (\textbf{h}) Evaporative cooling is based on the selective removal of high-energy atoms from the trap, which, after thermalization, reduces the temperature of the remaining cloud. 
  }
  \label{fig:CoolTrap}
\end{figure}

Laser cooling is the most widely used technique for cooling atomic gases to temperatures in the millikelvin range or below. The first proposal to use radiation pressure to slow atoms was formulated in~\citep{Hansch1975}, and was experimentally demonstrated soon thereafter~\citep{Phillips1982} in a supersonic Na beam. Although modern experiments generally make use of advanced schemes—designed to optimize cooling efficiency or to address a wider range of atomic species, the essential physics can still be illustrated by the simple two-level model of the original proposal. Consider an atom with an internal transition energy $\hbar \omega_0$, moving with velocity $\mathbf{v}$, and interacting with a monochromatic laser of frequency $\omega_L$ and wave vector $\mathbf{k}$. Due to its motion, the atom experiences the Doppler-shifted frequency $\omega_L'=\omega_L-\mathbf{k}\cdot\mathbf{v}$. If the laser is red-detuned, $\omega_L < \omega_0$, and the atom moves towards the beam, the Doppler effect shifts $\omega_L'$ closer to resonance, thereby enhancing the likelihood of photon absorption. Absorbing a photon imparts momentum $\hbar k$, slowing the atom, after which spontaneous emission returns it to the ground state with a photon emitted in a random direction. This sequence of absorption and emission cycle yields, on average, a net force along $\mathbf{k}$. Introducing a second, counter-propagating beam produces an opposing contribution, and it can be demonstrated~\citep{Stenholm1986} that the combined effect is a force proportional to $-\mathbf{v}$, a velocity-dependent dissipation mechanism that reduces kinetic energy. This setup, called \emph{Optical Molasses} (Fig.~\ref{fig:CoolTrap}a), can be straightforwardly generalized to three dimensions by applying three orthogonal pairs of counter-propagating beams~\citep{Chu1985}. 

The minimal temperature attainable in an optical molasses can be established through a more detailed analysis, as discussed in key references on laser cooling (e.g.~\citep{Metcalf1999}). This limit originates from the role of spontaneous emission and the finite excited-state lifetime $\tau$. Each spontaneous emission event imparts a recoil momentum to the atom, altering its velocity in a random direction. The process resembles a random walk in momentum space, with individual steps of size $\hbar k$ occurring at the rate $\gamma = 1/\tau$. This randomization of momentum introduces heating, in direct competition with the velocity damping provided by the molasses. From this balance one obtains the fundamental temperature limit, the Doppler temperature, $T_D = \frac{\hbar \gamma}{2 k_B}$. An alternative, more heuristic viewpoint is to interpret the finite lifetime $\tau$ as an uncertainty, which leads to an energy width of order $\hbar \gamma/2$, corresponding to the same temperature scale. In experimental implementations, the two counter-propagating beams are typically assigned different polarization states, either linear or circular. The resulting superposition of optical fields can give rise to polarization gradients and generate additional processes in polarization-dependent optical transitions. A notable example is sodium, where these effects allowed the atomic ensemble to reach temperatures below the theoretical Doppler limit $T_D$~\citep{Dalibard1989, Lett1988}. Such processes are collectively referred to as sub-Doppler schemes, and generally exploit the multilevel structure of atoms or make use of external magnetic or electric fields to enhance cooling.

A major limitation of the idealized two-level model is that the vast majority of atoms cannot be treated in such a simplified manner. Alkali elements are most frequently chosen for laser cooling because their spectra are comparatively straightforward. However, even in these cases the ground electronic state is split into two hyperfine components. Consequently, an atom excited from one hyperfine ground state may decay into the other, which lies too far detuned to remain resonant with the cooling laser. A clear example is~\isotope[87]{Rb}, where the D2 line couples the ground-state level $F=2$ to the excited-state level $F'=3$. Owing to mixing within the excited-state manifold, decay to the $F=1$ ground state is possible. Since the hyperfine splitting of the ground state in~\isotope[87]{Rb} is $\qty{6.8}{\giga\hertz}$, atoms in the $F=1$ level cannot interact with a laser tuned to the $\ket{g,F=2} \rightarrow \ket{e,F'=3}$ transition. To recover these atoms and reintegrate them into the cooling cycle, a second laser field is applied, resonant with the $\ket{g,F=1} \rightarrow \ket{e,F'=2}$ transition. This repumping scheme has been successfully implemented for all alkali atoms, including francium, despite the absence of stable isotopes~\citep{Lin1991, Chu1985, Williamson1995a, Ludvigsen1994, Salomon1990, Sprouse1997}.

Notably, among alkali atoms, the fermionic isotope of lithium,~\isotope[6]{Li}, presents a particular case in which the natural linewidth of the excited-state manifold of the D2 optical transition exceeds the hyperfine splitting of the manifold. As a consequence, coupling to the different hyperfine ground states is unavoidable, thus placing the repumping transition on equal footing with the cooling transition. This specific state mixing is known to be a limiting factor for achieving low temperatures in optical molasses~\citep{Bambini1997}. Indeed, most experimental implementations of optical molasses with~\isotope[6]{Li} are restricted to temperatures well above the Doppler limit, typically on the order of $\sim 4-6\,T_D$~\citep{HernandezRajkov2020, Burchianti2014}.

For atoms from other groups, further considerations apply, due to their richer spectrum; while this usually means a more complex laser system, it also opens further possibilities. As an example, alkali-earth elements, as strontium, and alkali-earth-like atoms, as ytterbium, have a highly forbidden intercombination transition that can be used to cool to lower temperature, due to its narrow width: with~\isotope[88]{Sr}, pure laser cooling schemes allow to reach temperatures of the order of $\qty{400}{\nano\kelvin}$~\citep{Katori1999}. Due to the breadth and depth of the subject, we will not go into exhaustive detail on all possible laser cooling techniques, but some reviews on the matter are~\citep{Adams1997, Perrin2011, Schreck2021}.

Laser cooling also finds application in atomic systems composed of antimatter, such as antihydrogen and positronium. These exotic atoms are studied to search for quantum electrodynamics tests~\citep{Adkins2022}, charge–parity–time reversal (CPT) symmetry violations and to investigate the effects of gravity on antimatter~\citep{Mariazzi2020, Anderson2023, Vinelli2023}. Such experiments face unique challenges: the available samples are extremely scarce and unstable, and the required laser sources often operate in the ultraviolet, a spectral region that is notoriously difficult to generate and stabilize.

In antihydrogen, cooling is performed on the 1S–2P (Lyman-$\alpha$, $\qty{121.6}{\nano\meter}$) transition using narrowband vacuum ultraviolet (VUV) pulses, red-detuned by a few hundred $\unit{\mega\hertz}$ and injected along a single axis of the magnetic trap. This technique has enabled cooling of trapped antihydrogen down to temperatures of about $\qty{10}{\milli\kelvin}$~\citep{Baker2021}.

For positronium (1S–2P at $\qty{243}{\nano\meter}$), two complementary strategies have been demonstrated. The first relies on a broadband Doppler cooling scheme ($\sim \qty{100}{\giga\hertz}$ bandwidth, $\sim \qty{70}{\nano\second}$ duration) that reduces the temperature from $\sim \qty{380}{\kelvin}$ to $\sim \qty{170}{\kelvin}$~\citep{Glogger2024}. The second uses a broadband chirped pulse train in counter-propagation, sweeping across the Doppler profile and cooling a fraction of the gas to $\sim\qty{1}{\kelvin}$ within $\sim \qty{100}{\nano\second}$~\citep{Shu2024}. This method detunes the laser dynamically during the cooling process, compensating for the recoil-induced frequency shifts from repeated absorption–emission cycles and enabling tens of cooling events.

This feature is particularly critical for positronium: being extremely light (just twice the electron mass), the recoil from each photon scattering event is large enough to significantly shift the resonance frequency, reducing the atom–laser interaction probability. At the same time, the very small mass of positronium implies that Bose–Einstein condensation can in principle be reached at temperatures on the order of a few kelvin~\citep{Shu2016,Mills2019}—further motivating the development of advanced cooling techniques for this unique system.

\subsubsection{Sub-doppler cooling}
\label{sec:trap:cool:subdoppler}

A remarkable consequence of the hyperfine splitting present in the ground state of many atomic isotopes is that an optical molasses can cool atoms to temperatures below the Doppler limit. This was first demonstrated in sodium molasses, where temperatures near $\qty{40}{\micro\kelvin}$ were observed in an optical molasses, roughly six times lower than the Doppler temperature of $\qty{240}{\micro\kelvin}$~\citep{Lett1988}. The phenomenon arises from the interaction between the polarization gradient of the cooling beams and the angular momentum of the ground state; this mechanism is thus called \emph{polarization gradient cooling} (PGC).

In the presence of polarization ellipticity gradients, the ground-state sublevels experience differential light shifts that vary periodically in space. 
Atoms optically pumped from a maximum of one potential to a minimum of another must move against the light-shift gradient, thereby losing kinetic energy. As this cycle repeats along the spatially periodic light-shift pattern, atoms continuously climb and lose energy—a process known as \emph{Sisyphus cooling}.
Alternatively, when the gradient is in the polarization angle, atoms tend to accumulate in the state that scatters light more strongly, producing a radiation pressure imbalance that opposes their motion and increases friction. A detailed treatment of both mechanisms can be found in~\citep{Dalibard1989, CohenTannoudji1990, Metcalf1999}. In more practical terms, the ellipticity gradient is found in counterpropagating beams with mutually perpendicular, linear polarization (lin $\perp$ lin), whereas the angle gradient happens when the beams have opposite circular polarizations ($\sigma^+\ -\ \sigma^-$), as shown in Fig.~\ref{fig:CoolTrap}g. It should be emphasized that these models are idealized: in realistic setups, the actual behavior is a superposition of the two mechanisms, especially in 3D configurations where three orthogonal beams cannot maintain parallel polarizations.

This cooling has been successfully implemented in \isotope[]{Na}, \isotope[]{Rb}~\citep{Sheehy1990}, \isotope[]{Cs}~\citep{Salomon1990}, \isotope[]{Sr}~\citep{Xu2003a}, \isotope[]{Yb}~\citep{Kostylev2014}, \isotope[]{Er}~\citep{Berglund2007}, and \isotope[]{Dy}~\citep{Youn2010}. However, it proves ineffective for elements such as \isotope[]{K} and \isotope[]{Li}, where the hyperfine structure overlaps with the natural linewidth. Here, atoms cannot be reliably cycled between two well-isolated levels, limiting the effectiveness of traditional polarization gradient cooling~\citep{Cooper1994, Bambini1997}. To address this issue, gray molasses cooling on the D1 optical transition has emerged as an effective sub-Doppler technique for alkali species such as potassium and lithium, where the small excited-state hyperfine splitting prevents efficient cooling on the D2 line~\citep{Grier2013, Fernandes2012, Landini2011, Burchianti2014, Sievers2015}. 

Gray molasses cooling is performed on the D1 line using a bichromatic scheme that couples two ground states to a common excited state in a $\Lambda$-type configuration. The two beams, commonly referred to as “cooler” and “repumper” due to their roles in conventional optical molasses, actually fulfill different functions in this configuration. Both beams are typically blue-detuned and tuned close to zero two-photon (Raman) detuning, $\delta = \Delta_r - \Delta_c \approx 0$, with $\Delta_{r,c}$ denoting the detunings of the repumper and cooler from their respective transitions (see Sec.~\ref{sec:manip:int:RamanRamsey}). The polarization-gradient standing wave generates spatially varying bright-state light shifts, while $\Lambda$-type coherence drives low-velocity atoms into the dark state, producing strong damping with minimal diffusion. This “gray” combination of bright and dark states cools large atomic clouds in a few milliseconds to temperatures of tens of microkelvin, thereby enhancing the phase-space density. 

Another method to cool atoms below the Doppler limit is \emph{Raman sideband cooling}~\citep{Hamann1998, Vuletic1998, Han2000, Kerman2000}. This technique applies to atoms confined in a sufficiently tight potential such that the vibrational levels of their motional states are well resolved; for this purpose, optical lattices~\citep{Weber2003} (see Sec.~\ref{sec:trap:optical_lattices}) or optical tweezers~\citep{Kaufman2012} (see Sec.~\ref{sec:trap:odt:tweezers}) are typically used. Let us consider an atom whose ground state possesses at least two Zeeman sublevels, $\ket{F, m_F}$ and $\ket{F, m_F'}$, and that occupies a vibrational state $\ket{n}$. 
By applying an external magnetic field, one can tune the energies such that $E(\ket{F, m_F}\ket{n}) = E(\ket{F, m_F'}\ket{n-1})$. Under this condition, a Raman transition (see Sec.~\ref{sec:manip:int:RamanRamsey}) can transfer population from $\ket{F, m_F}\ket{n}$ to $\ket{F, m_F'}\ket{n-1}$.
The atom is then optically pumped back to $\ket{F, m_F}$; however, since vibrational transitions are strongly forbidden during this process, the atom is ends up in $\ket{F, m_F}\ket{n-1}$. 
Repeating this sequence  removes one vibrational quantum per cycle, cooling the atom down to the motional ground state.

\subsubsection{Magneto-optical traps (MOTs)}
\label{sec:trap:cool:MOT}

Optical molasses and sub-Doppler techniques are highly effective for cooling atomic gases; however, they do not provide any confining mechanism, which is necessary to hold a gaseous cloud for extended periods of time. To overcome this limitation, the most widely used configuration is the \emph{Magneto-Optical Trap} (MOT). In a MOT, a weak magnetic field gradient $B(x) = bx$ induces a Zeeman shift of the atomic energy levels proportional to the spatial coordinate, $\Delta E \propto x$. While this shift alone generates a trapping potential, the fields used in typical MOTs do not create a very deep trap by themselves. Instead, the MOT operates by combining the Zeeman shift with radiation pressure~\citep{Raab1987, Metcalf1999}. In this configuration, the Zeeman shift causes atoms to preferentially absorb photons from the laser beam propagating opposite to their velocity, creating a harmonic potential that traps atoms near the trap center while simultaneously cooling them, as shown in Fig.~\ref{fig:CoolTrap}c. When pairs of counter-propagating red-detuned laser beams are sent along all three orthogonal directions a three-dimensional MOT (3D MOT) is realized, that cools the atoms moving into any direction and traps them in center.

The MOT was first demonstrated, and its theory explained, for sodium atoms~\citep{Raab1987}, but it has since become a cornerstone of almost any cold atom experiment. It has been realized in alkali atoms~\citep{Ludvigsen1994, Williamson1995a, Grego1996, Schunemann1998}, alkaline-earth atoms~\citep{Loo2004, Katori1999, Grunert2002}, metastable noble gases~\citep{Vassen2012}, lanthanides~\citep{Mcclelland2006, Youn2010a, Kuwamoto1999}, and many others. While providing a complete list of MOT implementations is beyond the scope of this review, the wide range of atoms that have been successfully trapped highlights the significance and versatility of this technique.

MOTs enable to obtain both low temperatures and relatively high atomic densities; however, they are subject to inherent limitations on the maximum achievable density. When the density exceeds approximately $\qty{e11}{\atom\per\cubic\centi\meter}$, radiation pressure from spontaneously emitted light within the cloud generates an outward force comparable to that of the trapping beams, preventing any further increase of the atomic ensemble density, a phenomenon known as radiation trapping~\citep{Walker1990}. One strategy to circumvent this limitation is to avoid radiation scattering from the inner core of the cloud, which can be accomplished by suppressing the repumping process in that region. This leads to the formation of a so-called \emph{dark spontaneous-force optical trap} (SPOT)~\citep{ketterle1993}. Experimentally, this is implemented by shaping the repumper beam with a hollow spatial profile, so that atoms in the center of the MOT are eventually pumped into a dark state, where they no longer contribute to radiation trapping. For certain atomic species, such as rubidium, which do not efficiently populate the dark state on their own, an additional depumping beam can be employed to achieve the same effect~\citep{Radwell2013}.

A 3D MOT can be loaded directly from a thermal atomic beam, but since the capture velocity is finite, it is usually advantageous to slow the atomic beam before directing it into the trap. A single resonant laser beam, counterpropagating with respect to the atomic motion, can exert sufficient radiation pressure to significantly decelerate the incoming flux of atoms. However, the Doppler cooling force is velocity dependent, and an atom quickly falls out of resonance, limiting further deceleration. A widely used solution is the \emph{Zeeman slower}, which employs a magnetic field decreasing linearly along the atomic trajectory, $B = B_0 - b'x$. The Zeeman shift induced by this field compensates the changing Doppler shift as the atoms slow down, keeping the scattering force nearly constant~\citep{Phillips1982, Lison1999, Ovchinnikov2008}.  

A complementary enhancement to the loading of a 3D MOT consists in cooling the atoms in the transversal directions to improve the collimation of the atomic beam. This is achieved using two pairs of (typically elongated) quadrupole coils, arranged orthogonally so that their longitudinal magnetic fields cancel out. In each transverse direction, a pair of counterpropagating laser beams is applied, forming a two-dimensional MOT (2D MOT) that cools the atoms in the transverse directions while leaving the longitudinal velocity largely unaffected~\citep{Weyers1997, Dieckmann1998}.

\subsection{Magnetic traps}
\label{sec:trap:magnetic_trap}

For atoms possessing a permanent magnetic dipole moment $\boldsymbol{\mu}$, an energy term $H = \boldsymbol{\mu} \cdot \mathbf{B}$ contributes to their Hamiltonian in the presence of a magnetic field. This allows the creation of a trapping potential using an inhomogeneous, static magnetic field $\mathbf{B}(\mathbf{r})$. Although the concept of magnetic trapping was proposed as early as the 1960s~\citep{Heer1963}, the relatively shallow trap depths achievable -- on the order of $\sim 1-\qty{10}{\kelvin}$ -- made the experimental realization feasible only after the development of laser cooling techniques~\citep{Pritchard1983}.

\begin{figure}[htb!]
  \centering
  \includegraphics[width=0.8\columnwidth]{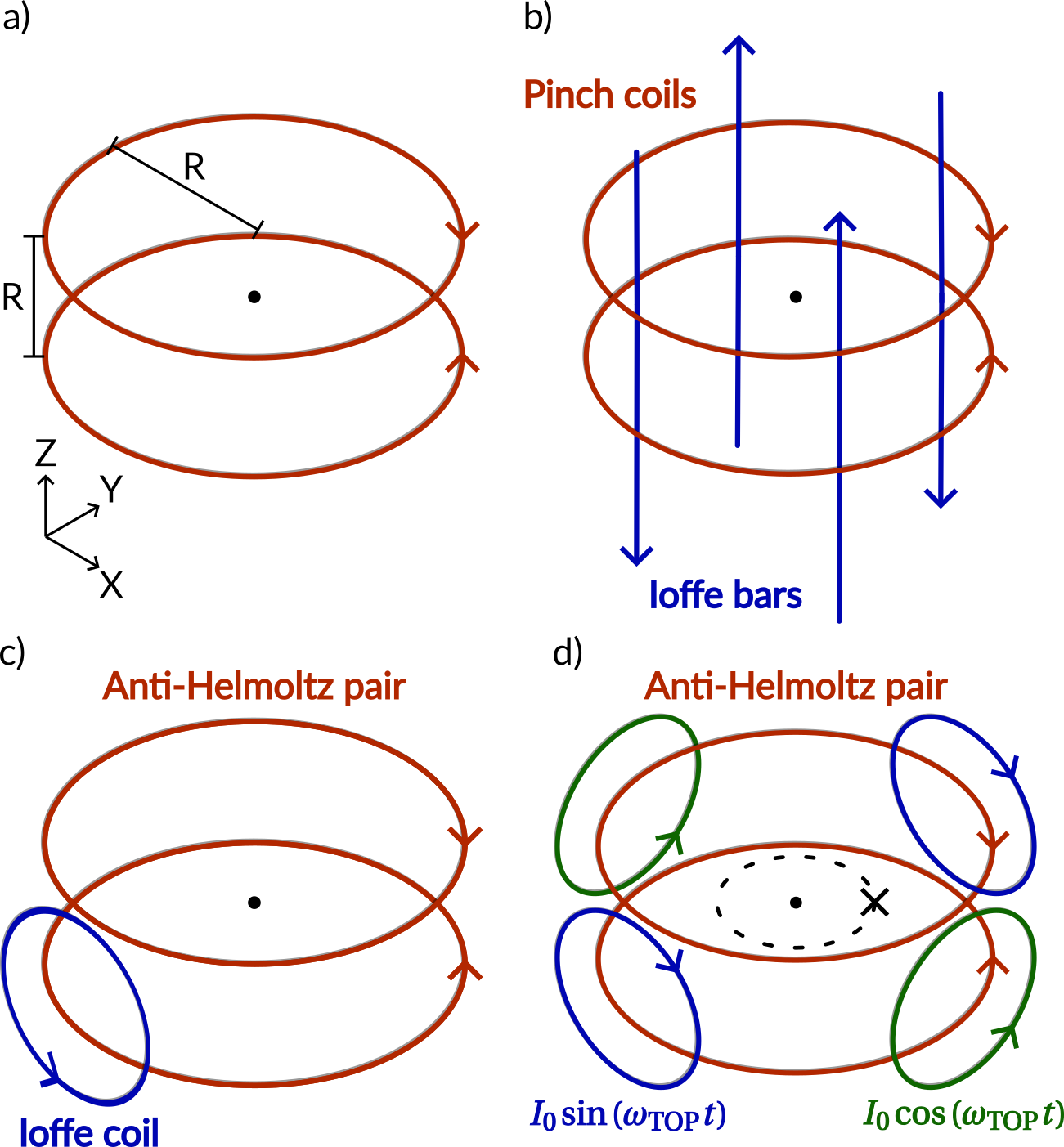}
  \caption{Common magnetic trap configurations: (\textbf{a}) an anti-Helmoltz quadrupole, (\textbf{b}) a Ioffe-Pritchard trap, (\textbf{c}) a QUIC trap, and (\textbf{d}) a TOP trap. Different coils are highlighted in different colors, according to their function, and the current verses are indicated by the arrows. A black circle indicates the position of the atoms. In sub-figure d, the X indicates the magnetic field minimum, and the dashed circle is its orbit.}
  \label{fig:magtraps}
\end{figure}

Magnetic trapping was first demonstrated with \isotope[]{Na} atoms~\citep{Migdall1985} and has since been employed, in one form or another, to confine nearly all laser-coolable atomic species. It also played a key role in the realization of the first BECs~\citep{Anderson1995, Davis1995, Bradley1995}. Magnetic trapping is particularly important for highly magnetic atoms, such as \isotope[]{Dy} or \isotope[]{Er}, for which it provides not only highly efficient confinement but also a versatile means of manipulation~\citep{Lu2010a,Chomaz2023}.

The main advantage of this type of trap, depicted in Fig.~\ref{fig:CoolTrap}b, compared to other neutral atom traps, lies in its depth and spatial extent, which allows to confine large clouds of cold atoms. Its primary limitation, however, is that only certain atomic states can be trapped. According to the magnetic version of Earnshaw's theorem, a local maximum of $B$ cannot be created in free space using only magnetostatic fields. The magnetic moment of an atom is given by $\boldsymbol{\mu} = -\mu_B g \boldsymbol{F}/\hbar$, so for a magnetic field along the $z$-axis, the corresponding energy contribution to the Hamiltonian is $E_Z = \mu_B g m_F B_z$, where $m_F = F_z/\hbar$. Consequently, only states with $g m_F > 0$ can be stably confined in a minimum of $B$; these are called \emph{low-field seeking states}, as they are attracted toward low magnetic field regions. In contrast, \emph{high-field seeking states} are expelled from the trap center and move toward regions of higher magnetic field~\citep{Perez-Rios2013}.

Another, related, limitation of magnetic traps is the finite lifetime of the trapped gases, particularly when the trap minimum is $B_{\rm min} = 0$. Near this region, atoms can undergo spin flips and transition to anti-trapped, high-field seeking states. In high-field regions, the Larmor precession frequency of the atomic magnetic moment, $\omega_L(B) = g \mu_B B/\hbar$, is much larger than the characteristic trap frequency $\omega_t$. Under these conditions, the Zeeman term in the Hamiltonian can be approximated by a time-averaged $\boldsymbol{\mu}$ with a fixed orientation relative to $\mathbf{B}$. However, near the trap center, this approximation breaks down: the relative orientation of $\boldsymbol{\mu}$ and $\mathbf{B}$ changes as the atom moves, generating an effective vector potential that couples to the momentum through a term proportional to $\mathbf{A} \cdot \mathbf{p} + \mathbf{p} \cdot \mathbf{A}$~\citep{Sukumar1997}. This term includes a factor $\sigma_z L_z$, which couples the Zeeman sub-levels, a phenomenon known as a Majorana spin flip. In the first magnetic trapping experiment, the measured lifetime of trapped atoms was only $\qty{0.8}{\second}$~\citep{Migdall1985}. While such a loss rate can be acceptable in some contexts, suppressing Majorana losses to achieve longer-lived atomic clouds requires a trap minimum with a sufficiently large magnetic field. This condition can be realized in certain trap configurations.

One of the simplest types of magnetic traps is the quadrupole trap, made with two identical coaxial coils carrying currents in opposite directions~\citep{Bergeman1987}. Typically, the coils are arranged such that the distance between their centers equals the coil radius
(Fig.~\ref{fig:magtraps}a). This configuration is known as \textit{anti-Helmholtz}, because if the currents flowed, instead, in the same direction, the arrangement would form a Helmholtz pair, which is a configuration used to produce a nearly uniform magnetic field (up to second order). The anti-Helmholtz arrangement, on the other hand, generates the magnetic field gradient used in a magneto-optical trap (MOT, see section~\ref{sec:trap:cool:MOT}) Near the trap center, the quadrupole field is axially symmetric, with the longitudinal component given by $B_z\propto z$ and the transverse component by $B_\rho\propto-\rho/2$. This produces a linear trapping potential, rather than a harmonic one, meaning that the period of atomic motion depends on the temperature. Crucially, the magnetic field vanishes at the trap center ($B=0$), making the trap vulnerable to Majorana losses, particularly at low temperatures (i.e., high trapping frequencies). Historically, one effective method to mitigate these losses was the plugged quadrupole trap, in which a repulsive optical potential is placed at the trap center to keep atoms away from the zero-field region~\citep{Davis1995}.

Another commonly used configuration is the Ioffe-Pritchard trap~\citep{Pritchard1983}. In this design, four parallel wires are arranged along the long edges of an imaginary rectangular prism, with alternating current directions at each corner, providing confinement in the transverse directions. Longitudinal confinement is achieved by a pair of quadrupole coils, whose separation is chosen to produce the desired trapping characteristics (Fig.~\ref{fig:magtraps}b). Consequently, the field magnitude close to the minimum can be approximated as  $B_z=b_0+b_2(z^2-\rho^2/2)$, $B_\rho=-b_2 z\rho + c_1 \rho \cos(2\phi)$, $B_\phi=-c_1 \rho \sin(2\phi)$~\citep{Bergeman1987}. This results, near the minimum, in a field magnitude $B \approx b_0 + \alpha z^2 + \beta \rho^2$, corresponding to an approximately harmonic potential in both the radial and longitudinal directions, with a non-zero minimum field.

Given its favorable properties, it is unsurprising that the Ioffe-Pritchard trap has inspired numerous variants over the years. For instance, by adding a lateral coil, known as a Ioffe coil, to an anti-Helmholtz quadrupole trap, one can generate a magnetic field similar to that produced by the four parallel wires, resulting in a Quadrupole-Ioffe-Configuration (QuIC) trap~\citep{Esslinger1998} (Fig.~\ref{fig:magtraps}c). A key advantage of this design is that by ramping the current in the Ioffe coil, a quadrupole trap—such as that used in a MOT—can be smoothly transformed into an Ioffe-Pritchard trap, greatly facilitating the loading of cold atoms. 

Another variant is the cloverleaf trap, in which the four wires are replaced by four smaller coils arranged in a square perpendicular to the longitudinal axis, resembling a cloverleaf, which gives the configuration its name~\citep{Streed2006}. This design provides excellent optical access in the radial direction. Yet another example is the baseball trap~\citep{Bergeman1987}, where a single coil is shaped along a pattern resembling the seams of a baseball. Although the complex conductor geometry prevents analytic expressions for the magnetic field, this configuration offers both deep trapping potentials and high trapping frequencies~\citep{Wang2007}.

An ingenious way to circumvent Earnshaw's theorem is to employ time-dependent magnetic fields. A Time-Orbiting-Potential (TOP) trap consists of two quadrupole coils along the $z$-axis and two pairs of Helmholtz coils along the $x$ and $y$ axes~\citep{Petrich1995} (Fig.~\ref{fig:magtraps}d). These coils provide a transverse bias field that shifts the position of the trap minimum. By driving each pair of Helmholtz coils with sinusoidal currents that are $\qty{90}{\degree}$ out of phase, the trap minimum can be made to orbit in a closed circular trajectory. The rotation frequency must be carefully chosen: it should be much smaller than the Larmor frequency, but much larger than the trapping frequency. This trap is historically important, as it was used in one of the first experiments to achieve Bose-Einstein condensation~\citep{Anderson1995}.

\subsection{Optical dipole traps}
\label{sec:trap:odt}

\subsubsection{Far-off-resonance optical dipole traps}
\label{sec:trap:odt:fort}

As with magnetic traps, \emph{optical dipole traps} (ODTs) rely on adding a field-dependent term to the atomic Hamiltonian. However, instead of a Zeeman shift, the interaction arises from the AC Stark shift of the ground state~\citep{Grimm1999}. In the simplest case of a two level atom with transition frequency $\omega_0$ and natural linewidth $\gamma$, subjected to a laser beam with intensity profile $I(\mathbf{r})$ and frequency $\omega_L$, the resulting potential is
\begin{equation}
U_\text{ODT}(\mathbf{r}) = \frac{3\pi c^2}{2 \omega_0^3}
\frac{\Gamma}{\omega_L-\omega_0} I(\mathbf{r})\ .
\end{equation}
This implies that the potential energy at position $\mathbf{r}$ is proportional to the local light intensity, with its sign determined by the sign of $\omega_L - \omega_0$: for \emph{blue detuning} ($\omega_L > \omega_0$) the potential is repulsive, while for \emph{red detuning} ($\omega_L < \omega_0$) it is attractive, as illustrated in Fig.~\ref{fig:CoolTrap}d. The proportionality constant $\alpha$ is known as the (real) atomic polarizability. For a more realistic atomic structure, the dipole potential also depends on the polarization and the magnetic quantum number of the ground state, but the general considerations regarding sign and spatial dependence remain valid. This highlights a major advantage of optical traps: by shaping $I(\mathbf{r})$, one can engineer potentials of virtually any spatial form and, by choosing an appropriate laser frequency, make them either attractive or repulsive.

At first glance, this force may appear very different from the radiation pressure force involved in laser cooling. However, the two are actually different manifestations of the same interaction: the conservative dipole force corresponds to the real part, while the dissipative radiation pressure corresponds to the imaginary component~\citep{Metcalf1999}. It can be shown~\citep{Grimm1999} that the heating rate is proportional to $\Gamma^2/(\omega_L-\omega_0)^2 I$, so that for sufficiently large detunings the heating becomes negligible, albeit at the cost of a shallower
trapping depth - typically compensated by using higher-power laser beams. From this preference for large detunings, either red or blue, comes the name \emph{Far Off-Resonance Trap} (FORT)~\citep{Miller1993}.

The most commonly used beam is a Gaussian beam, whose intensity profile is~\citep{Svelto2010}:
\begin{equation}
I(\rho, z) = I_0 \frac{w_0^2}{w(z)^2} e^{-2\frac{\rho^2}{w(z)^2}},
\end{equation}
where the beam propagates along the $z$-axis, $w(z) = w_0 \sqrt{1+z^2/z_R^2}$ is the beam diameter, and $z_R = \omega_0^2 n \omega_L / (2 c)$ is called the \emph{Rayleigh length}. Assuming that the trapped atoms only explore a region near the potential minimum $U_0 = U(0,0) = \alpha I_0$, the trap frequencies for an atom of mass $m$ can be estimated as $\omega_\rho^2 = 4U_0/(m w_0^2)$ and $\omega_z^2 = 2U_0/(m z_R^2)$~\citep{Grimm1999}. Given that $I_0 = 2 P / (\pi w_0^2)$, where $P$ is the optical power of the laser beam, we can see that the trap frequencies scale as $P^{1/2}$, and with $w_0^{-2}$ and $z_R^{-2}$ in the radial and longitudinal directions, respectively. Typically, $w_0 \ll z_R$, so that $\omega_\rho \gg \omega_z$: this results, in practice, in a poor trapping performance in the longitudinal direction for a single beam. To compensate for this, a second beam is usually added at an angle to the first, creating additional longitudinal confinement in a configuration known as a \emph{crossed dipole trap}.

\subsubsection{Tailored optical potentials}
\label{sec:trap:odt:tailored_odt}

The direct dependence of the optical potential on the laser intensity provides an exceptional tunability on the trap geometry via beam shaping techniques. Multiple laser beams can be combined to produce diverse trap configurations, from crossed dipole traps of tunable aspect ratio, through arbitrarily shaped arrays of small dipole traps and to optical lattices with counterpropagating beams of same frequency~\citep{Grimm1999}. In the last decades, the possibility of beam shaping has greatly advanced thanks to the advent of beam shaping devices, namely spatial light modulators (SLMs)~\citep{Gauthier2021}. These devices use diverse technologies to grant precise control over a laser beam intensity or phase, statically or dynamically, and can be thus employed in cold atoms experiment to create \textit{ad hoc} optical dipole potentials of arbitrary geometry and with high resolution. We can broadly group SLMs into two classes, \emph{intensity} and \emph{phase} devices, depending on how they manipulate the illuminating beam.

Intensity SLMs act on the intensity of the laser beam, allowing to locally shape $I(x, y)$, like Digital Micromirror Devices (DMDs) or Acousto- and Electro-Optic Deflectors (AODs and EODs) and acoustic-optic modulators (AOMs). The intensity profile arbitrarily shaped with these devices is typically imaged on the atomic sample with a high resolution imaging system, so to sculpt an arbitrary time-depend optical dipole trap with high precision. 

DMDs consist of an array of micrometer-sized mirrors, the tilt of which can be tuned between two positions (ON and OFF), so to locally shape the reflected intensity. They effectively work as controllable light masks and allow for a binary control of the local intensity, i.e. each mirror is either in an ON or OFF position, which, once combined with the finite resolution of the imaging system that groups few of them as acting on a single resolved spot, allows to tune the intensity in greyscale.
    
Beam deflectors like AODs and EODs exploit diffraction or refraction of an incoming laser beam into a crystal of tunable properties to precisely control its output position, that once imaged on the atomic sample creates the potential in the desired position~\citep{Duocastella2020}. In an AOD the beam is deflected by Bragg scattering from phonons of tunable frequency and amplitude driven in the crystal by a radio-frequency (RF) signal, using the same working principle of AOMs~\citep{Goutzoulis1994}. Instead, in an EOD an electric field of tunable strength is applied to change the refractive index of the crystal, effectively steering the incoming laser beam. In both cases, the beam deflectors operate along one spatial direction, so to draw a 2D trap two orthogonally oriented deflectors are typically used in cascade. The 2D control over the laser beam is then transformed into an effective optical trap by time-painting the intensity on the atomic sample by scanning the beam position over timescales faster than the atomic response time. 3D pointing and random-access control is also possible by using four AODs~\citep{Ricci2022, Ricci2024, Picard2025, Lu2025a} or in combination with electrically tunable lenses (ETLs) that can change the addressed focal plane~\citep{Chen2021b}.

Both the DMD and the beam deflector family of devices allow also for a dynamical control of the intensity profile, with timescales of tens of $\unit{\kilo\hertz}$ for DMDs, $\unit{\mega\hertz}$ for AODs and $\approx 10-\qty{50}{\mega\hertz}$ for AOMs. Some applications of time-dependent tunable optical potentials are presented in Sec.~\ref{sec:trap:odt:tweezers} and~\ref{sec:manip:ext:dynamicMan} and are leveraged also in other scientific domains~\citep{Duocastella2020, Pang2022, Ricci2022, Ricci2024}. 

Phase SLMs locally modify the phase $\phi$ (x, y) of the laser beam, and many are based on liquid crystal devices (LCD)~\citep{Esmer2015, Schroff2023, Ammenwerth2025} and a common type is called Liquid Crystal on Silicon SLMs (LCOS-SLM). The modified phase pattern is then projected through an optical system into the atomic sample to holografically create the desired optical dipole trap. LCDs employ the response of liquid crystals to an external electric or magnetic field to create controllable polarization switches. The anisotropic molecules creating the liquid crystal are induced to form crystal-like structure by the electric field thanks to the interplay between the electromagnetic and hydrodynamic forces. The external field controls the orientation of the crystalline structure, that in turns changes the local birifrangence properties of the LCD, effectively allowing for a local tuning of the polarization of an incoming laser beam. These devices are typically slower than the intensity SLMs counterpart, with typical timescales in the range of $10-\qty{100}{\hertz}$ for the nematic LCDs up to a few tens of $\unit{\kilo\hertz}$ for the ferroelectric LCDs, but allow for realizing more stable static potentials, avoiding the refresh rate of DMDs or the intrinsic discrete nature of beam deflectors. Moreover, by controlling the phase profile of the laser beam, these devices effectively work as lenses of tunable focal length (like ETLs), allowing for controlling the beam intensity profile also along its direction of propagation, i.e. along z.

\subsubsection{Optical tweezers}
\label{sec:trap:odt:tweezers}

Optical tweezers are highly focused laser optical dipole traps, generated typically with optics such as high-numerical-aperture objectives~\citep{Grier2003, Gruenzweig2010, Nogrette2014, Endres2016, Barredo2016, Giardini2025}, microlens arrays~\citep{Schaeffner2020, Schlosser2023, Pause2024} or metasurfaces~\citep{Hsu2022, Huang2023, Huang2024, Holman2024}, capable of confining and manipulating single neutral atoms with high precision using the dipole force~\citep{Ashkin1997, Andersen2022}. Optical tweezers rely on the interaction between the induced electric dipole moment of an atom and the electric field of a FORT. When tightly focused, the gradient of the light intensity creates a potential well that attracts and confines the atom near the intensity maximum, as illustrated in Fig.~\ref{fig:CoolTrap}e. In experiments, pre-cooled atoms from a MOT are spatially overlapped and loaded into these optical potentials, where they may be further cooled by sub-Doppler techniques, such as PGC, Sisyphus and Raman sideband cooling (see Sec.~\ref{sec:trap:cool:subdoppler}), to achieve near-ground-state state occupation~\citep{Kaufman2012, Andersen2022}.

The trapping of single atoms in optical tweezer arrays is a foundational technique for quantum information and simulation platforms~\citep{Browaeys2020, Gieseler2021, Kaufman2021, Daley2023}. Each optical tweezer site is typically loaded with a small random number of atoms (0, 1, or more). To ensure that only one atom remains, the ensemble is illuminated with light that induces light-assisted two-body collisions. Pairs of atoms undergo hyperfine-changing collisions and are ejected from the trap, leaving behind either a single atom or an empty tweezer. This process is often called "collision-induced blockade" and results in a maximum single-atom loading probability of approximately $50\%$ per site~\citep{Frese2000, Schlosser2001, Schlosser2002}. The filling fraction can be enhanced up to $90\%$ by Dark-State Shelving and Gray Molasses schemes~\citep{Kuhr2001, Gruenzweig2010, Lester2015, Fung2015, Brown2019, Shaw2023}. A perfect occupation of a selected set of target traps is achieved by rearranging atoms across the array of tweezers~\citep{Beugnon2007, Miroshnychenko2006, Schlosser2012, Nogrette2014, Kim2016, Endres2016, Barredo2018, Cooper2018, OhldeMello2019, Schymik2020, Glicenstein2021, Schymik2022, Bloch2023}, once or iteratively. After loading, fluorescence imaging (see Sec.~\ref{sec:det:ensemble:fluo}) is used to detect which tweezers contain single atoms and a sorting algorithm calculates the rearrangement steps that are required. Then the desired target sites are populated by moving atoms from occupied auxiliary traps using a set of fast, dynamically steerable traps, allowing defect-free rearrangement and filling of arbitrary array geometries.

\begin{figure}[htb!]
  \centering
  \includegraphics[width=1.0\columnwidth]{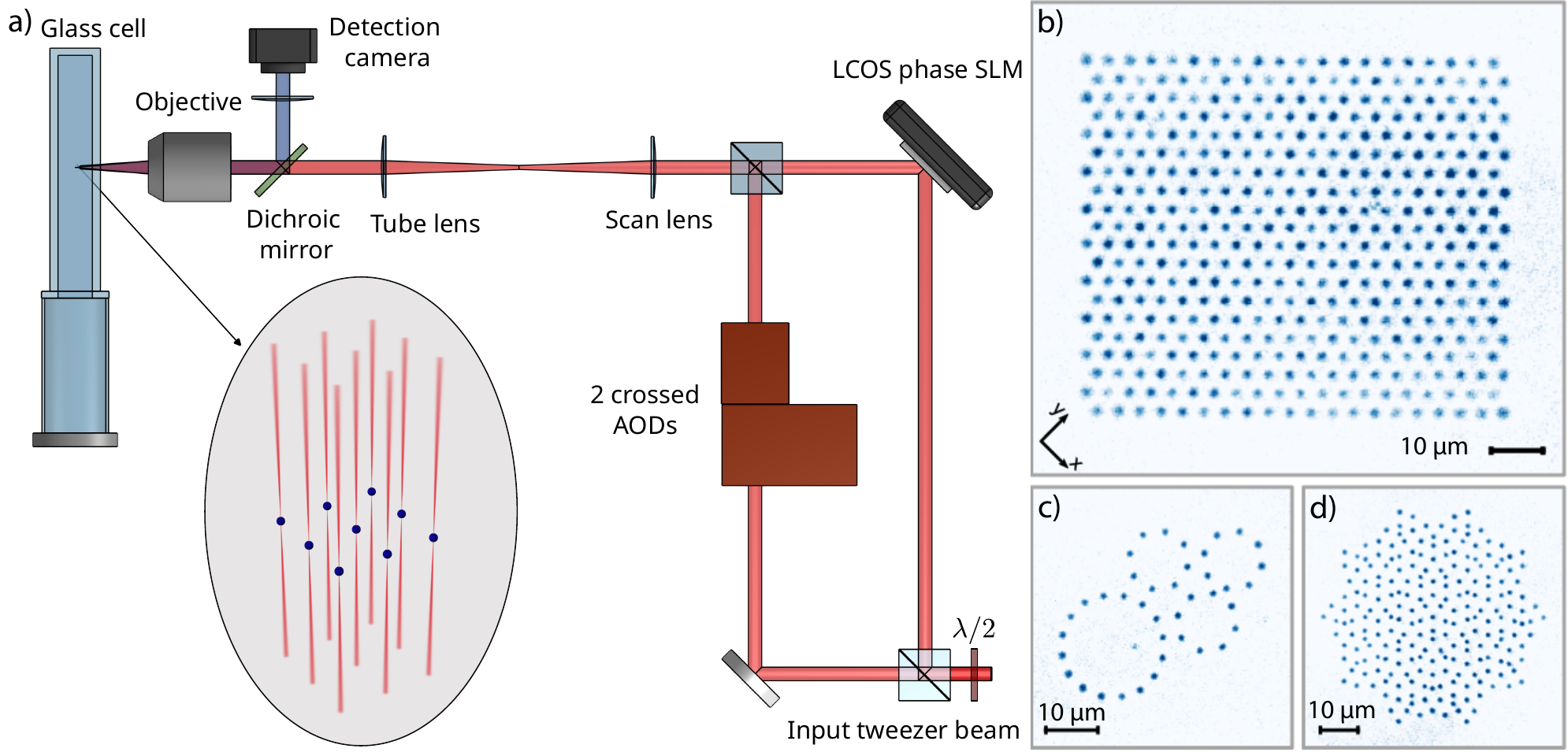}
  \caption{Common optical lattice and tweezer setup: (\textbf{a}) Optical layout for tweezer generation and imaging, (\textbf{b}) Experimental absorption image of a 400 site triangular lattice, where each dark spot corresponds to a microscopic ensemble of $\approx 30$ ultracold \isotope[39]{K} atoms. 
  (\textbf{c}) 40 site ring structure and (\textbf{d}) 226 site Penrose quasicrystal lattice. Panels (\textbf{b,c,d}) are adapted from~\citep{Wang2020}.}
  \label{fig:tweezers}
\end{figure}

The scalability and flexibility of optical tweezer arrays depends on the ability to generate and control multiple traps. This can be achieved primarily by using AODs and phase SLMs (typically LCDs). By modulating in an AOD the amplitudes and frequencies of the RF drive, multiple diffracted beams at different angles can be produced, with each corresponding beam focus serving as an optical tweezer. Real-time adjustments to the RF drive allow dynamic control over tweezer number, positions and depths at $\unit{\mega\hertz}$ rates, which is key for realizing atom rearrangement. A limitation of AODs is that the trap patterns are limited to square or rectangular grid arrays due to their operation principle. On the other hand, phase SLMs can generate arbitrary trap topologies by imparting spatially varying and suitably calculated phase holograms to an incoming laser beam. The resulting diffraction forms a desired pattern of focal spots (tweezers) in the microscope’s focal plane, and can concurrently minimize optical aberrations. Thus, SLMs enable highly flexible and programmable trap geometries, from 1D chains to 2D or even 3D arrays~\citep{Barredo2018, Qiu2020, Kim2020, Kim2022, Schlosser2023, Kusano2025, Gavryusev2024}, with precise control over the number, spacing, and spatial arrangement of sites. The principle of SLM operation allows to move or change concurrently and independently all traps, but typical devices have a phase mask refresh rate of $\qty{60}{\hertz}$, which is not competitive for atom rearrangement relative to the AOD-based method. Recent technical progress has enabled refresh rates up to $\qty{1.4}{\kilo\hertz}$~\citep{Lin2025a, Knottnerus2025}, which provides an increasing advantage over AODs for moving more than a few hundred atoms. Further details on tweezer-based atom motion and their applications are provided in Sec.~\ref{sec:manip:ext:moving}.


Combining AODs and SLMs can yield hybrid platforms with both high-speed dynamical control and arbitrary geometrical versatility. Recent advances
allow for the creation of super-resolved tweezers with subwavelength separation by exploiting light-structuring techniques such as superoscillatory fields, which push beyond standard diffraction-limited configurations~\citep{Rivy2023, Florshaim2025}.




Single-atom trapping using optical tweezers provides exceptional spatial and quantum-state control. Large tweezer arrays, configured using spatial light modulators or microlens arrays, have realized hundreds of individually addressable atoms with site-resolved manipulation and detection capabilities~\citep{Kaufman2021, Schaeffner2024}. Tweezer-based sensor arrays have demonstrated quantum-enhanced measurements, such as optical magnetometry with spatial resolution at the micrometer scale, highlighting their role in next-generation quantum sensors~\citep{Schaeffner2024}. 

Innovations such as steerable tweezers and mid-sequence error detection pave the way toward defect-free quantum registers and reliable quantum operations~\citep{Ebadi2022, Ruttley2024}. A method to extend the connectivity between qubits is based on coherent transport of entangled atom arrays ~\citep{Bluvstein2022} that allows to realize neutral-atom quantum computers with reconfigurable high-fidelity parallel entangling gates~\citep{Evered2023, Bluvstein2023}

Another important application is the realization of spin models, leveraging strong and tunable Rydberg interactions between individually trapped atoms. In such systems, two-level atomic states can be encoded as effective spin-$1/2$ degrees of freedom, with coherent coupling provided by laser driving and interactions mediated by van der Waals or resonant dipole–dipole interactions~\citep{Saffman2010}. This approach enables the direct implementation of spin Hamiltonians such as the transverse-field Ising model, XXZ models, and extensions relevant for quantum magnetism.

One of the most important breakthroughs was the demonstration of quantum Ising dynamics in Rydberg atom arrays, where controlled interactions lead to correlated phases and collective excitation dynamics~\citep{Labuhn2016}. By arranging tweezers in flexible geometries, including one- and two-dimensional arrays, experiments have realized Ising-type spin models with programmable connectivity, allowing the study of long-range antiferromagnetic order and frustrated magnetism~\citep{Bernien2017,Lesanovsky2012}. Subsequent experiments extended these ideas to larger systems of hundreds of atoms, enabling the exploration of quantum critical dynamics and benchmarking against theoretical predictions for quantum phase transitions in strongly interacting spin models~\citep{Ebadi2021, Scholl2021}.

The high degree of control in tweezer arrays also allows for the simulation of Hubbard-type physics, where effective tunneling processes can be engineered through Rydberg dressing or controlled atom rearrangement~\citep{Browaeys2020}. Moreover, recent work has combined single-site addressability with long-range interactions to explore exotic quantum phases, including spin liquids and constrained lattice gauge theories~\citep{Samajdar2021}.

The current state of the art is characterized by rapid scaling to larger system sizes, improved coherence through advanced laser and trapping techniques, and the development of hybrid approaches that integrate tweezer-based platforms with cavity QED or superconducting interfaces. Future applications include the simulation of frustrated quantum magnets, exploration of non-equilibrium spin dynamics, implementation of quantum optimization protocols, and potential use of spin models in neutral-atom quantum processors. By combining microscopic control with programmable interactions, optical tweezers provide a unique path toward addressing outstanding problems in many-body quantum physics and quantum information science.

\subsection{Optical lattices}
\label{sec:trap:optical_lattices}


Optical lattices, formed by the interference between two or more laser beams, create periodic trapping potentials for neutral atoms that closely mimic the crystalline environment of electrons in solids. Thanks to the long coherence time and the absence of defects -- contrary to a real solid -- it is possible to observe the transport of quantum particles.

The simplest implementation is a one-dimensional lattice formed by a retroreflected laser beam, yielding a standing wave potential of the form $V(x)=V_0 \cos^2( 2 \pi x/\lambda)$ that has a spacing of $\lambda/2$, where $\lambda$ is the wavelength of the trapping light, as depicted in Fig.~\ref{fig:CoolTrap}f. More complex geometries are obtained by intersecting two or more beams at defined angles, enabling the creation of square, triangular or three-dimensional cubic lattices with tunable periodicity~\citep{Bloch2005}. Superlattices, quasicrystals, and other engineered structures are also within reach through superposition of multiple wavelengths or relative phase adjustment~\citep{SebbyStrabley2006}.

Atoms confined in these periodic potentials are well described by the Hubbard model, with the key ingredients being the tunneling amplitude $J$ between neighboring lattice sites and the on-site interaction energy $U$ between atoms that occupy the same site. The lattice depth~\citep{Zhou2018a} is controlled by the laser intensity and light detuning from the atomic transition, tunes the balance between kinetic and interaction energies, allowing access to diverse quantum regimes. Using Feshbach resonances (see  Sec.~\ref{sec:manip:interactions:Feshbach}), $U$ can be precisely controlled, and direct exploration of strongly correlated phases is made possible~\citep{Jaksch1998}. 
If the interaction between particles is repulsive ($U>0$), the Hubbard model features a quantum phase transition from a superfluid phase (when $U\ll J$) to an insulating regime (for $U\gg J$). For the superfluid regime the atoms are extended in the single lattice sites, whereas for the insulating regime the wavefunctions are localized in the individual lattice sites. The insulating regime is known as the Mott insulator.
The mapping of optical lattices onto Hubbard physics formed a robust basis for quantum simulation. 

A notable experimental achievement was the observation of the superfluid–Mott insulator transition in a gas of ultracold bosons confined in a three-dimensional lattice~\citep{Greiner2002}. This represented one of the initial realizations of a paradigm condensed matter model in a clean and highly tunable atomic system. Subsequent studies extended this framework to fermionic atoms, allowing one to study fermionic Mott insulators and antiferromagnetic correlations~\citep{Jordens2008, Mazurenko2017}. The advent of quantum gas microscopy further revolutionized the field by allowing site-resolved detection and manipulation of atoms in a lattice, providing direct access to local observables and correlation functions~\citep{Bakr2009, Sherson2010}.

Besides fundamental research, optical lattices are the backbone for some of the most precise quantum technologies, such as optical lattice clocks with frequency uncertainties below $\qty{e-18}{}$~\citep{Nicholson2015}, and atom interferometers based on lattice beam splitters for high precision metrology~\citep{Clade2006}. Future applications will include quantum information processing, quantum-enhanced sensing, and the exploration of strongly correlated matter under synthetic fields, placing optical lattices at the forefront of modern atomic, molecular, and optical physics. 

\subsubsection{Optical lattices for Hubbard physics.}

Optical lattices, which we presented in sec.~\ref{sec:trap:optical_lattices}, provide an extremely well-controlled environment within which paradigmatic examples of condensed matter physics are implemented. Of these, the Hubbard and Fermi–Hubbard models have a particularly important role to play because they reproduce the interplay of interactions and tunneling in lattice systems.
The first observation of this crossover in a Bose gas in an optical lattice~\citep{Greiner2002} was a breakthrough for quantum simulation, and optical lattices are now firmly established as clean and controllable realizations of strongly interacting lattice systems.

For fermions, application of the Fermi–Hubbard model has enabled studies close to the center of condensed-matter physics, such as antiferromagnetism and the physics of high-temperature superconductivity. Initial experiments revealed the observation of fermionic Mott insulating phases in optical lattices~\citep{Joerdens2008,Schneider2008}, demonstrating the first evidence of strongly correlated fermionic matter in ultracold atomic gases. Subsequent experiments investigated short-range magnetic correlations~\citep{Greif2013}, and quantum gas microscopes enabled single-site-resolved imaging of fermions in optical lattices, allowing direct observation of antiferromagnetic correlations in two dimensions~\citep{Mazurenko2017}.

These advances have made ultracold fermions in optical lattices a powerful probe for studying quantum magnetism, non-equilibrium dynamics, and exotic quantum phases~\citep{Tusi2022}. State preparation on demand has allowed nearest-neighbor spin correlations and thermalization to be observed in closed quantum systems~\citep{Chiu2019}. Furthermore, novel techniques like Floquet engineering and synthetic gauge fields have opened up the range of accessible Hamiltonians, allowing experimental realization of topological lattice models~\citep{Jotzu2014}.

The field is currently dominated by the manipulation of fermionic atoms in optical lattices under control, which makes it possible to study doped Hubbard systems, stripe order, and spin-charge separation. Experiments provide insights into high-temperature superconductivity strongly correlated phenomena and quantum magnetism. In the future, prospective applications include the application of Hubbard physics in quantum simulators for probing computational methods, study of out-of-equilibrium quantum many-body physics, and development of novel paradigms for quantum materials engineering. Optical lattices thus remain an interface between atomic and condensed matter physics, with sole privilege of studying models that are at the forefront of quantum science.

\subsection{Evaporative cooling}
\label{sec:trap:evaporative_cooling}
After having introduced the most common trapping techniques, let us now go back to the different cooling methods. While laser cooling, either in a MOT or in a molasses, is capable of bringing atoms to remarkably low temperatures, it is still not sufficient to create a quantum degenerate gas. In order to reach quantum degeneracy, two conditions must be, indeed, satisfied simultaneously: not only must the temperature $T$ be sufficiently low, but also the density $n$ must be sufficiently high, such that the phase space density $n\lambda_{dB}^3$, defined as $n\lambda_{dB} = n\sqrt{2\pi\hbar^2/(mk_BT)}$, is above a critical value of order 1 (more precisely, for obtaining a BEC the critical condition is met when $n/\lambda^3_{db} \approx 2.612$)~\citep{Ketterle1999}. The limitations of laser cooling do not allow to achieve, simultaneously, both the low temperatures and the high densities needed. For this reason, in order to achieve quantum degeneracy, a new cooling method had to be devised, namely evaporative cooling~\citep{Hess1986, Masuhara1988, Davis1995a}.

Evaporative cooling works by selectively removing the most energetic  particles from a trapped, thermal ensemble, and letting the remaining ones thermalize to a lower temperature~\citep{Luiten1996} (Fig.~\ref{fig:CoolTrap}h).  In practice this “knife”-like removal truncates the high-energy tail of the thermal (Boltzmann) distribution; subsequent elastic collisions redistribute energy so that the temperature and the mean energy of the remaining sample fall.

For the evaporation to work, therefore, three different parameters, expressed as characteristic times, have to be balanced: the elastic collision $t_{el}$, the evaporation rate $1/t_{ev}$, and the trap loss rate $1/t_{loss}$. In an evaporation sequence, we want to make sure that the elastic collision rate is at least not decreasing, or, rather, even increasing (this regime is called \emph{runaway evaporation}) after each removal step, as otherwise the evaporation would slow down. It can be shown~\citep{Ketterle1996} that there is a certain threshold value $R_{min}$ of the ratio $R = t_{loss}/t_{el}$ that has to be exceeded in order to have runaway evaporation. The value of $R_{min}$ crucially depends on the trap geometry, expressed a parameter $\delta$, defined as the exponent for which the volume $V$ of the trapped gas obeys $V \propto T^\delta$; the higher $\delta$, the lower $R_{min}$ and, therefore, the easier to achieve is runaway evaporation. For a potential box, $\delta=0$, for an harmonic potential $\delta=3/2$, and for a linear potential $\delta = 3$. The highest runaway evaporation efficiency ensures the temperature decrease per atom lost is optimal; however, the aim of cooling to degeneration is to gain phase space density. It can be also shown that the optimal evaporation trajectory is the one that maximizes, at each instant, the figure of merit $\gamma = \dot{D} N / (D \dot{N})$, where $N$ is the number of atoms and $D$ is the phase space density.

\paragraph{Evaporative cooling in magnetic traps}
Experimentally, evaporative cooling is performed in different ways depending on the type of trap. It was first implemented~\citep{Anderson1995, Davis1995, Bradley1995} in magnetic traps, where atoms are confined in low-field seeking Zeeman sublevels by inhomogeneous magnetic fields, as explained in Sec.~\ref{sec:trap:magnetic_trap}. The fact that the magnetic field is spatially inhomogeneous causes the atoms to have a spatially dependent Zeeman shift in their energy levels.

A radio-frequency (RF) field can be applied to drive spin-flip transitions between different internal states, as described in~\ref{sec:manip:int}, transferring atoms from magnetically trapped Zeeman states to untrapped (or anti-trapped) ones. Because of the spatially-dependent Zeeman shift, this \textit{RF knife} will act in a spatially selective way, removing only those atoms that are in a position of the trap such that their Zeeman-shifted transition frequency matches the RF frequency. Because the Zeeman shift maps position to energy in the trap, a chosen RF frequency corresponds to a precise energy cutoff; sweeping the RF downward continuously lowers this cutoff and forces evaporation. This method was central to the original evaporative-cooling sequences that led to the achievement of BEC in the mid-1990s~\citep{Anderson1995, Davis1995, Bradley1995}.

The advantages of RF evaporation include a precise control of the energy threshold, the ability to shape the evaporation surface in space via the control of the trap geometry and a straightforward implementation for magnetic traps. This accurate control of the evaporation energy threshold allows to precisely control the evaporation rate, and consequently the rate at which the temperature is decreased. This has enabled, for instance, investigations of how the timescales for BEC onset and condensate fraction growth are affected by the rate at which the critical temperature is crossed in the cooling ramp~\citep{Miesner1998, Wolswijk2022}, as well as the study, as a function of the cooling rate, of the formation of different phase domains within the system when the transition is crossed rapidly, eventually leading to the nucleation of topological defects such as vortices~\citep{ Campo2011, Lamporesi2013, Weiler2008, Serafini2015, Donadello2016, Liu2018e, Wolswijk2022}. 


Limitations to RF evaporation in magnetic traps arise from spin-flip collision losses near zeros of the magnetic field (e.g. Majorana losses in quadrupole traps, see Sec.~\ref{sec:trap:magnetic_trap}), from heating due to technical noise in the RF field and from magnetic field fluctuations due, for instance, to variations of the environmental magnetic field (even the movement of a vehicle outside of the laboratory could cause a shift of the magnetic field corresponding to the bottom of the trap, and, with that, a shift of the energy-cutoff addressed by a certain RF frequency). If an accurate control of the final temperature and atomic density is wanted, one has thus to accurately control the RF field frequency and amplitude, as well as the magnetic field felt by the atoms, applying compensation magnetic fields to compensate the variations. Passive magnetic shielding can significantly help to achieve a controlled, low-noise environment, useful especially for experiments performed at low magnetic fields~\citep{Farolfi2019, Rogora2024}. A fundamental limitation of evaporative cooling in magnetic traps, moreover, lies in the requirement that the atoms must be in a magnetically trappable spin state, with an accessible untrapped Zeeman partner. This issue can be tackled by performing evaporation in optical traps.

\paragraph{Evaporative cooling in optical traps: all optical cooling}
In an optical dipole trap (see Sec.~\ref{sec:trap:odt:fort}) atoms are confined in the intensity maximum of the electromagnetic field of a far red-detuned laser. In such a trap, evaporative cooling can be implemented by lowering the trap depth $U$, which can be performed directly by reducing the laser optical power~\citep{Adams1995, Chaudhuri2007, Li2017}. 
This forced lowering may be combined with spatial shaping (e.g. changing beam waists) to control density and collision rates during the evaporation ramp.

Compared to RF evaporation, optical evaporation can work for atoms in any internal state, including states that are insensitive to magnetic fields. Evaporative cooling in optical traps was first demonstrated in 1995~\citep{Adams1995} and in 2001 it enabled \emph{all-optical} formation of an atomic BEC~\citep{Barrett2001}. Using a tightly confining trap to maintain a high phase space density, optical evaporation has proven to be a very efficient technique to rapidly reach quantum degeneracy, in timescales often much shorter than for comparable RF evaporative cooling ramps in magnetic traps~\citep{Barrett2001}. Moreover, while in magnetic traps only specific spin states can be confined, evaporative cooling in optical traps is spin-independent, therefore allowing to simultaneously cool all the spin projections of a given atomic state. This is particularly relevant in experiments on spin mixtures, where effects involving different spin states of the same atom are investigated~\citep{Cominotti2023, Zenesini2024}.

\paragraph{Sympathetic cooling}
The evaporative cooling techniques described above rely on the fact that the atoms rethermalize via \textit{s-wave} collisions (see Sec.~\ref{sec:manip:interactions:short}). This is a problem when considering fermions: identical fermions at ultracold temperatures, indeed, cannot scatter via such collisions due to the Pauli exclusion principle, which strongly suppresses thermalization. However, by immersing fermions into a bosonic reservoir that can be directly evaporated, efficient cooling can nevertheless be achieved. This method of cooling one atomic species thanks to the collisions with particles of a different atomic species is called \textit{sympathetic cooling}.
The underlying principle is that interspecies collisions redistribute energy between the two components until they share a common temperature. The actively evaporated species continuously loses its high-energy atoms, and the sympathetic partner is dragged along in temperature through thermal contact. The efficiency of the process depends critically on elastic cross sections, the relative densities and trap geometries of the two species, and the absence of strong inelastic processes. 

Sympathetic cooling enabled the first observations of Fermi degeneracy in trapped atomic gases, such as~\isotope[40]{K} sympathetically cooled with~\isotope[87]{Rb}~\citep{Truscott2001, Roati2002} and~\isotope[6]{Li} cooled with~\isotope[23]{Na}~\citep{Schreck2001}. Earlier milestones also include the onset of Fermi degeneracy observed directly in~\isotope[40]{K}~\citep{DeMarco1999}, where efficient cross-species scattering permitted thermalization to temperatures well below the Fermi temperature.

\subsection{Miniaturizing the setup: atom chips}
\label{sec:trap:atom_chips}

\begin{figure}[htb!]
  \centering
  \includegraphics[width=0.7\columnwidth]{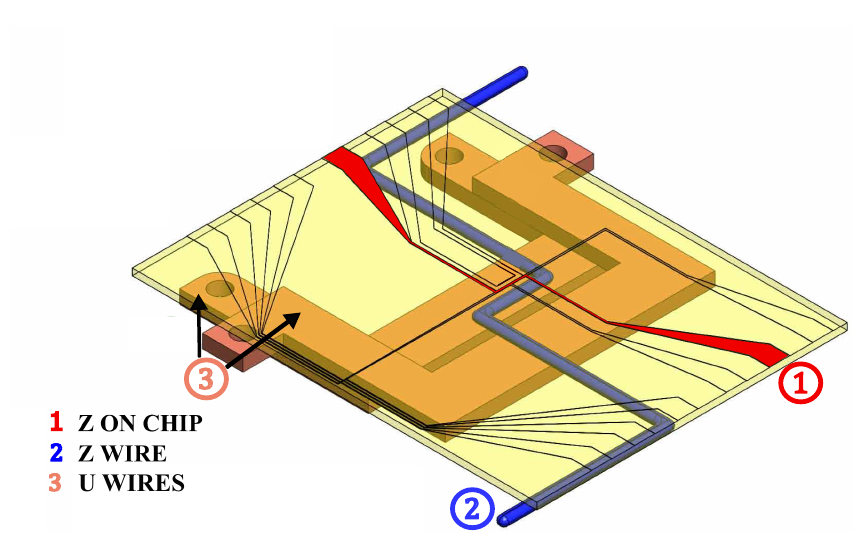}
  \caption{Example of an atom chip, reproduced from~\citep{Petrovic2013}. A microfabricated Z-shaped wire (1) generates a tight Ioffe–Pritchard trap near the surface, while a macroscopic Z-wire (2) provides a larger auxiliary trap for loading. Additional U-shaped wires (3) serve as radiofrequency antennas and generate quadrupole fields.}
  \label{fig:atomchip}
\end{figure}

While traditional cold and ultracold atom experiments usually rely on bulky apparatus, the cooling and trapping techniques discussed above can also be realized on integrated, miniaturized platforms known as atom chips.
Atom chips are microstructures designed to trap, cool, manipulate, and detect cold and ultracold atoms with high precision, using magnetic and, in some cases, electric or optical fields~\citep{Folman2002,Reichel2011}. They are typically produced with semiconductor processing techniques such as photolithography, electron-beam lithography, and thin-film deposition, on planar substrates including SiO\textsubscript{2}, GaAs, sapphire, or alumina~\citep{Reichel2011}. These methods allow the realization of microscale wire patterns that generate spatially varying magnetic fields, forming localized trapping potentials capable of confining atoms only a few micrometers away from the chip surface. 
Recent advances in nanofabrication technology allow for the integration of current-carrying conductors with optical components—such as micro-optics, photonic crystals, microcavities and optical fibers—on-chip sensors, microwave structures, arrays of microtraps, matter-wave interferometers and double-well potentials~\citep{Wang2005,Schumm2005}. Besides, there is growing experimental work showing that laser beams for addressing, as well as optical tweezers can be integrated alongside the magnetic trapping structures~\citep{Vyalykh,Xu}. This integration allows atom chips to function as compact, multifunctional platforms where trapping, manipulation, and detection can occur within the same device. The term ``atom chip'' reflects indeed the strong analogy with electronic integrated circuits, emphasizing the high degree of miniaturization and versatility that these systems offer~\citep{Folman2002}. 

The operating principle of an atom chip exploits the interaction between the magnetic dipole moment of an atom and an external magnetic field. Compared to conventional macroscopic magnetic coils, atom chips offer several advantages. First, the proximity of the trapping region to the wires (typically a few micrometers) allows for very large magnetic field gradients and curvatures to be achieved with modest currents, leading to extremely tight confinement potentials and correspondingly small ground-state sizes. Second, the compact planar geometry and reduced power dissipation simplify experimental setups used in traditional BEC experiments~\citep{Folman2002,Reichel2011}. A notable consequence of this stronger confinement is the significant acceleration of BEC production, with condensation times reduced by up to one order of magnitude compared to conventional techniques~\citep{Hansel2001a}. This improvement results from the increased atomic density, higher collision rates, and faster thermalization, which together make evaporative cooling far more efficient.

Figure~\ref{fig:atomchip} illustrates an atom chip fabricated on a silicon substrate coated with a patterned gold layer using photolithography~\citep{Petrovic2013}. At the center of the chip, a micrometer-scale Z-shaped wire (1 in the figure) generates a three-dimensional Ioffe–Pritchard trap with a nonzero magnetic minimum. Near the trap center, the potential is approximately harmonic, while the strong field gradients near the wire ensure tight confinement. Due to the dimensions of the wire, this microtrap has only a limited capture volume.  Atoms are, therefore, first loaded into a larger trap created by a millimeter-scale Z-shaped wire (2 in the figure) mounted beneath the chip. Thanks to its larger cross-section, this wire can sustain higher currents and provides a trap of larger volume, located farther from the chip surface. Once precooled in an optical molasses, atoms can be transferred from this auxiliary trap into the tighter on-chip microtrap. Additional U-shaped conductors (3 in the figure) play a dual role: when combined with an external bias field, they approximate a quadrupole configuration, replacing the conventional quadrupole coils, and they also serve as radiofrequency  antennas for evaporative cooling and coherent manipulation of atomic states.

It is worth emphasizing that the proximity of atoms to the chip surface also gives rise  to additional sources of decoherence, leading to technical limitations.
Since tight confinement requires atoms to be trapped very close to the chip surface, typically within a few micrometers or even hundreds of nanometers, atom–surface interactions become significant. These include decoherence effects induced by Johnson noise, Casimir–Polder forces, and magnetic field fluctuations~\citep{Henkel1999,Henkel2003,Lin2004}. Such effects impose a lower limit on the atom–surface distance and ultimately constrain the timescales of coherent quantum manipulation~\citep{Treutlein2008}. Moreover, heat dissipation in metallic wires limits the achievable currents and, consequently the maximum trap depths. Nonetheless, through careful optimization of chip materials and fabrication methods, it has become possible to trap atoms at submicron distances from the surface while maintaining coherence~\citep{Lin2004,Wang2005,Schumm2005}.

The concept of employing planar microfabricated wires to create tightly confining magnetic traps for cold atoms was first proposed in 1995 by Weinstein and Libbrecht~\citep{Weinstein1995}. A significant experimental milestone followed in 1999, when Reichel, Hänsel, and Hänsch demonstrated the first trapping of cold atoms using surface-mounted magnetic microtraps~\citep{Reichel1999}. The term \emph{atom chip} was shortly afterward introduced by Folman et al. in 2000~\citep{Folman2000}, and in 2001, the first Bose–Einstein condensates of~\isotope[87]{Rb} atoms on an atom chip were independently realized by Hänsel et al. in Munich~\citep{Hansel2001} and Ott et al. in Tübingen~\citep{Ott2001}. 
Since then, atom-chip technology has progressed well beyond simple magnetic trapping, emerging as a versatile platform for both fundamental studies and applied quantum technologies.

One important development has been the integration of optical cavities directly onto the chip, enabling strong coupling between ultracold atoms and photons and allowing cavity quantum electrodynamics to be explored in compact setups~\citep{Purdy2008}. Another significant advance is the use of superconducting materials, which suppress magnetic noise from the chip surface and allow atoms to maintain quantum coherence for much longer times than near normal metal conductors~\citep{Kasch2010}. On the application side, atom chips are now employed in compact interferometers, where ultracold atoms serve as extremely sensitive probes of gravity and inertial forces; experiments have demonstrated that such devices can measure gravitational acceleration with high precision in a volume of only a few cubic centimeters~\citep{Abend2016}. Furthermore, chip-based architectures have enabled multi-state interferometers and precision studies of atom–surface interactions, highlighting the potential of these systems for both quantum sensing and fundamental physics~\citep{Petrovic2013}. More recently, atom-chip technology has been extended to microgravity environments, where compact setups have successfully produced Bose–Einstein condensates in space, demonstrating the feasibility of creating and manipulating ultracold atoms in microgravity~\citep{Becker2018}. Together, these advances illustrate the ongoing evolution of atom chips toward integrated, noise-resilient, and application-ready quantum devices.

\begin{table}[ht!] 
\centering
\caption{Overview of cooling and trapping techniques and their applications.}
\label{tab:CoolTrap}
\small
\begin{tabularx}{\textwidth}{X X X X}
\toprule
\textbf{Methods} & \textbf{Benefits} & \textbf{Limitations \& Requirements} & \textbf{Main applications} \\
\midrule

\textbf{Laser cooling} & & & \\
\multirow[t]{3}{*}{Doppler cooling}
& - Cool atoms from $\sim$K to $\sim$mK temperatures & - Suitable transition & - Basic building block for cold atoms experiments~\citep{Schreck2021}\\
& - Ease of implementation & - Relatively high achievable temperatures (Doppler limit) &  \\
\multirow[t]{3}{*}{Sub-doppler cooling}
& - Lower achievable temperatures (below Doppler limit) & - Recoil temperature & - Cool to low temperatures~\citep{Lett1988}\\
& & - Dedicated laser lines for some configurations (e.g. a D1 laser for Li) &  \\
\multirow[t]{3}{*}{Magneto-optical traps}
& - Simultaneous cooling and trapping & - Magnetic field coils & - First stage for cold atoms experiments~\citep{Raab1987}\\
& - High capture velocity & - Limited achievable densities (radiation trapping) &  \\
\midrule

\textbf{Magnetic traps} & & & \\
\multirow[t]{3}{*}{}
& - High depth & - Only low-field seeking states trappable & - Trapping atomic clouds~\citep{Migdall1985}\\
& - Large spatial extension & - Majorana losses &  - RF evaporation~\citep{Anderson1995}\\
& - Enable RF cooling & - Difficulty of reconfiguration &  \\
\midrule

\textbf{Optical dipole traps} & & & \\
\multirow[t]{3}{*}{Far-off-resonance traps}
& - Ease of realization & - Shallow trap depth & - Trapping and coarse manipulation of atomic clouds~\citep{Grimm1999}\\
& - Attractive or repulsive potentials & - High-power laser sources &  \\
\multirow[t]{3}{*}{Tailored optical potentials}
& -Tunable trap profile & - Limited trap height & - Tailored and dynamical optical potentials~\citep{Gauthier2021}\\
& - Dynamical shaping of the potential & - Finite refresh rate &   \\
\midrule

\textbf{Optical lattices} & & & \\
\multirow[t]{3}{*}{Common properties}
& -Tunable periodic potentials for neutral atoms & -Stable lasers in frequency and alignment & \\
& -Long coherence time & -Susceptible to heating and decoherence &  \\
& - High controllability in depth, geometry and dimensionality & -Depending on the geometry, atoms can move freely in transverse directions respect to the lattice &  \\
\multirow[t]{3}{*}{1D}
&  &  & Atomic lattice clock~\citep{Takamoto2005}\\
& & & Quantum transport~\citep{Paredes2004}\\
\multirow[t]{3}{*}{2D}
&  &  & Quantum Hall effect~\citep{Zhou2023e}\\

\multirow[t]{3}{*}{3D}
&  &  & Mott insulator transition~\citep{Greiner2002}\\
\midrule

\textbf{Evaporative cooling} & & & \\
\multirow[t]{3}{*}{RF-induced}
& - Precise control of temperature & - Needs good collisional properties & - Cooling to quantum degeneracy~\citep{Anderson1995}\\
& - Possibility of shaping evaporation surface  & - Needs magnetic trap &  \\
& - Easy to implement in a magnetic trap & - RF noise induced heating &  \\
\multirow[t]{3}{*}{All-optical}
& - Insensitive to internal state & - Needs optical trap & - Cooling to quantum degeneracy~\citep{Barrett2001}\\
& - Faster than RF-induced & - Limited by optical trap depth &  \\
\bottomrule
\end{tabularx}
\end{table}


\section{Manipulation}
\label{sec:manip}




In this section, various modern ultracold atom manipulation techniques are illustrated, highlighting the extreme level of control that can be achieved. Atomic state manipulation can be remarkably extended to essentially all degrees of freedom. It is for example possible to control the motion of the atomic center of mass at the sub-micron scale using optical lattices and optical tweezers, while preserving the coherence of the ensemble wavefunction. Similarly, momentum states can be manipulated at the quantum level through Bragg diffraction.
At the same time, it is possible to produce virtually any superposition of internal states using coherent processes involving Rabi oscillations through stimulated Raman transitions or single-photon transitions. These enable, for example, advanced sensing techniques through Ramsey and Mach-Zehnder interferometers. These in turn allow for the development of next-generation atomic clocks, inertial sensors that can probe fundamental physics or the creation of synthetic gauge fields.
Furthermore, the ability to tune atomic interactions, most notably through Feshbach resonances, makes the study of complex many-body quantum phenomena possible. 
Table~\ref{tab:StatManip} summarizes the key information on these topics.


Section~\ref{sec:manip:ext} treats techniques for the manipulation of external degrees of freedom through optical lattices and tweezers (Section~\ref{sec:manip:ext:moving}), appropriately shaped laser beams (Section~\ref{sec:manip:ext:dynamicMan}) and Bragg transitions (Section~\ref{sec:manip:ext:bragg}). Manipulation techniques for the atomic internal states are discussed in Section~\ref{sec:manip:int} with particular reference to Rabi oscillations (Section~\ref{sec:manip:int:Rabi}), Raman transitions, Ramsey interferometry and spin-echo techniques (Section~\ref{sec:manip:int:RamanRamsey}). The simultaneous control of internal and external degrees of freedom is discussed in Section~\ref{sec:manip:intext}. This involves stimulated Raman transitions (Section~\ref{sec:manip:intext:Raman}) and single-photon transitions (Section~\ref{sec:manip:intext:1ph}). The important application to the study of synthetic gauge fields is illustrated in Section~\ref{sec:manip:intext:SynthGauge}. Atom interferometry techniques for inertial sensing and timekeeping are outlined in Section~\ref{sec:manip:AtomInterf} with a focus on matter-wave interferometry, atomic clocks (Section~\ref{sec:manip:AtomInterf:atomclock}), clocks and interferometers in space (Section~\ref{sec:manip:AtomInterf:space}). Finally, section~\ref{sec:manip:interactions} illustrates the control and exploitation of interatomic interactions, namely short-range interactions (Section~\ref{sec:manip:interactions:short}), interaction tuning through Feshbach resonances (Section~\ref{sec:manip:interactions:Feshbach}), long-range and anisotropic interactions (Section~\ref{sec:manip:interactions:long}), and light-mediated interactions (Section~\ref{sec:manip:interactions:cavity}).

\subsection{Control of external degrees of freedom}
\label{sec:manip:ext}

Precise control over the external degrees of freedom of atoms—namely, their position and momentum—is an indispensable tool for the development of quantum technologies. Manipulating the motional state allows a wide range of applications, from building scalable quantum computer architectures to simulating complex condensed-matter phenomena and creating high-precision sensors.

In this section, we will explore the main techniques developed to exert accurate control over the motion of ultracold atoms. We will first discuss methods for the controlled transport of single atoms and atomic ensembles, such as moving optical lattices and optical tweezers, which are fundamental for dynamically assembling and reconfiguring quantum systems. Subsequently, we will cover dynamic manipulation techniques, such as phase imprinting and the use of shaped optical potentials, which allow the motional state to be perturbed and engineered in situ to study phenomena like superfluidity. Finally, we will analyze Bragg transitions, a powerful mechanism for transferring momentum with high precision, which is essential for atom interferometry and many-body spectroscopy.

\subsubsection{Moving lattices, optical tweezers, transport protocols}
\label{sec:manip:ext:moving}


The controlled transport of ultracold atomic ensembles and individual atoms is a fundamental capability for quantum science and technology. Efficient and coherent transport enables flexible preparation of initial states, dynamic rearrangement of qubits, and integration of cold atoms with other quantum platforms.

Early demonstrations relied on moving magnetic traps, where atoms confined in quadrupole or Ioffe--Pritchard configurations were displaced mechanically or by varying magnetic field gradients. Such traps enabled transport over centimeter distances with minimal heating, providing a pathway to connect spatially separated regions in atom-chip experiments and hybrid interfaces~\citep{Haaensch2001,Fortagh2007}. Optical lattices, created by interfering laser beams, introduced another powerful transport mechanism. By dynamically shifting the relative phase of the lattice beams, ensembles of atoms can be coherently translated across the lattice sites. This approach has been used to realize atom conveyors and to implement spin-dependent transport, laying the groundwork for experiments in quantum simulation and quantum walks~\citep{Schmid2006,Mandel2003}.

The advent of optical tweezers brought unprecedented control over single atoms. Individual atoms trapped in tightly focused laser beams can be moved by steering the tweezer position using AODs, SLMs or fast optical modulators. This technique allows for the reconfiguration of arrays of neutral atoms with single-site resolution~\citep{Barredo2016,Endres2016}. Deterministic rearrangement protocols, in which defects in initial random loading are corrected by moving atoms to empty sites, now achieve nearly defect-free filling of large arrays. Recent work has demonstrated efficient transport of hundreds of atoms with negligible heating and loss~\citep{Ebadi2021}.

Single-atom transport has also reached the regime of quantum coherence preservation. By carefully optimizing transport protocols---for example, using adiabatic ramps, polynomial trajectories, or shortcuts to adiabaticity---it is possible to minimize motional excitation during motion~\citep{Kaufman2014}. Such optimization is critical for quantum information processing, where coherence and motional ground-state population must be preserved during repeated operations.

Optimal control in phase space can be applied to transfer of thermal atoms in optical traps in a minimum time~\citep{Morandi2025}.

Among the most important results are the development of large-scale defect-free tweezer arrays~\citep{Barredo2016,Endres2016,Ebadi2021}, coherent transport of single atoms for entangling gate operations~\citep{Kaufman2014}, and hybrid transport schemes combining optical and magnetic methods for interfacing with solid-state or photonic devices~\citep{Thompson2013}. These advances highlight the versatility of atomic transport as both a technical tool and a scientific subject.

Currently, state-of-the-art platforms routinely transport hundreds of atoms across micrometer to millimeter scales with high efficiency and low heating. Active areas of research include extending transport to larger system sizes, integrating transport with entangling operations, and exploring collective transport protocols in many-body settings. Future applications will likely involve scalable defect-free architectures for quantum computing, programmable reconfiguration of quantum simulators, and precision assembly of hybrid quantum systems.

The transport of ultracold trapped atoms has been extensively studied in recent years, with particular emphasis to  optimization protocols and efficient high-fidelity transfer.
Several approaches have been explored for trapped ions, with a focus on strategies that minimize the vibrational excitation and the  decoherence during the transfer process. Transport optimization has been implemented in ion traps to obtain fast shuttling with minimal heating effects~\citep{Huber2008,Bowler2012,Walther2012}. Optimal transport has also found several applications to ultracold atoms in optical dipole traps~\citep{Couvert2008,Chen2010} and lattices, including BECs. Experimental studies have demonstrated the possibility to achieve fast transport of single atoms while preserving quantum coherence, reducing motional excitations, and optimizing loading efficiency in optical lattices~\citep{Rosi2013}. In the context of BECs, non-adiabatic transport of condensates has been explored, revealing strategies to mitigate excitations induced by trap movement~\citep{Jaeger2014}. Theoretical models have analyzed the role of anharmonic potentials and optimized control protocols are now available to enhance transport fidelity and suppress unwanted excitations~\citep{Buecker2013}. Optimal control techniques have been proposed to achieve fast, high-fidelity transport of BECs, ensuring minimal energy cost and robustness against experimental imperfections~\citep{Hohenester2007,Morandi2025}.

\subsubsection{Dynamic manipulation}
\label{sec:manip:ext:dynamicMan}

Far off-resonant light can be used to control the motion of cold thermal atoms and to induce tailored excitations in a quantum gas of degenerate atoms. In particular, atomic motion can be excited either by introducing a density perturbation or a phase perturbation using appropriately shaped laser profiles.

A repulsive or attractive optical dipole potential can be employed to perturb the density. Tightly focused laser beams are used to steer ultracold atoms to probe their superfluid behavior~\citep{Raman1999, Miller2007, Weimer2015}, to excite vortices in bulk systems~\citep{Anderson2010, Kwon2021a}, or to generate  currents in ring traps~\citep{Wright2013a, Cai2022b}. The optical potential can be dynamically controlled using  acousto-optic modulators (AOMs), acousto-optic deflectors (AODs), or by programming a spatial light modulator (SLM) in a time-dependent manner (see Sec.~\ref{sec:trap:odt:tailored_odt}). 

Laser light can also be employed to directly modify the phase of the atomic wavefunction $\psi$ through \emph{phase imprinting}~\citep{Burger1999, Yefsah2013}. When an external conservative potential $U$ is applied for a duration $\Delta t$ that is short compared to the atomic system's response time, the atomic wavefunction is subject to a simple unitary evolution, namely:

\begin{equation}
    \psi (\Delta t) = \psi_0 e^{\Delta t U/\hbar},
\end{equation}

where $\psi_0$ is the unperturbed wavefunction. In this regime, thus, the external potential only introduces a phase shift $\Delta \phi = \Delta t U / \hbar$. By suitably controlling the potential $U$ and the imprinting time $\Delta t$, the phase of the wavefunction can be arbitrarily tailored, which in turn allows control over the particle velocity. In fact, the velocity $v$ of a particle with wavefunction $\psi = |\psi|e^{i\phi}$ is connected to the phase gradient by the relation:
\begin{equation}
    v = \frac{\hbar}{m} \nabla \phi.
\end{equation}
By locally shaping the phase of an atomic sample, the velocity of the individual particles can be, thus, precisely controlled. This phase imprinting technique can also be applied to coherently modify the phase of degenerate ultracold gases, such as Bose-Einstein condensates of Fermi superfluids in the BEC-BCS crossover, which are described by a single macroscopic wavefunction. By imposing homogeneous flat potentials, the phase of the atomic system can be locally twisted to excite solitons~\citep{Burger1999, Yefsah2013} or to trigger Josephson oscillations between tunnel-coupled reservoirs~\citep{Luick2020, Biagioni2024}. Alternatively, by shaping $U$ as a linear gradient, a constant velocity can be imparted to the system, for instance to generate persistent currents in a ring trap ~\citep{Zheng2003, Kumar2018c, DelPace2022}.


\subsubsection{Bragg transitions}
\label{sec:manip:ext:bragg}

Bragg transitions are multi-photon transitions that provide an efficient method to control the momentum state of an atomic ensemble. The atoms are illuminated by two counter-propagating laser beams with nearly identical frequencies, creating a moving optical lattice that imparts a controlled momentum kick to the atoms. This process, known as Bragg scattering, is analogous to the scattering of X-rays and neutrons from crystalline planes: the optical lattice plays the role of the crystal planes, from which the atomic matter wave scatters in a manner similar to photons in the solid-state counterpart~\citep{Martin1988,Muller2008}.

To understand the working principle of Bragg transition, let us consider two laser beams with frequencies $\omega_1$ and $\omega_2$, detuned from the atomic resonance and separated in frequency by $\delta = \omega_1 - \omega_2$, interacting with atoms of mass $m$. During the interaction, an atom absorbs a photon from the $\omega_1$ beam and de-excites via stimulated emission into the $\omega_2$ beam, acquiring a net energy of approximately $\simeq\hbar\omega_r$, where $\omega_r=\frac{\hbar k_L^2}{2m}$ is the recoil frequency, and $k_L$ is the laser wavevector. To satisfy energy conservation, the detuning between the Bragg beams should be set to be $\delta = 2 \hbar k_L^2 /m$, with the sign of $\delta$ determining the direction of the momentum transfer. The net effect of the Bragg transition is a change in kinetic energy without affecting the internal state, resulting in a direct acceleration of the atoms.
To minimize the probability of transitions to other atomic internal states, the laser detuning from the ground–intermediate transition should be chosen sufficiently large to suppress single-photon scattering.

Bragg scattering can be employed to accelerate atoms by several $\hbar k_L$ through a sequence of repeated Bragg transitions, as well as to coherently split and recombine atomic wavefunctions in interferometric setups~\citep{Giltner1995,Altin2013}, as discussed in Sec.~\ref{sec:manip:AtomInterf}. Additionally, it can be used for precision spectroscopy of quantum gases, taking advantage of the energy- and momentum-selective nature of Bragg transitions to probe the excitation spectrum with high accuracy~\citep{Stenger1999, Veeravalli2008, Biss2022}. 

\paragraph{Opto-mechanics}
\label{sec:manip:ext:bragg:optomech}

A field where various techniques developed in atomic physics are used and further advanced is optomechanics. Its goal is as ambitious as it is fundamental: to bring mechanical systems of vastly different scales, with masses ranging from grams to zeptograms, to behave according to the laws of quantum mechanics~\citep{Aspelmeyer2014}.

One of the greatest challenges is cooling these objects, such as vibrating membranes or small levitated spheres, to temperatures close to absolute zero, which is essential to suppress thermal motion and make them sensitive to quantum effects. To this end, cavity laser cooling techniques - analogous to those used for cold atoms - are employed, where the interaction between the laser and the mechanical oscillator provides an effective optical friction that slows the motion of the resonator towards its motional ground state. This approach has been successfully applied to both vibrating membranes~\citep{Peterson2016,Rossi2018,Chowdhury2019, Galinskiy2020, Saarinen2023} and levitating systems, such as nanospheres confined in optical cavities. In particular, coherent scattering cooling~\citep{Delic2019, Delic2020,Ranfagni2020} has proven to be an extremely effective method for lowering the temperature of levitated nanospheres, enabling the exploration of their quantum behavior with an unprecedented degree of control. Further control is provided by optical feedback techniques~\citep{Magrini2021}.

Beyond simple cooling, optomechanics has enabled the preparation of purely quantum states in objects much larger than individual atoms. An example is the realization of squeezed states of a mechanical oscillator's motion~\citep{Chowdhury2020,Vezio2020}. These techniques have also allowed to observe the regime of strong coupling between mechanical modes and optical cavity modes, creating novel quasi-particles known as vectorial polaritons~\citep{Ranfagni2020}, confirming that quantum interactions can dominate even at the mesoscopic scale. Such advances pave the way for new applications in ultra-precision metrology and for experimental tests of fundamental physics, including the realization of matter-wave interferometers with levitated nanometric systems.

\subsection{Control of internal states}
\label{sec:manip:int}

The coherent manipulation of quantum internal states is central to ultracold atom and molecule experiments, enabling both the preparation and control of quantum states for fundamental studies and technological applications. Internal states can be manipulated using laser, microwave, or radio-frequency (RF) radiation, depending on the relevant energy level structure of the system under study. These tools form the basis for the the implementation of protocols such as simple two-level Rabi oscillations, Ramsey and spin-echo techniques, together with more complex dynamical decoupling methods. 

State manipulation also intersects closely with detection techniques, such as state-selective imaging (See Section~\ref{sec:det:statesel}).

Recent years have seen dramatic advances in this field, including the realization of high-fidelity single- and multi-qubit gates in optical tweezer arrays~\citep{Saffman2016}, the control of nuclear spin states in alkaline-earth atoms for optical clock applications~\citep{Nicholson2015}, and the development of robust error-resilient control sequences in noisy environments. The current state of the art demonstrates coherence times exceeding seconds, gate fidelities approaching the fault-tolerant threshold, and flexible protocols for simulating spin models and many-body dynamics.

\subsubsection{Rabi oscillations}
\label{sec:manip:int:Rabi}

\begin{figure}[htb!]
    \centering
    \includegraphics[width=1\textwidth]{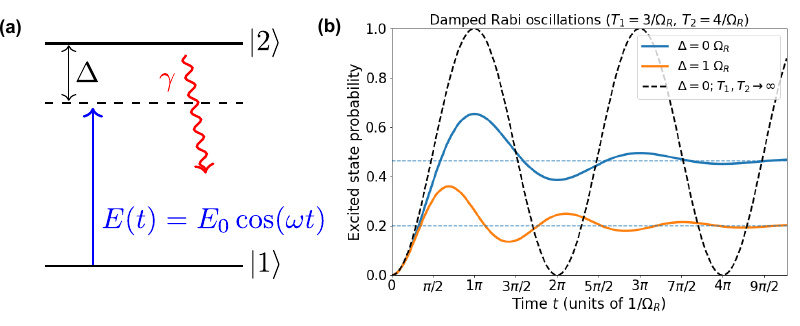}
    \caption{Two-panel figure: (\textbf{a}) schematic of the two-level system, (\textbf{b}) Rabi oscillations showing detuning and damping effects.}
    \label{fig:Rabi_combined}
\end{figure}

At the most fundamental level, the manipulation of a two-level quantum system can be described by coherent driving with resonant radiation~\citep{Foot_Book, AllenEberly1975,  ScullyZubairy1997}. When such system is driven by a near-resonant field, as illustrated in Figure~\ref{fig:Rabi_combined}(a), the population oscillates between the two states at the so-called Rabi frequency.

A description of the problem can be treated in the semi-classical picture, with a Hamiltonian which is naturally divided into the atomic part and the atom–field interaction. The total wavefunction can be expressed as a superposition of the two unperturbed atomic states $\ket{1}$ and $\ket{2}$. When the system is driven by the oscillating electric field, the coupling is determined by the dipole operator, and the strength of the coherent interaction is characterized by the Rabi frequency $\Omega_R$, which depends on the dipole matrix element and the field amplitude. 

A detailed derivation can be found in standard textbooks~\citep{CohenTannoudji1998}, but the essential point is that, after applying the rotating-wave approximation (RWA), one obtains coupled equations for the atomic amplitudes that lead to population oscillations between the two states (see black dashed lines in Fig.~\ref{fig:Rabi_combined} (b)). These oscillations are governed by the detuning $\Delta=\omega-\omega_0$ between the field frequency $\omega$ and the atomic resonance $\omega_0$. Assuming the atom is initially in the ground state, the probability of finding it in the excited state at time $t$ is given by

\begin{equation}
    P_2(t)= \frac{\Omega_{R}^2}{\Omega_{R}^2+\Delta^2}\,\sin^2\!\left(\tfrac{\sqrt{\Omega_{R}^2+\Delta^2}}{2}\,t\right)
    \label{eq:solution pop2}
\end{equation} 

while the ground-state probability is $P_1(t)=1-P_2(t)$. This coherent exchange of population between the two states is known as \emph{Rabi oscillation} or \emph{Rabi flopping}.

Rabi oscillations form the basis of coherent quantum control, but their visibility is limited by decoherence processes, such as spontaneous emission, inhomogeneous broadening, and magnetic or optical field fluctuations. Understanding and mitigating these effects has been central to improving the fidelity of state manipulation. To account for decoherence and relaxation, one introduces the optical Bloch equations, which include dissipation in addition to the coherent driving~\citep{CohenTannoudji1998}. The longitudinal relaxation time $T_1$ describes population decay, while the transverse relaxation time $T_2$ accounts for dephasing of the atomic coherences. Their combined effect results in damped Rabi oscillations, in which the amplitude gradually decreases and the system approaches a steady state, as illustrated in Fig.~\ref{fig:Rabi_combined} (b). 

More complex pulse sequences, such as numerical optimal control strategies like dCRAB (dressed chopped random basis) optimization~\citep{Caneva2011}, allow for tailored control of decoherence and robust state manipulation in noisy environments.

\begin{figure}[htb!]
  \centering
  \includegraphics[width=0.75\textwidth]{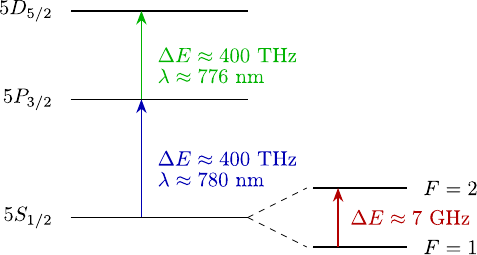}
  \caption{Simplified electronic structure of \isotope[87]{Rb} showing orders of magnitude for optical (blue and green) and ground-state hyperfine (red) transitions.}
  \label{fig:RabiFig4}
\end{figure}

The formalism just presented can be applied across different frequency ranges, from optical transitions driven by lasers to microwave or RF transitions used in nuclear magnetic resonance (NMR) or electron spin resonance (ESR) experiments~\citep{NielsenChuang2010}. In atomic physics, optical dipole transitions are addressed by laser fields coupling ground and excited electronic states, whereas in the microwave regime one typically drives hyperfine or Zeeman transitions in the ground state manifold, as illustrated for the example case of \isotope[87]{Rb} in Fig.~\ref{fig:RabiFig4}.

\subsubsection{Raman transitions, Ramsey interferometry and spin-echo technique}
\label{sec:manip:int:RamanRamsey}

\begin{figure}[htb!]
    \centering
    \includegraphics[width=1\textwidth]{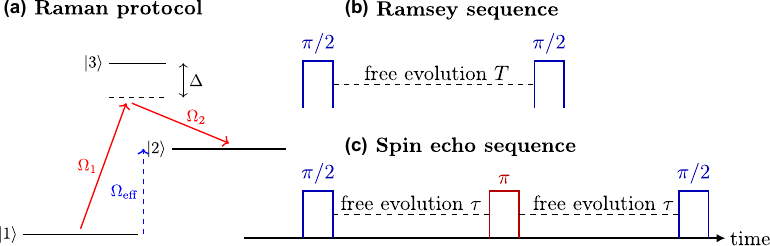}
    \caption{(\textbf{a}) Three-level $\Lambda$ scheme for a Raman transition. Two ground states $\ket{1}$ and $\ket{2}$ are coupled via an off-resonant excited state $\ket{3}$ by two laser fields detuned by $\Delta$. (\textbf{b,c}) Schematic sketch of pulse sequences for Ramsey interferometry and spin echo.}
    \label{fig:RamanRamseySpinecho}
\end{figure}

In addition to single-photon transitions, \textit{Raman transitions} schemes are widely employed, especially when direct optical transitions are impractical or forbidden. In this method, two laser fields couple two long-lived levels via an off-resonant excited state, yielding an effective two-level system with a tunable effective Rabi frequency~\citep{Kaufman2012}

\begin{equation}
    \Omega_\text{eff} \propto \frac{\Omega_1 \Omega_2}{\Delta}
\end{equation}
where $\Omega_{1,2}$ are the single-photon Rabi frequencies and $\Delta$ is the detuning from the intermediate state. This scheme suppresses spontaneous emission while allowing coherent control of transitions that may be dipole-forbidden, such as the intercombination transitions of alkali-earth atoms. Moreover, Raman transitions enable the creation of coherent superpositions with distinct momenta due to the optical momentum transfer 
\(\hbar \Delta k = \hbar (k_1 - k_2)\), where $k_1$ and $k_2$ are the photon wavevectors, see Figure~\ref{fig:RamanRamseySpinecho}(a). This effect is routinely exploited in atom interferometry to spatially separate internal states, as in alkali atoms where a direct microwave transition between hyperfine levels would transfer negligible momentum because of its long wavelength~\citep{KasevichChu1991}. 

Raman transitions offer excellent flexibility, including the ability to drive spin-dependent momentum transfer, enabling atom interferometry, spectroscopy~\citep{KasevichChu1991} and quantum simulation of synthetic gauge fields. However, their limitations include spontaneous emission due to off-resonant scattering, technical complexity of phase stabilization, and the requirement of high laser power for efficient coupling.

Ramsey’s method of \textit{separated oscillating fields}~\citep{Ramsey1950} is one of the most effective techniques for high-resolution spectroscopy, providing a powerful means of probing coherence times and measuring small energy shifts.. 
In its simplest form, two $\pi/2$ pulses separated by a free evolution time $T$ (see Figure~\ref{fig:RamanRamseySpinecho}(b)) create an interference pattern in the excitation probability,
\begin{equation}
    P_2 = \tfrac{1}{2}\left[1+\cos(\Delta T)\right]
\end{equation}
where $\Delta$ is the detuning from resonance. The resulting Ramsey fringes have a separation scaling as $1/T$, which enables frequency measurements of exceptional precision and forms the basis of modern atomic clocks. In atom interferometry, Raman transitions are routinely employed to implement the $\pi/2$ pulses that define the interferometer arms~\citep{KasevichChu1991}.



A complementary use of coherent pulses is found in the \emph{spin echo} sequence, first introduced in nuclear magnetic resonance~\citep{Hahn1950}. When an ensemble of atoms experiences inhomogeneous frequency shifts, coherence decays due to phase dispersion. Insertion of a $\pi$ pulse at time $\tau$ (see Figure~\ref{fig:RamanRamseySpinecho}(c)) reverses the phase accumulation, such that at $2\tau$ the system rephases and coherence is recovered. The amplitude of the recovered signal follows
\begin{equation}
    M(2\tau) \approx M(0)\, e^{-2\tau/T_2}
\end{equation}
where $T_2$ characterizes the homogeneous coherence time. Spin echo sequences are therefore employed to mitigate dephasing mechanisms, extending coherence well beyond the free induction decay time.


\subsection{Simultaneous control of internal and external degrees of freedom}
\label{sec:manip:intext}

Although in many contexts one wishes to control either the internal atomic states or the external (motional) degrees of freedom, the basic atom-light interaction, of course, involves both internal and external degrees of freedom. In laser cooling this feature is used to reduce the ensemble temperature. Here instead we consider the case where the relevant atomic levels are long-lived. Laser radiation that couples two of these atomic levels couples, for free atoms, to both internal and external degrees of freedom. It is then possible to perform a coherent control of state superpositions of the form $\alpha\ket{a,p_a}+\beta\ket{b,p_b}$, where $a$ and $b$ label internal atomic states and $p_a,p_b$ are the respective momenta. It is interesting to note that (a) the internal and external degrees of freedom are entangled and (b) the selective detection of the internal state therefore automatically reveals also the momentum state. This last feature is very useful in applications involving matter-wave interferometers as we will see.

\subsubsection{State manipulation with stimulated Raman transitions}
\label{sec:manip:intext:Raman}

Since optically excited atomic levels usually have lifetimes of the order of a few nanoseconds, a direct single-photon transition would cause the superposition to collapse, in this time, to the internal ground state. Stimulated Raman transitions offer a convenient alternative since they couple two lower-energy states through absorption of a photon from one beam and stimulated emission into another other beam while both fields are sufficiently detuned from a single-photon transition. If the two Raman beams are aligned to propagate in opposite directions and define their wavenumbers as $k_1$ and $k_2$, the Raman transition couples a state of the form $\ket{g_1,p}$ with $\ket{g_2,p+\hbar(k_1+k_2)}$, where $g_1$ and $g_2$ are two internal states corresponding, for example, to the ground state hyperfine structure of an alkali atom. This kind of manipulation closely resembles that obtained through Bragg diffraction (Section~\ref{sec:manip:ext:bragg}), with the important difference that also the internal atomic state changes. 

The long lifetime of the states involved in a Raman transition means that, in practice, the linewidth of the transition is proportional to the reciprocal of the Raman pulse duration $\tau$. Because of the Doppler effect, only a certain velocity class of width $\propto \tau^{-1}$ interacts with the Raman beams. This mechanism can be used to perform velocity selection in a sample of cold ($\mu$K) atoms~\citep{Moler1992}, where the central velocity component can be tuned at will, owing to the Doppler effect, by varying the frequency difference between the Raman lasers. This difference can usually be referenced to RF or microwave sources and therefore be made very stable. 

As in standard Raman transitions, the fact that two internal states are involved means that AC Stark shifts potentially affect the level energies differently. By appropriately choosing the detuning from the optically excited states it is possible to cancel this effect to first order~\citep{Sorrentino2014}.

Stimulated Raman transitions have played a central role in the development of matter-wave interferometers for intertial sensing as is discussed in Section~\ref{sec:manip:AtomInterf}.

\subsubsection{State manipulation with single-photon transitions}
\label{sec:manip:intext:1ph}

Despite the fact that in most atomic transitions the excited state lifetime is too small to preserve the coherence for scientifically interesting applications, some special transitions exist in which the excited state has a lifetime on the order of a few $\mu$s or even minutes. This allows for the manipulation of the combined internal and external states with the simplest single-photon transition. While this appears as the basic textbook topic on matter-light interaction, its actual implementation, particularly for freely falling atoms, is technically challenging. The long lifetime of the excited state implies that the Rabi frequency, proportional to the square root of the natural linewidth $\Gamma$ is also very small. Since this in turn requires relatively long pulses, as seen for Raman transitions, the velocity selectivity reduces the overall efficiency. To counteract these losses it is then necessary to either increase the optical intensity or reduce as much as possible the atomic temperature. Additionally, in order to efficiently address such a narrow single-photon transition, it is necessary to operate with low-phase noise lasers. Interestingly, the same kind of velocity selection that is possible in Raman transitions can be realized also with single-photon transitions. Since, however, the frequency stability of optical sources is rarely as good as that of the microwave counterpart, frequency fluctuations usually translate into changes in the selected velocity through the Doppler effect.  This kind of manipulation can be performed with ultra-narrow transitions such as the $^1 S_0$-$^3 P_0$ optical clock transition of alkaline-earth or alkaline-earth-like atoms, in conjunction with a laser referenced to a high-finesse stable cavity. 

\subsubsection{Synthetic gauge fields for neutral atoms}
\label{sec:manip:intext:SynthGauge}

Beyond reproducing and investigating conventional condensed-matter phenomena, a primary goal of quantum simulations with ultracold atoms is to engineer \emph{synthetic gauge fields} that emulate the action of magnetic fields on charged particles~\citep{Dalibard2011,Goldman2016,Cooper2019}. This capability enables the exploration of novel topological states of matter, such as strongly correlated topological phases, in a clean and highly tunable environment. Despite being electrically neutral, synthetic magnetic fields can be implemented in ultracold atomic systems through a variety of protocols.

The rotation of ultracold atomic gases provided the earliest route toward the realization of effective magnetic fields. In this scheme, the Coriolis force in the rotating frame plays the role of the Lorentz force, enabling the emulation of Landau levels once the rotation frequency matches the trapping frequency. This approach led to landmark observations such as Abrikosov vortex lattices in Bose--Einstein condensates, where interactions favor a regular triangular arrangement of vortices~\citep{Abrikosov1957,Madison2000}. More recently, advances in imaging and potential shaping have enabled renewed studies of rotating gases, including the preparation of states confined to the lowest Landau level~\citep{Fletcher2021}, the observation of chiral edge modes in sharp box potentials~\citep{Yao2024}, and the investigation of interaction induced vortex crystallization~\citep{Mukherjee2022}. A recent study has demonstrated the realization of a two-atom Laughlin state, achieved with a pair of rapidly rotating fermionic atoms~\citep{Lunt2024}. Note that an off-resonant laser field coupled to atomic internal states can be also employed to generate synthetic magnetic fields in the continuum, where the laser-dressed states become position-dependent due to the spatial variation of the laser electric field~\citep{Lin2009,Nascimbene2025}.

Atomic gases confined in optical lattices offer a highly versatile platform for simulating a wide range of discrete lattice models~\citep{Bloch2008,Bloch2012}. Of particular interest are lattices featuring nontrivial topological bands, which require complex-valued hopping amplitudes~\citep{Goldman2016,Cooper2019}. In conventional optical lattices formed by the dipole trapping of off-resonant standing waves, atoms tunnel between sites via spontaneous tunneling processes, characterized by real-valued transition amplitudes, which do not generate effective magnetic flux~\citep{Jaksch1998}.

Laser-assisted tunneling provides a way to overcome this limitation. A linear energy offset $\hbar \Delta$ is introduced between neighboring sites to suppress tunneling along a chosen direction. Resonant tunneling is then restored by applying a pair of far-detuned running-wave laser beams having a frequency difference that matches precisely the energy offset $\Delta$. Since the two beams propagate along orthogonal directions, they also imprint a spatially varying phase pattern onto the atoms~\citep{Aidelsburger2013,Miyake2013,Goldman2014,Goldman2014a}. The resulting hopping amplitude acquires a complex Peierls phase proportional to the momentum difference of the laser beams~\citep{Jaksch2003,Dalibard2011}. This phase acts as a synthetic magnetic flux per plaquette, which can be tuned by adjusting the lattice geometry or laser configuration. Such schemes have enabled the experimental realization of the Harper--Hofstadter model, providing controlled access to topological band structures~\citep{Aidelsburger2015}.

An alternative route to engineering synthetic gauge fields in optical lattices relies on \emph{Floquet engineering}, which uses time-dependent lattice potentials. In this approach, periodic modulation of lattice parameters — such as lattice shaking or on-site energy modulation - induces effective gauge fields by producing a time-averaged Hamiltonian where hopping amplitudes acquire complex phases, simulating magnetic fluxes or spin-dependent interactions~\citep{Goldman2014,Eckardt2017}. Early experiments demonstrated tunable gauge potentials for neutral atoms, enabling control over hopping phases in simple lattice geometries~\citep{Struck2012}, and were extended to implement frustrated spin models in triangular lattices~\citep{Struck2013}. Floquet methods further allowed the realization of topological lattice models, including the Haldane honeycomb lattice~\citep{Jotzu2014} and anomalous Floquet topological systems~\citep{Wintersperger2020}.

A particularly powerful approach for generating artificial gauge fields is based on the concept of \emph{synthetic dimensions}~\citep{Boada2012,Celi2014}, where internal atomic states act as additional lattice sites along a fictitious dimension. Coupling between these internal states is typically realized via Raman transitions (see Sec.~\ref{sec:manip:intext:Raman}), which imprint complex phases on the hopping along the synthetic dimension~\citep{Mancini2015,Stuhl2015}. This method also enables the realization of higher-dimensional topological lattice models within lower-dimensional real-space setups, creating effective spin-dependent gauge potentials and uniform synthetic magnetic fields in the combined real–synthetic lattice~\citep{Lohse2018,Bouhiron2024}. By precisely controlling the Raman coupling strength, detuning, and phase, a wide variety of topological models can be implemented. The synthetic dimension approach provides exceptional flexibility, as it allows independent manipulation of both the internal (spin) and motional degrees of freedom~\citep{Ozawa2019}. This makes it a key platform for investigating interaction-driven topological phases and other many-body topological phenomena in a highly controllable setting~\citep{Nascimbene2025}.

Synthetic gauge fields open a pathway toward realizing strongly correlated topological phases with neutral atoms, where interaction-driven and universal manifestations of the Hall effect were recently unveiled~\citep{Zhou2023e,Leonard2023,Zhou2024,Impertro2025}. These advances not only deepen our understanding of exotic states of matter but also bring us closer to practical goals in quantum technologies, such as the simulation of high-energy gauge theories and the pursuit of fault-tolerant quantum computation via non-Abelian excitations.

\subsection{Atom interferometry techniques}
\label{sec:manip:AtomInterf}

Atom interferometry has undergone remarkable development over the past decades and has emerged as a powerful tool for both fundamental and applied physics. Atom interferometric systems can be seen as the “matter-wave” counterpart of classical optical interferometers, where the mirrors and beam splitters are realized by laser pulses that manipulate the quantum state and motion of atoms. Atom interferometers are highly sensitive to gravitational forces and external fields that modify the potential energy, the internal quantum states, or the kinetic energy of matter waves traveling along different paths within the interferometer. Atomic interferometers are therefore very precise metrological sensors for measuring gravitational acceleration g (gravimeter) or gradients of g (gradiometer).

\begin{figure}[htb!]
  \centering
  \includegraphics[width=0.95\textwidth]{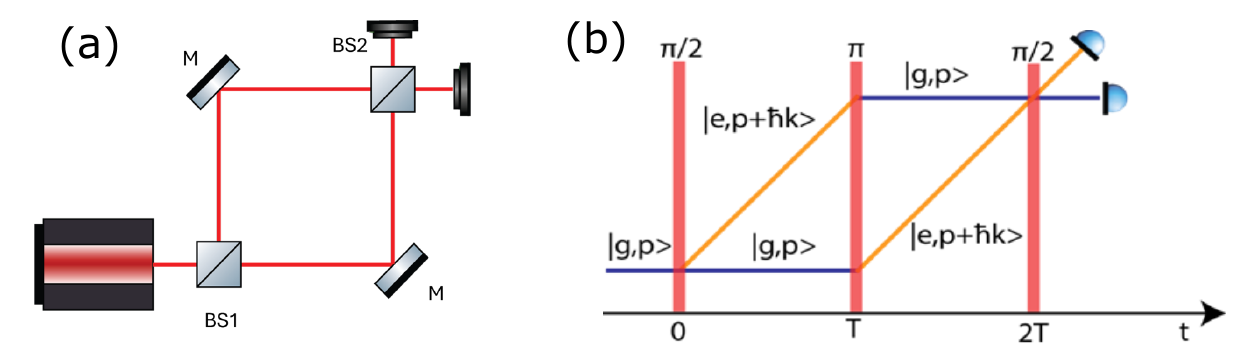}
  \caption{Analogy between the well-known optical interferometer (\textbf{a}), where a laser beam is split and recombined by mirrors (M) and beam splitters (BS1 and BS2), and an atomic Mach–Zehnder interferometer (\textbf{b}), where laser pulses manipulate matter waves, acting as mirrors and beam splitters.
  }
  \label{fig:mz_interferometer}
\end{figure}

Common pulsed configurations, based on Raman, Bragg or single photon transitions, operate in Mach–Zehnder geometries. In the basic configuration a first laser pulse ($\pi$/2 pulse) coherently splits the atomic wavefunction, a second pulse ($\pi$ pulse) redirects it, and a third pulse ($\pi$/2 pulse) recombines the two paths, as illustrated in Fig.~\ref{fig:mz_interferometer}. After recombination, the phase difference accumulated along the two paths is translated into an atom number difference between the two output ports, which can be measured experimentally. This phase shift depends, among other factors, on the gravitational acceleration and scales with the square of the interrogation time~\citep{Storey1994, Young1997, Chu2002, Tino2014, Abend2020, Tino2021}:
\begin{equation}\label{eq:interferometer}
\Delta\phi=k_{\rm eff}gT^2
\end{equation}
where $\hbar k_{\rm eff}$ is the effective momentum, i.e. the momentum difference between the two interferometer branches, $g$ is the gravitational acceleration and $T$ is the time interval between the $\pi$/2 pulses and $\pi$ pulse.

To increase the interferometric signal, large-momentum transfer (LMT) interferometers capable of transferring hundreds of photon momenta have been demonstrated~\citep{Parker2018,Gebbe2021,Canuel2020}. However, during an LMT interferometric sequence, parasitic interferometers often arise. These are unwanted interferometric paths generated by multiple or spurious transitions, which produce fringes in addition to those in the main path. These signals reduce contrast, introduce phase noise and limit the overall sensitivity of the interferometer.

A variety of strategies have been developed to suppress parasitic interferometers. One approach is to carefully design the timing of interferometric pulse sequences, so that unwanted paths interfere destructively and do not contribute to the final signal~\citep{Béguin2023}. Another method makes use of dichroic “mirror” pulses, which act as selective reflectors: they redirect only the desired interferometric trajectories while transmitting parasitic ones, thereby isolating the signal of interest~\citep{Pfeiffer2025}. A significant contribution comes from the use of advanced quantum control techniques, such as composite pulse sequences. These methods, which rely on the application of a sequence of light pulses rather than a single pulse, offer precise and robust control of matter waves. Specifically, composite pulses have been employed to improve the fidelity of state inversion operations in thermal atom clouds, halving the infidelity caused by experimental inhomogeneities~\citep{Dunning2014}. Similarly, Floquet atom optics enables engineering of the system's time evolution to achieve high pulse efficiency even in the presence of large detuning errors and differential Doppler shifts~\citep{Wilkason2022}. More recently, quantum optimal control has enabled the design of highly robust pulse shapes, which significantly reduce sensitivity to laser noise, intensity fluctuations, and velocity spread, further enhancing contrast and stability in high-order Bragg interferometry~\citep{Louie2023}.

Atom interferometry finds prominent applications in gravitational physics, enabling highly accurate measurements of gravitational acceleration~\citep{Peters1999,Gillot2014,Dickerson2013}, gravity gradients~\citep{McGuirk2002,Sorrentino2014}, gravitational waves~\citep{Graham2013, Canuel2020,Badurina2020,Abe2021}, gravity curvature~\citep{Rosi2015,Asenbaum2017}, and the Newtonian gravitational constant~\citep{Rosi2014}. Experiments with ultracold atoms, including Bose–Einstein condensates, have further enhanced quantum coherence and improved sensitivity~\citep{Abend2020,Jamison2014}. A rapidly growing area of interest concerns the use of atomic species beyond alkali metals, in particular alkaline-earth or alkaline-earth-like atoms such as \isotope[]{Ca}, \isotope[]{Sr}, \isotope[]{Yb} and \isotope[]{Cd}, which offer unique advantages for precision measurements~\citep{Riehle1991,Graham2013,Hartwig2015,Jamison2014,delAguila2018}. These atoms, widely employed in state-of-the-art optical clocks~\citep{Ludlow2015}, possess a $^1$S$_0$ ground state with zero electronic angular momentum, making them less sensitive to magnetic-field perturbations compared to alkali atoms. Moreover, they feature a variety of both dipole-allowed and narrow intercombination transitions, which allow for the implementation of efficient multiphoton Bragg diffraction as well as single-photon interferometric schemes~\citep{Hu2017b, Hu2020,delAguila2018,Rudolph2020}. Furthermore, some of their resonance transitions lie in the blue or near-UV spectral range—for example, $\qty{461}{\nano\meter}$ for strontium and $\qty{399}{\nano\meter}$ for ytterbium—providing larger momentum transfer per diffraction order compared to alkali atoms, and thus offering higher potential sensitivity for interferometric devices~\citep{Mazzoni2015}.

Atom interferometry has become a key tool for testing the Weak Equivalence Principle (WEP) at the quantum level~\citep{Rosi2017}. Following the first demonstration by Fray et al.~\citep{Fray2004}, several experiments have compared the free fall of different atomic species, including~\isotope[85]{Rb},~\isotope[87]{Rb},~\isotope[39]{K}, bosonic~\isotope[88]{Sr} and fermionic~\isotope[87]{Sr}, and also atoms in different spin orientations~\citep{Bonnin2013,Zhou2015,Schlippert2014,Tarallo2014, Tino2020}. The current accuracy of these measurements, in the 10$^{-12}$ range~\citep{asenbaum2020}, is expected to improve significantly owing to rapid advances in atom-optical elements enabling LMT~\citep{Muller2009}, along with experimental platforms offering several seconds of free fall during interferometer sequences~\citep{Abe2021} or allowing a large spatial separation between wave packets in quantum superposition~\citep{Kovachy2015b}.

Several tests of the WEP and the Einstein Equivalence Principle (EEP) have been proposed using atom interferometry. Some of these involve measurements performed on space-based platforms, which will be discussed in the next section, while others focus on antimatter systems, in particular antihydrogen~\citep{Hamilton2014} and positronium atoms~\citep{Oberthaler2002,Vinelli2023,McCaul2025}. Such precision tests are not only crucial to distinguish general relativity from alternative theories of gravity but are also of particular interest within the context of quantum gravity, since violations of the equivalence principle are commonly predicted by several low-energy quantum gravity frameworks, despite their different motivations and mathematical structures~\citep{Dvali2002,Maartens2010,Damour2012,Kostelecky2015}. We refer the reader to~\citep{Tino2020, Mancini2025} for extensive reviews and a summary of the current state of the art.

\subsubsection{Atomic clocks}
\label{sec:manip:AtomInterf:atomclock}

\begin{figure}[htb!]
  \centering
  \includegraphics[width=\textwidth]{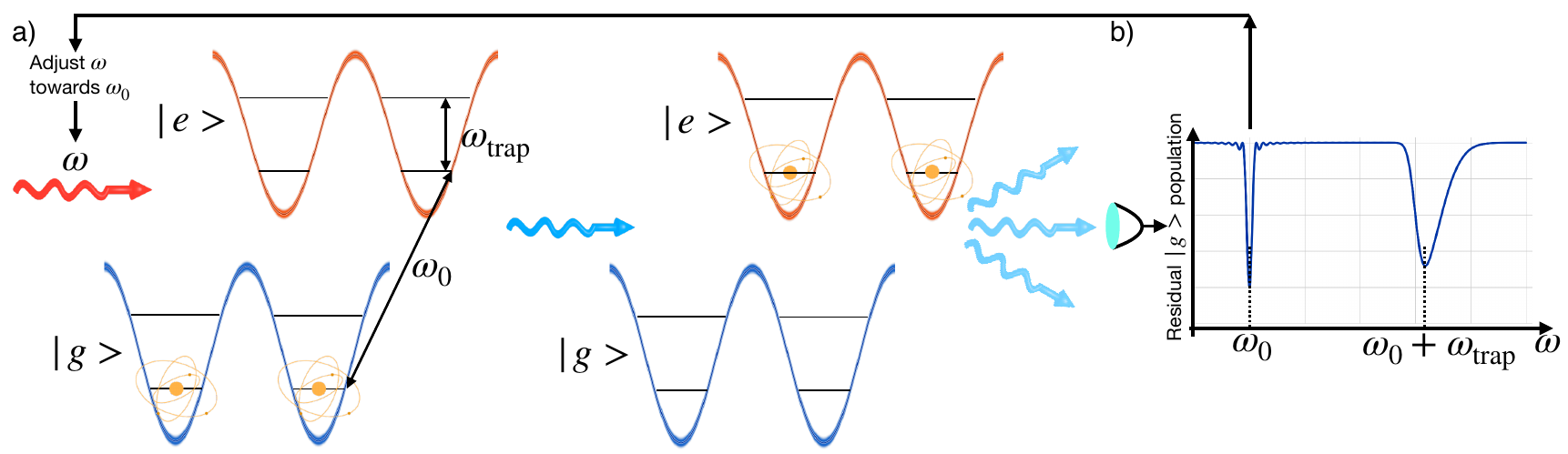}
  \caption{Illustration of an optical lattice clock. (\textbf{a}) a laser with frequency $\omega$ (red) excites the optical clock transition $\ket{g}\rightarrow\ket{e}$ at $\omega_0$ that does not change the motional state in the magic wavelength optical lattice. Center: atoms are illuminated with light that is resonant with a strong dipole-allowed transition to count the atoms left in the ground state after clock excitation. (\textbf{b}) Excitation spectrum showing also the blue sideband at $\omega_0+\omega_{\rm trap}$ which corresponds to the addition of one motion quantum at frequency $\omega_{\rm trap}$. This line is broadened due to radial motion in the optical lattice. This signal is used to correct the laser frequency $\omega$ in a feedback loop in order to reach $\omega = \omega_0$.}
  \label{fig:atomic_clock_figure}
\end{figure}

The intrinsic stability of atomic transition frequencies has, since the 1950s, motivated significant efforts to interrogate these transitions for the realization of highly stable oscillators. Such oscillators are fundamental to the definition of the SI second, as well as to the development of global positioning systems and the execution of precision tests of fundamental physics~\citep{Ludlow2015}. A relevant development in this respect was the Ramsey interferometer sequence~\citep{Ramsey1949}, which consists of two radiation pulses that drive a two-level system, separated by a controlled dark period. The pulses are tuned to produce a $\pi/2$ interaction that places the atomic system in a superposition of two levels, chosen so that their energy separation is highly stable. During the dark period, the atomic wave function, in superposition state, accumulates a phase shift proportional to the detuning between the radiation and the atomic transition frequency, as well as to the duration of the dark time. Increasing this time interval therefore increases the sensitivity to the detuning while a feedback mechanism can be applied on the oscillator to set the detuning itself to zero. This makes the interrogating oscillator as stable as the atomic transition frequency. Modern atomic clocks exploit this principle. The current primary standards - cesium fountain clocks developed at NIST~\citep{Gerginov2025}- achieve fractional frequency stabilities of the order of $\qty{1.5e-13}{}/\sqrt{\tau}$, where $\tau$ is the averaging time in seconds. Although these clocks rely on microwave transitions, even higher performance can be achieved from optical transitions involving metastable excited states, such as the widely used $^1S_0-^3P_0$ transition in~\isotope[87]{Sr}. Owing to their frequencies being about  $\approx \qty{e5}{}$ times higher than those of microwave transitions, such optical transitions promise a corresponding improvement in stability. However, in the optical domain, the Doppler effect in thermal atomic ensembles imposes fundamental limits on the achievable precision and stability of atomic clocks. A major breakthrough in overcoming this limitation was the development of optical lattice clocks~\citep{Takamoto2005}. 

The core idea is to trap atoms in an optical lattice while interrogating them with a laser of exceptionally high frequency stability, which serves as the local oscillator. When the confinement is strong enough, the system enters the so-called \emph{Lamb-Dicke regime}, in which a carrier transition, leaving the motional state unchanged, can be spectrally isolated from the sidebands that do change the motional state, as shown in Fig.~\ref{fig:atomic_clock_figure}. The result is that the atoms are essentially frozen, thus eliminating the Doppler effect altogether. It is important that the wavelength of the optical lattice is carefully selected in order to produce equal AC Stark shifts on the two clock levels. A deviation from this "magic" wavelength (or frequency) by $\qty{1}{\mega\hertz}$ produces no more than $\qty{1}{\milli\hertz}$ shift in the optical clock transition frequency. Over the past two decades, this technology has achieved record fractional frequency stabilities of $\qty{1.5e-18}{}$ at an averaging time of $\qty{1}{\second}$ ~\citep{Kim2025}.

As seen, a crucial element of optical atomic clocks is a highly phase-coherent and frequency-stable optical source, i.e. an ultra-stable laser. Laser frequency stabilization is typically achieved using a properly designed passive optical cavity, typically a two-mirror Fabry-P\'erot resonator with a very high finesse (of the order of $10^5-10^6$). Several techniques exist to tightly lock the laser frequency to the cavity resonance, the most widely used for high performance being the Pound-Drever-Hall (PDH) method~\citep{Drever1983}. A recently developed and promising alternative method relies on detecting changes in the ellipticity of the reflected beam as it passes through the cavity resonance~\citep{Diorico2024}. In order to achieve a highly coherent laser source the cavity must be extremely stable and well isolated from environmental noise sources, particularly fluctuations that alter the mirror spacing and hence the resonance frequency. In Ref.~\citep{Parke2025}, a level of $\qty{e-19}{}$ fractional frequency stability has been reached with a $\qty{68}{\centi\meter}$ long cavity. See also~\citep{Matei2017}. The most fundamental noise source stems from thermomechanical noise of the cavity spacer, the mirror substrates, and the optical coating~\citep{Ludlow2015}. 

The difficulty of measuring very high optical frequencies has been overcome thanks to the frequency comb, which compares and links optical frequencies and radio frequencies. A frequency comb consists in a laser source that produces a train of regularly spaced, phase-coherent spectral lines at the femtosecond time scale, acting as a precise ruler, in order to provide a direct link between optical frequencies and radio frequencies~\citep{Hall2006, Hansch2006}. By down-converting optical frequencies into the radio frequencies range, one obtains an electronic signal that can be processed. The comb can introduce noise, which has to be measured and cancelled.

Achieving ultimate stability and accuracy in optical lattice clocks requires careful evaluation and control of a variety of systematic effects. One of the most significant among these arises from AC Stark shifts induced by the lattice laser. The level shifts caused by the lattice light arise from three contributions: the scalar, vector and tensor polarizabilities. Although the scalar contribution to the clock transition frequency is largely canceled by operating near the magic wavelength, vector and tensor polarizabilities may still contribute. There are strategies to effectively cancel the vector shift, while the tensor contribution usually remains but can be controlled to better than the $\qty{e-17}{}$ fractional frequency uncertainty level. Additional effects include the hyperpolarizability, for instance due to two-photon resonances coupling the excited clock state to nearby levels, as well as contributions from multipole transitions.
Other important systematic effects arise from Zeeman shifts of both first and second order. First-order Zeeman shifts can be kept under control while the second-order Zeeman shift is typically much smaller in alkali-earth-based setups than for the alkali-based microwave clocks. Arguably the most important systematic effect in optical clocks arises from the Stark shift induced by blackbody radiation (BBR) emitted by the finite temperature environment surrounding the atoms. Improving the stability of clocks then requires an accurate knowledge of the polarizability of the clock states as well as of the thermal environment. Calibrated thermal probes allow to correct for the BBR shift with uncertainties at the $\qty{e-18}{}$ level. 
AC Stark shifts also occur from the spectroscopy laser itself, since its wavelength is far from the magic wavelength. However, the ever increasing coherence times of ultra-stable lasers allow one to use longer spectroscopy pulses, thus reducing the required clock laser intensity.
Static electric fields may also induce static Stark shifts. These typically arise from insulating surfaces that are exposed to the atomic sample and can accumulate electric charges. Although this shift can indeed be dominant, it can be accurately measured with uncertainty at the $\qty{e-18}{}$ level.
Changes in the clock transition frequency can occur due to cold collisions. These can be strongly suppressed by reducing the occupation of the optical lattice sites as is the case, for example, in a 3D optical lattice~\citep{Campbell2017}. 
Finally, relative vibrations between the lattice and clock lasers effectively turn into an additional Doppler shift that is usually mitigated by passive isolation of the relevant optomechanical components.

A promising candidate for overcoming many of these systematic limitations is the low-lying nuclear transition of~\isotope[229]{Th}. A nuclear clock based on thorium-229 exploits an exceptionally low-energy nuclear transition (around 8.4 eV, equivalent to about $\qty{148}{\nano\meter}$), the lowest known nuclear excitation, accessible with vacuum-ultraviolet (VUV) lasers. Because nuclear transitions are far less susceptible to external perturbations than electronic ones, this system has the potential for stability and accuracy beyond that of today’s optical atomic clocks. During the past two decades, the determination of the energy and properties of~\isotope[229]{Th} has steadily advanced and, in 2024, the first laser excitation and frequency measurement of the thorium nuclear transition in a solid-state setup was achieved, using a VUV frequency comb to link the~\isotope[229]{Th} transition to a~\isotope[87]{Sr} atomic clock~\citep{Zhang2024l}.

\subsubsection{Atomic clocks and interferometers in space}
\label{sec:manip:AtomInterf:space}

Atomic clocks and interferometers have found important applications in space, where they can achieve performances unattainable on Earth. As shown in Eq.~\ref{eq:interferometer}, the phase accumulated in a Mach–Zehnder atom interferometer depends on the square of the free evolution time $T$ between the laser pulses of the interferometric sequence. In microgravity, evolution times can be extended from about 1 s on Earth to tens of seconds in orbit, enhancing sensitivity by two orders of magnitude. Moreover, space-based sensors also benefit from a very quiet environment with low vibrations, accelerations, and other perturbations. For an extensive review, see Ref.~\citep{Alonso2022}. 

The idea of performing cold-atom experiments in space dates back to the early 1990s, with the first proposals in France and Italy~\citep{Tino2021}. The first demonstrations came in 2017, when the Chinese CACES (Chinese atomic clock ensemble in space) mission operated a cold rubidium atomic clock in space~\citep{Liu2018d}, and the MAIUS-1 sounding rocket achieved laser cooling, trapping, and Bose–Einstein condensation of~\isotope[87]{Rb}~\citep{Becker2018b}. Since 2018, NASA’s Cold Atom Laboratory (CAL) has been operating on the International Space Station (ISS), producing ultracold samples of~\isotope[87]{Rb},~\isotope[39]{K}, and~\isotope[41]{K}, with results including sub-nK effective temperatures and expansion times exceeding 1 s~\citep{Elliott2018, Aveline2020}. Its successor, BECCAL (a NASA–DLR collaboration), is to be installed on the ISS in the coming years and will offer various improvements~\citep{Frye2021}.

In April 2025, ESA finally launched the ACES mission to the ISS. ACES is a cold cesium clock designed to reach a fractional frequency stability and accuracy of $1 \times 10^{-16}$, aiming to test the gravitational redshift at the 2(1) ppm level—ten times better than the most recent tests. ACES will also search for time variations of fundamental constants by comparing ground clocks based on different atomic transitions, and search for new physics~\citep{Cacciapuoti2009,Cacciapuoti2024}.

Looking further ahead, large-scale ESA mission concepts such as STE-QUEST (Space-Time Explorer and QUantum Equivalence Space Test, 2011)~\citep{Gaaloul2022, Struckman2024}, SAGE (2016)~\citep{Tino2019}, MOCASS (Mass Observation with Cold Atom Sensors in Space, 2018)~\citep{Migliaccio2019}, and AEDGE (Atomic Experiment for Dark matter and Gravity Exploration, 2019), envision constellations of satellites carrying strontium optical clocks and atom interferometers. Their goals include precision tests of the WEP at the $10^{-17}$ level, measurements of gravitational redshift, searches for gravitational waves and new physics. The realization of such missions may still take a couple of decades, with the main challenges being the miniaturization of laser systems and the compact design of vacuum chambers.

There is also great interest in measuring magnetic-field signals from space, as exemplified by the Swarm mission, part of ESA’s Earth Explorer Programme launched in 2004~\citep{Friis-Christensen2006}. These measurements have not yet been performed with ultracold atoms, but given the promising technologies presented in Sec.~\ref{sec:det:QND:magnetometry}, future space missions based on atomic magnetometers could be envisioned.

\subsection{Interatomic Interactions in Ultracold Quantum Gases}
\label{sec:manip:interactions}

Precise control over interatomic interactions~\citep{Chin2010} has been a crucial feature in pushing the field of quantum gases to the forefront of quantum simulation~\citep{Georgescu2014}, quantum metrology~\citep{Ludlow2015, Marciniak2022, Colombo2022}, and quantum technology~\citep{Browaeys2020, Kaufman2021, Cornish2024}. 

By tuning interactions from weak to strong regimes, experiments have explored a wide range of phenomena. These include the Bose-Einstein condensation of atomic species with naturally inefficient scattering properties, such as ~\isotope[7]{Li}~\citep{Khaykovich2002},~\isotope[85]{Rb}~\citep{Cornish2000},~\isotope[133]{Cs}~\citep{Weber2003},~\isotope[39]{K}~\citep{Roati2007}; and the observation of quantum phase transitions~\citep{Bloch2008}, like the paradigmatic superfluid to Mott insulator transition in optical lattices~\citep{Greiner2002}.
Tuning the interactions either in the repulsive or attractive regime, enables the study of phenomena such as the collapse of BECs~\citep{Roberts2001}, the formation of solitons~\citep{Strecker2002}, and the observation of the BEC-BCS crossover~\citep{Inguscio2007}, where the system smoothly evolves from a condensate of weakly bound molecules to a superfluid of long-range Cooper pairs, thus connecting the physics of bosonic and fermionic superfluidity~\citep{Zwierlein2005}.
The control of the interactions plays also a crucial role in the manipulation of cold molecules, obtained by associating two atoms~\citep{Donley2002, Bigagli2024}, opening the door to ultracold chemistry~\citep{Karman2024} and new perspectives in quantum simulations~\citep{Cornish2024}.
Increasing the range of the interactions, for example, exploiting long-range dipolar interactions in atoms and molecules~\citep{Lahaye2009, Chomaz2022}, has led to the discovery of exotic states such as quantum droplets~\citep{Ferrier-Barbut2016} and supersolids~\citep{Tanzi2019, Boettcher2019,Chomaz2019}. Leveraging highly excited Rydberg states has enabled to even tune the interaction strength over several orders of magnitude and adjust their character between Van der Waals and Dipole-dipole, allowing several applications, such as non-linear optical-matter transistors~\citep{Peyronel2012,Pritchard2013,Shao2024a} and demonstration of Noisy Intermediate Scale Quantum Computing (NISQC) architectures~\citep{Bluvstein2023,Xu2024d,Maskara2025}. 

Important applications are also found in the field of quantum metrology and quantum information science: ultracold atoms in optical lattices have enabled the world's most precise atomic clocks~\citep{Ludlow2015},  while the ability to engineer entangled states of two or many atoms is crucial to building powerful quantum sensors and computers based on qubits~\citep{Saffman2016, Demille2002, Cornish2024}.

In this section, we cover the basics of interatomic interactions, starting with short-range contact interactions and their tunability through Feshbach resonances. We then extend the picture to long-range interactions, such as dipole-dipole interactions in atomic, molecular, and Rydberg atom ensembles, and cavity-mediated all-to-all interactions. This overview aims to describe the key tools and platforms constituting the foundation of the control of quantum gases in experiments.

\subsubsection{Short-Range Contact Interactions}
\label{sec:manip:interactions:short}

Understanding interatomic interactions in ultracold gases starts from the quantum mechanical scattering of two atoms. When two particles interact via a central, short-range potential, their scattering at very low energies is dominated by the so-called \textit{s-wave} component~\citep{Dalibard1999}. Higher terms in the partial wave expansion of the scattered wavefunction can be neglected~\citep{Dalibard1999, Chin2010}. 

In this regime, the scattering is effectively isotropic, meaning the scattering probability is the same in all directions. Importantly, the total scattering cross section becomes
\begin{equation}\label{eq:cross_section}
    \sigma(k) = \frac{4\pi a^2}{1+k^2a^2}
\end{equation}
which tends to the constant value $4\pi a^2$ for small relative collision momentum $k$, therefore it is solely determined by the single parameter $a$, called the \textit{scattering length}. This parameter encapsulates the essential features of the complicated two-body interaction potential into a single number, accounting for the phase shift acquired by the relative wavefunction of the atoms during the collision~\citep{Dalibard1999}.
With this simplification, we can replace the detailed interaction potential with a zero-range \textit{pseudopotential} proportional to the scattering length~\citep{Dalibard1999, Chin2010}:
\begin{equation}\label{eq:pseudopotential}
    V(\mathbf{R}) = \frac{4\pi \hbar^2 a}{m} \,\delta(\mathbf{R}) \frac{\partial}{\partial R}R\,\cdot,
\end{equation}
where $m$ is the atomic mass and $\mathbf{R}$ the interparticle distance.
Note that the particular choice for a pseudopotential is arbitrary, as long as it gives the same exact scattering length $a$ of the real interacting potential. Thanks to this approach, we can describe in a simple way the interactions in a many-body system without losing accuracy in the description of the low-energy physics that dominates the experiments.
To give an intuitive physical interpretation to the scattering length, we can see that it appears in $V(\textbf{R)}$ as a coupling constant, so a larger (smaller) $a$ would result in stronger (weaker) interaction strength. Moreover, the sign of $a$ indicates whether the effective interaction is repulsive ($a > 0$) or attractive ($a < 0$). 

Probably the most important breakthrough in the field is that $a$ itself can be tuned experimentally. This is most commonly achieved through the use of Feshbach resonances~\citep{Chin2010}, as it will be presented in the following section.

\subsubsection{Feshbach Resonances and Interactions Tunability}
\label{sec:manip:interactions:Feshbach}

The key idea behind a Feshbach resonance is that the scattering of two atoms generally involves multiple interaction potentials corresponding to different internal (hyperfine) states, leading to different \textit{channels}. When two or more channels are close in energy, the scattering amplitude is dramatically modified (with respect to the single channel case), enhancing, suppressing, or even changing the sign of the interaction strength. Since different channels can be shifted in energy by changing specific experimental parameters (i.e., an external magnetic field), the resonances can be controlled experimentally to tune interactions.

To understand this mechanism better, consider a pair of colliding atoms with total energy $E$. Depending on their internal state (spin) and relative motion (angular momentum), they interact via a certain scattering potential $V(R)$, defining a certain scattering channel. Changing the spin or the angular momentum of either one of the atoms, will result in a different channel, hence a different scattering potential $V'(R)$. In general, $V$ and $V'$ are coupled by the hyperfine interaction, and they host molecular states, bound states of two atoms. If no molecular state is energetically close to $E$, the scattering process is non-resonant, and a certain scattering length $a$ describes it. If instead a molecular state approaches, the scattering amplitude is dramatically modified
and $a$ effectively depends on the relative energy $E_{res}$ of the resonant bound state with respect to $E$. The shape of the resonance is described by the Breit-Wigner formula for the scattering amplitude~\citep{Breit1936}, which in terms of the scattering length gives
\begin{equation}\label{eq:scattering_length}
    a = -\frac{\hbar \tilde{\gamma}}{\sqrt{2\bar{m}}E_{res}}\,,
\end{equation}
where $\tilde{\gamma}$ is the coupling strength between the scattering channels, and $\bar{m}$ is the reduced mass. It is worth noting that the scattering length $a$ depends on the sign of $E_{res}$, then tuning the energy of the molecular state one can pass from attractive ($a<0$) to repulsive ($a>0$) interactions, passing trough the \textit{pole} of the resonance, where the scattering length diverges.


In the most common scenario, the resonant state is a molecular state supported by a scattering potential higher in energy (for this reason it is called \textit{closed channel}), with respect to the scattering potential seen by the two colliding atoms (called \textit{open channel}). Since the closed and open channels have different magnetic moments $\mu'$ and $\mu$ respectively, changing the external magnetic field induces an energy shift $\Delta E = (\mu - \mu')B$ between the two scattering channels. Changing the magnetic field, we are able to tune $\Delta E$, hence it is then possible to approach resonances in the experiments. Close to a Feshbach resonance, the scattering length is described by:
\begin{equation}\label{eq:magnetic_Feshbach_resonance}
    a(B) = a_{\mathrm{bg}}\left(1 - \frac{\Delta}{B - B_0}\right).
\end{equation}
Here, $a_{\mathrm{bg}}$ is the background scattering length (away from resonance), $B_0$ is the resonance position in the magnetic field, and $\Delta$ characterizes the width of the resonance, which depends also on the coupling $\tilde{\gamma}$ between the channels. As $B$ approaches $B_0$, the scattering length diverges and can even change sign, enabling access to both strongly repulsive and attractive regimes. Sitting near the resonance, the inelastic losses are strongly enhanced, providing an experimental tool to locate $B_0$ and extract $\Delta$. Indeed, the most common way to characterize Feshbach resonances is the so-called loss spectroscopy, featured in the first experimental observation of this phenomenon~\citep{Inouye1998}, in a sodium BEC. For further insights on the resonance mechanism and a more detailed review of the experiments, see~\citep{Chin2010}.\\

Besides magnetic Feshbach resonances, alternative mechanisms involve, for example, the use of laser light to couple atomic scattering states to excited molecular states (optical resonances)~\citep{Chin2010} in a dynamic and spatially resolved fashion, enabling the control of the scattering length even for atomic species lacking strong magnetic resonances~\citep{Theis2004}. Another type of resonances are the so-called shape resonances, where typically the closed channel is a quasi-bound state trapped behind the repulsive barrier (at low momenta) of a higher partial wave (p- or d-wave) scattering potential~\citep{Chin2010}.

\subsubsection{Long-Range and Anisotropic Interactions}
\label{sec:manip:interactions:long}


Short-range isotropic collisions are not the only way for two atoms to interact. Setting aside the case of charged atoms, where the Coulomb force mediates long-range (and still isotropic) interactions between ions, an important case for neutral atoms is the dipole-dipole interaction. Indeed, when atoms or molecules possess a non-zero magnetic or electric dipole moment, they interact via a long-range and anisotropic force, which depends on the magnitude of the dipole moments and their orientations. Typically, in experiments a bias field polarizes the dipoles along a fixed direction $\textbf{e}$. In such a case, the dipole-dipole interaction can be written as~\citep{Lahaye2009}:
\begin{equation}\label{eq:dipolar_interaction}
    V_{dd}(\textbf{R}) = \frac{C_{dd}}{4\pi}\frac{1-3\cos^2(\theta)}{R^3}\,,
\end{equation}
where $\textbf{R}$ is the interparticle distance, $\theta$ indicates the orientation of the dipoles with respect to their relative distance, such that $\cos(\theta) = \textbf{e}\cdot \textbf{R}/(R)$, and $C_{dd}$ is the coupling constant containing information about the magnitude of the dipole moment. For particles with magnetic dipole moment $\mu$, we have $C_{dd}=\mu_0\mu^2$. For electric dipoles with dipole moment $d$, we have $C_{dd}= d^2/\varepsilon_0$. Here $\varepsilon_0$ and $\mu_0$ are the vacuum permittivity and permeability, respectively. Note that the dependence on $\theta$ makes this interaction strongly anisotropic: for $\theta=\pi/2 $ the dipoles sit side-by side and the force is repulsive; for $\theta=0$ the dipoles are instead in head-to-tail configuration, and the interaction is attractive with $V_{dd}(R,0)=-2V_{dd}(R,\pi/2)$. A direct consequence of the long-range nature of such $\sim 1/R^3$ interaction is that it cannot be reduced to a single partial wave contribution~\citep{Dalibard1999,Lahaye2009,Chomaz2022}, hence it cannot be recast in more convenient forms (as in the case of contact interactions). This leads to a non-local type of interaction, where all the particles in the ensemble are coupled, and we cannot isolate two-body processes. As a result, even if a \textit{magic angle} $\theta_m =\arccos(1/\sqrt{3})$ yields zero interaction for two particles, in a three-dimensional cloud the total interaction is always non-zero. The control of dipole-dipole interactions tuning the orientation of the external field is although possible in reduced dimensions~\citep{Chomaz2022}.

For practical reasons, it is convenient to compare the strength of dipolar interactions to that of contact interactions (which are always present in the system), defining an interaction parameter $\epsilon_{dd}$ via the so-called \textit{dipolar length} $a_{dd}$ as
\begin{equation}\label{eq:dipolar_length}
    \epsilon_{dd}= \frac{a_{dd}}{a} =\frac{1}{a}\left(\frac{C_{dd}\,m}{12 \pi \hbar^2}\right)\,,
\end{equation}
so that we can quantify how dipolar a gas is. To control $\epsilon_{dd}$, few methods have been experimentally explored~\citep{Lahaye2009, Chomaz2022}. The most common one is to exploit a Feshbach resonance to tune the scattering length $a_s$ at the denominator of $\epsilon_{dd}$, effectively changing the relative role of dipolar interactions in the system. This scenario is typically used for dipolar Bose gases of magnetic atoms, where the magnetic dipole moment is fixed (cannot be tuned via the amplitude of the bias field).\\

In the last part of this section, we present the different types of dipolar systems, highlighting their advantages and disadvantages.

\begin{table}[ht!] 
\centering
\caption{Overview of the dipolar interaction among a selection of experimental systems. Ranges, indicated with squared brackets, refer to the possibility of tuning the interactions. For a more detailed overview see~\citep{Chomaz2022} for magnetic atoms,~\citep{Karman2025} for polar molecules and~\citep{Gallagher2008, Paris-Mandoki2016, Sadeghpour2023} for Rydberg atoms.}
\label{tab:Dipolar}
\small
\begin{tabularx}{\textwidth}{X X X X X}
\toprule
\textbf{Species} & $\boldsymbol{d}$ or/and $\boldsymbol{\mu}$ & $\boldsymbol{a_{dd}\,(a_0)}$ & $\boldsymbol{\epsilon_{dd}}$ & \textbf{Ref.} \\
\midrule
Rb      & $1\,\mu_B$  & $0.7$           & $0.007$       &~\citep{Lahaye2009} \\
Cr      & $6\,\mu_B$  & $16$            & $0.16$        &~\citep{Lahaye2009} \\
Er      & $7\,\mu_B$  & $66$            & $[0.3,1.0]$   &~\citep{Chomaz2022} \\
Dy      & $10\,\mu_B$ & $130$           & $[0.9,1.9]$   &~\citep{Chomaz2022} \\
KRb     & $0.6$ D     & $2.0\times10^3$ & $20$          &~\citep{Lahaye2009} \\
RbCs    & $[0.3,0.6]$ D &               &               &~\citep{Gregory2024}\\
LiCr    & $3.3$ D and $5\,\mu_B$  &     &               &~\citep{Finelli2024}\\
CaF     & $3.12$ D  &   &                               &~\citep{Childs1986} \\
NaCs    & $4.75$ D    & $[-2,+2]\times10^3$ & $[-3,3]$  &~\citep{Bigagli2024}\\
\bottomrule
\end{tabularx}
\end{table}

\paragraph{Magnetic atoms}
\label{sec:manip:interactions:long:mag}

Although typical magnetic moments for alkali atoms are small, around $1\,\mu_B$, lanthanides like Erbium ($7\mu_B$) and Dysprosium ($10\mu_B$) feature rather large magnetic interactions. The first dipolar BEC was realized with Chromium ($6\mu_B$)~\citep{Beaufils2008}, leading to the observation of the first dipolar effects, like the modification of the density distribution due to the attractive part of the interaction (\textit{magnetostriction})~\citep{Stuhler2005} or the collapse of a dipolar BEC~\citep{Lahaye2008}. During the last decade, also Dysprosium~\citep{Lu2011, Tang2015} and Erbium~\citep{Aikawa2012} have been Bose condensed, gaining a prominent role in the field of dipolar quantum gases thanks to the possibility to reach $\epsilon_{dd}>1$, where dipolar effects dominate. The peculiarity of lanthanides resides in their electronic structure, featuring an inner open \textit{f}-shell, screened by an outer \textit{s}-shell hosting two electrons. While the external \textit{s} electrons recall the simpler picture of alkali-earth atoms, the inner ones contribute to the total orbital angular momentum $L$ of the ground state, providing a large total magnetic moment. The rich internal structure of lanthanides also gives many optical transitions for the cooling, trapping, and manipulation of the dipolar gas~\citep{Chomaz2022}.

With the bosonic isotopes of Er and Dy, it has been possible to investigate the roton instability~\citep{Santos2003, Chomaz2018}, the formation of self-bound droplets~\citep{Ferrier-Barbut2016}, and the elusive supersolid phase of matter~\citep{Tanzi2019, Bottcher2019, Chomaz2019}. All these phenomena stem from the attractive part of the dipole-dipole interaction counterbalancing the repulsive contact interactions, and giving rise to an instability of the BEC, which leads to a clusterization (or crystallization) of the system~\citep{Biagioni2022}, stabilized by the quantum fluctuations~\citep{Petrov2015, Ferrier2016}. 

Magnetic atoms offer a reliable platform to study dipolar physics in the bulk, where the interparticle spacing is small enough to ensure large dipole-dipole interactions. Realizing quantum simulators based on optical lattices with sufficiently large magnetic interactions to couple nearest neighbors, is on the other hand challenging, since it requires the use of quantum gas microscopes with single-site resolution and control, but gives the possibility to explore a variety of quantum solid phases and exotic insulators~\citep{Su2023}.\\

\paragraph{Polar molecules}
\label{sec:manip:interactions:polar}

In the presence of an external electric field, heteronuclear molecules in their ro-vibrational ground state possess a permanent electric dipole moment which is typically very large, of the order of the Debye, entering the regime of strongly dipolar systems where $\epsilon_{dd}$ is one order of magnitude bigger than the magnetic atoms case. Despite this advantage in terms of strength of the dipolar interactions, to realize a dipolar molecular gas at low temperatures is incredibly more complex with respect to the atomic case. Indeed, creating cold molecules in their absolute ground state has been an exciting challenge overarching the last decade, with recent important key steps. 

The direct laser cooling of molecules, like CaF~\citep{Zhelyazkova2014,Anderegg2018} for example, is challenging due to their immensely rich internal structure, which makes finding effectively closed transitions suitable for cooling a formidable task. For this reason, a common strategy is to associate heteronuclear molecules starting from a double species ultracold gas, realizing weakly bound states very close the input scattering channel~\citep{Kohler2006, Chin2010}. To develop a non-zero electric dipole moment, such molecules need to be optically transferred to their absolute ground state exploiting the so-called stimulated Raman adiabatic passage (STIRAP)~\citep{Bergmann1998}.

Finally, the molecular gas must be stable against collisional loss, which prevents (in bosonic systems) or makes inefficient (for fermionic molecules~\citep{De2019}) the evaporative cooling of the sample to reach a quantum degenerate gas. The solution to this long-standing problem is to shield the molecules by engineering long-range repulsive interactions that prevents two molecules to approach one another, reducing both two-body inelastic losses and three-body recombinations. A promising technique is the microwave shielding~\citep{Karman2018, Anderegg2021}, recently optimized and exploited to realized the first molecular BEC~\citep{Bigagli2024, Karman2025}.\\

\paragraph{Rydberg atoms}
\label{sec:manip:interactions:long:rydberg}

When an electron is promoted to a very high principal quantum number $n$, the resulting permanent electric dipole moment scales as the size of the orbital, hence $d\sim n^2 a_0$, giving rise to enormous tunable (via $n$) dipolar interactions characterized by a dipolar length $a_{dd}$ by far larger than the interparticle spacing. On top of that, Rydberg states feature large polarizabilities and long lifetimes, making them ideal candidates for quantum simulation and computation ~\citep{Saffman2010,Browaeys2020}.

The interaction between two Rydberg atoms are dominated by van der Waals forces $V_{\mathrm{vdW}} \propto C_6/R^6$ at large distances $R$, and by 
dipole-dipole interactions $V_{\mathrm{dd}} \propto C_3/R^3$ at short distances. The coefficients depend on the principal quantum number, as $C_6\propto n^{11}$ and $C_3 \propto n^4$. By choosing specific Rydberg levels and tuning the external fields, it is possible to engineer the crossover between the two regimes experimentally~\citep{Comparat2010}.

A central concept arising from these strong interactions is the \emph{Rydberg blockade}. When one atom in a given volume is excited to a Rydberg state, the interaction shift prevents simultaneous excitation of nearby atoms. This blockade mechanism has been harnessed to entangle pairs of atoms~\citep{Wilk2010,Isenhower2010} and to implement high-fidelity two-qubit quantum gates. It also underpins collective phenomena in many-body ensembles, where the blockade radius sets the characteristic scale of spatial correlations.

Another important feature of Rydberg systems is the occurrence of F{\"o}rster resonances, where pair states of Rydberg atoms become nearly degenerate and resonant dipole--dipole interactions dominate. By tuning electric fields (Stark tuning), interactions can be switched from van der Waals to resonant dipolar form, enabling precise control over interaction strengths and anisotropies~\citep{Ravets2014}.

Over the past decade, experimental progress has demonstrated several landmark results. Defect-free arrays of hundreds of Rydberg atoms have been assembled using optical tweezers, enabling the exploration of quantum magnetism, frustrated spin models, and topological phases~\citep{Bernien2017,GuardadoSanchez2018,Scholl2021,Ebadi2021}. High-fidelity two-qubit gates with fidelities exceeding 97\% have been realized~\citep{Levine2019,Madjarov2020}, pushing neutral atoms closer to the threshold for fault-tolerant quantum computing. Moreover, programmable quantum simulators with tunable geometries and interaction graphs now provide a powerful tool for analog quantum simulation of complex many-body dynamics~\citep{Browaeys2020}.

At present, Rydberg atom arrays stand as one of the leading platforms for quantum technologies. Current research directions include scaling to thousands of qubits, integrating Rydberg systems with cavity QED or photonic interfaces, and exploring hybrid architectures with molecules or solid-state systems. Future applications range from universal digital quantum computation to the analog simulation of gauge theories, lattice models, and exotic phases of matter. Precision metrology, quantum networking, and quantum-enhanced sensing may also benefit from the unique properties of Rydberg interactions, ensuring their central role in the next generation of quantum technologies.

\subsubsection{Light and Cavity Mediated Interactions}
\label{sec:manip:interactions:cavity}


Light provides a uniquely versatile handle to engineer and control interactions in ultracold atomic systems, complementing and extending the traditional use of magnetic Feshbach resonances. Two distinct approaches have emerged that illustrate the breadth of possibilities: optical Feshbach resonances and cavity-mediated interactions. These methods demonstrate how photons can act both as microscopic mediators of collisional phase shifts and as macroscopic carriers of global many-body interactions, thereby opening complementary pathways to the control of quantum matter.

Optical Feshbach resonances offer a compelling route to modulate atom–atom interactions optically by dressing colliding atom pairs with laser light close to a molecular resonance, thereby altering the effective scattering length on ultrafast timescales. In the context of collisional dynamics of ultra-cold gases~\citep{Dalibard1999}, the scattering length fundamentally characterizes low-energy binary collisions and thus determines the macroscopic properties of condensates. By employing a near-resonant laser tuned to a bound molecular level, one  introduces both a shift and an effective width to the scattering amplitude, allowing tunable interaction strength and sign via a so-called optical Feshbach resonance. This method enables interaction control on a faster timescale and in a spatially more flexible manner than magnetic techniques, though limited by heating and losses due to scattering with the near-resonance light. It is particularly attractive in systems like spin-singlet bosons where magnetic Feshbach resonances are unavailable, and aligns with the theoretical framework of two-body scattering formalism essential to collisional dynamics~\citep{Dalibard1999}. Optical Feshbach resonances thus extend the collisional toolbox for ultracold quantum gases in inventive ways.

Cavity-mediated interactions leverage collective coupling between atoms and a quantized light field inside an optical resonator to induce effective long-range interatomic interactions. These kind of interactions arise when atoms couple to the same quantized mode of an optical resonator. Each atom, through its electric dipole, interacts with the intracavity field. The cavity mode acts as a shared communication channel: the light scattered by one atom can be reabsorbed by others, creating an effective long-range interaction between all atoms coupled to the mode. Such an effect is fostered by the presence of the cavity, that allows the light scattered by each atom to be stored for a long time, until the successive rearsorption, introducing thus a coherent interaction of infinite range as the photons in the cavity are fully delocalized.
The long-range interactions are typically introduced by illuminating the atomic sample with a side pump beam, red-detuned with respect o the cavity mode by $\Delta_c$.
The strength and sign of the cavity-mediated interactions depend on $\Delta_c$, the intensity of the total light filed created by the pump and the cavity mode and on the spatial overlap of the atomic density distribution with the latter.
A recent paradigm shift in this domain is described by a novel “laser-painted” method, where a cavity-enhanced pump pattern sculpts fully tunable spatial interactions in range, shape, and sign~\citep{Bonifacio2024}. These advances build on foundational experiments in cavity quantum electrodynamics, such as the Dicke quantum-phase transition observed with a Bose–Einstein condensate in a cavity~\citep{Baumann2010} and methods for coherent multiqubit operations mediated by Rydberg-like interactions via cavity modes in neutral atom arrays~\citep{Levine2019}. The combination of real-space interaction sculpting and cavity-coupled coherence heralds a versatile platform for engineered many-body Hamiltonians, programmable long-range coupling, and exploration of collective phases and quantum information protocols.

\begin{table}[ht!] 
\centering
\caption{Overview of quantum state manipulation techniques and their applications.}
\label{tab:StatManip}
\small
\centering
\begin{tabularx}{\textwidth}{X X X X}
\toprule
\textbf{Methods} & \textbf{Benefits} & \textbf{Limitations \& Requirements} & \textbf{Main applications} \\
\midrule

\textbf{Control of external DOF} & & & \\
\multirow[t]{3}{*}{Bragg transitions}
& -No internal state losses & -Higher pulse power & Interferometer with Bragg scattering~\citep{Stenger1999}\\
& -High momentum transfer & -Momentum state losses & Bragg spectroscopy of a BEC~\citep{Giltner1995} \\
& -Reduced AC Stark shift effects & -Pulse shape optimization &  \\
\midrule

\textbf{Control of internal DOF} & & & \\
\multirow[t]{3}{*}{Rabi, Ramsey, Spin Echo}
& -Coherent control & -Pulse shaping & Qubit manipulation~\citep{CohenTannoudji1998}\\
& -Long coherence times & -Stable fields & Precision spectroscopy~\citep{CohenTannoudji1998}  \\
&  & -Decoherence-sensitivity & Decoherence studies~\citep{CohenTannoudji1998} \\
\multirow[t]{3}{*}{Raman}
& -Low Losses & -Multiple lasers required & State preparation, cooling~\citep{Thompson2013a,Vitanov2017}\\
& -Specific state addressable & -Careful detuning & Quantum sensors~\citep{Vitanov2017}  \\
& -Control of external DOF & -Intensity control &  \\
\multirow[t]{3}{*}{Atomic Clocks}
& -High stability & -Stability of local oscillator & Metrology~\citep{Ludlow2015}\\
& -High accuracy & -Black body radiation& Fundamental physics~\citep{Ludlow2015}\\
& -Versatility & -Stark and Zeeman shifts & Space metrology~\citep{Alonso2022}\\
\midrule

\textbf{Control of internal \& external DOF} & & & \\
\multirow[t]{3}{*}{Matter-wave interferometers}
& -Sensitivity to inertial forces & -Signal-to-noise ratio & Fundamental physics~\citep{Tino2021}\\
& -Quantum and gravitational & -Laser noise & Gravity measurements~\citep{Tino2020} \\
& -Transportability & -Systematics & Antimatter gravitation~\citep{Hamilton2014,Oberthaler2002,Vinelli2023}\\

\multirow[t]{3}{*}{Synthetic Gauge Fields}
& -High controllability & -Precise laser control & Quantum Hall physics~\citep{Dalibard2011,Nascimbene2025}\\
& -Clean systems and Versatility & -Heating and losses & Novel topological matter~\citep{Goldman2016,Cooper2019}\\
& -Gateway to strongly correlated phases & -Finite system size &  \\

\bottomrule
\end{tabularx}
\end{table}


\section{Detection}
\label{sec:det}

One of the most important parts of an ultracold atom experiment is the detection process, by means of which the relevant properties of the atomic ensemble can be measured. Achieving high-resolution, low-noise detection is essential for obtaining accurate measurements of physical observables in an experiment. Most detection techniques rely on the interaction between atoms and laser light, which can be precisely controlled and tailored to the needs of the experiment.  
Over the years, a wide variety of detection techniques have been developed, each optimized for specific applications, ranging from imaging large atomic ensembles to detecting individual atoms. 

In this section, we review the most widely used methods. We begin with one of the earliest and still most common techniques: \emph{resonant absorption imaging} (Sec.~\ref{sec:det:ensemble:abs}), where the resonant interaction between the atoms and a probe beam is exploited.  A major limitation of this method is the fact that it typically perturbs the probed system in a destructive way, preventing multiple images of the same atomic sample. To overcome this issue and enable, for instance, the observation of time-dependent dynamics or the preparation of a quantum system in a specific initial state, various minimally destructive (or non-destructive) detection methods have been developed. These include \emph{partial transfer absorption imaging} (Sec.~\ref{sec:det:ensemble:ptai}), and various \emph{dispersive imaging} techniques, such as \emph{phase contrast imaging} (Sec.~\ref{sec:det:ensemble:phasecontr}) and \emph{polarization contrast imaging} (Sec.~\ref{sec:det:ensemble:polcontr}). 
We also briefly review \emph{fluorescence imaging} (Sec.~\ref{sec:det:ensemble:fluo}), a method that enables high-fidelity detection down to the single-atom level and plays a central role in quantum computation experiments. In section~\ref{sec:det:statesel} we address the issue of \emph{state-selective detection} through \emph{Stern-Gerlach separation}, which achieves spin-state spatial separation through magnetic field gradients (Section~\ref{sec:det:statesel:stern-gerlach}) or, in its optical variant, through laser intensity gradients. Section~\ref{sec:det:statesel:optpump} treats \emph{shelving} and \emph{optical pumping} methods to improve detection using metastable or dark states and in Sec.~\ref{sec:det:statesel:optpump:nondistr} we describe \emph{non-destructive state-selective} imaging techniques. Section~\ref{sec:det:corrmeas} explains how it is possible to probe \emph{many-body correlations}, which reveal important quantum statistical effects in quantum gases. An important class of measurements that allow for repeated and predictable measurements, \emph{quantum non-demolition (QND) measurements}, is outlined in Section~\ref{sec:det:QND}, with their application to atomic ensembles  (Sec.~\ref{sec:det:QND:intro}) along with important sensitivity enhancements that can be obtained in an optical cavity (Sec.~\ref{sec:det:QND:cavity}). These methods are relevant also in the conditional preparation of states with reduced quantum uncertainty as shown in Section~\ref{sec:det:QND:squeezing}. In Section~\ref{sec:det:QND:magnetometry} we discuss the important application of non-demolition measurements to \emph{magnetometry}. Finally, Section~\ref{sec:det:highres} presents \emph{high-resolution detection} techniques that attain high fidelity and reach single-site and single-atom resolution in deep optical traps.
An overview table~\ref{tab:Detection} summarizes the key information on these topics.

\begin{figure}[htb!]
  \centering
  \includegraphics[width=0.8\textwidth]{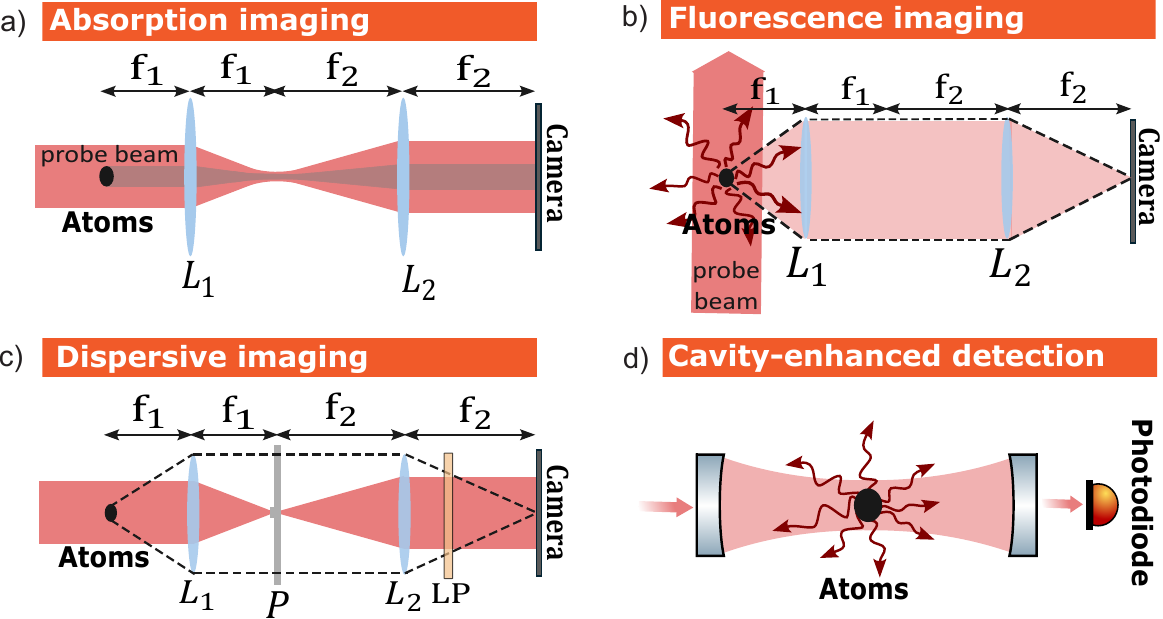}
  \caption{Overview of detection and imaging methods. 
  (\textbf{a}) \textbf{Absorption imaging:} The incoming resonant probe beam is absorbed by the atomic cloud, which creates a shadow on the transmitted light. The probe beam, after propagating through the atomic cloud, is collected by the lens $L_1$, and a second lens $L_2$ forms an image, with magnification $M = \text{f}_2 / \text{f}_1$, of the shadow of the atomic distribution at the camera plane. 
  (\textbf{b}) \textbf{Fluorescence imaging:} A near-resonant probe beam excites the atoms, which scatter photons isotropically. The emitted fluorescence is collected by the lens system ($L_1$ and $L_2$) and imaged onto a camera. 
  (\textbf{c}) \textbf{Dispersive imaging:} The incoming far-detuned probe beam is split into a diffracted component (black dashed lines) and a non-diffracted component (red shaded region). The setup for Phase-contrast imaging, includes a phase plate $P$, placed in the Fourier plane of $L_1$, shifting only the non-diffracted probe component. For Polarization-contrast imaging, the diffracted component of light suffer a rotation of polarization induced by the atoms; then this setup includes a linear polarizer $LP$ that allows the interference between diffracted and non-diffracted components. 
  (\textbf{d}) \textbf{Cavity-enhanced detection:} The presence of atoms can be detected with extremely high resolution when they are located in the circulating field of an optical cavity. This method is based on the modification of the field properties such as absorption, dispersion or polarization rotation which can be revealed, for example, by detecting the transmitted light. As in other detection methods, cavity-enhanced detection can be made state-selective.}
  \label{fig:Detect}
\end{figure}

\subsection{Ensemble detection methods}
\label{sec:det:ensemble}

\subsubsection{Absorption imaging}
\label{sec:det:ensemble:abs}

When a resonant laser beam propagates through an atomic cloud, atoms repeatedly scatter photons through cycles of absorption and spontaneous emission. 
Each absorption event removes a photon from the probe beam, and the subsequent spontaneous emission occurs in a random direction. 
As a result, the transmitted beam intensity decreases proportionally to the local atomic density. 
This principle forms the basis of \emph{resonant absorption imaging} (RAI), one of the most widely used techniques for probing ultracold atomic samples~\citep{Ketterle1999}. 
The photons emitted during spontaneous decay, on the other hand, can be collected to perform fluorescence imaging, discussed later in this section.

We now focus on absorption imaging, considering the case of a probe beam that is resonant (or nearly resonant) with a closed two-level atomic transition and propagates through a dilute gas of cold atoms, as schematically illustrated in Fig.~\ref{fig:Detect}~(a).
The attenuation of the probe intensity $I$ as it travels through the atomic medium can be described as~\citep{Reinaudi2007, Pappa2011}:
\begin{linenomath}
\begin{equation}
\frac{dI}{dz}=-\hbar\omega R n, 
\end{equation}
\end{linenomath}
where $n$ is the local atomic density, $\omega$ the laser frequency, and $R$ the scattering rate between light and atoms, given by:
\begin{linenomath}
\begin{equation}
\label{eq:scattering_rate}
R=\frac{\Gamma}{2}\frac{I/\left(\alpha I_{sat}\right)}{1+\frac{I}{I_{sat}}+\left(\frac{2\Delta}{\Gamma}\right)^2}.
\end{equation}
\end{linenomath}
Here, $I_{sat}=\hbar\omega^3_0\Gamma/(12\pi c^2)$ is the saturation intensity for a closed two-level atomic transition, $\Gamma$ the natural linewidth of the excited atomic state, $\Delta$ the detuning of the probe laser frequency from the atomic resonance frequency $\omega_0$ and $\alpha\geq1$ is a corrective factor accounting for the probe polarization, the magnetic field orientation and the presence of other atomic levels.
Substituting this expression for $R$, the absorption of the probe beam can be rewritten as
\begin{linenomath}
\begin{equation}
\frac{dI}{dz}=-n\frac{\sigma_0}{\alpha}I_{sat}\frac{I/I_{sat}}{1+\frac{I/I_{sat}}{\alpha}+\left(\frac{2\Delta}{\Gamma}\right)^2}
\end{equation}
\end{linenomath}
where $\sigma_0=3\lambda^2/(2\pi)$ is the resonant absorption cross-section. 
Integration along the line of sight (i.e., along the probe propagation direction), assuming resonant light ($\Delta=0$), yields the expression for the \emph{optical density} (OD): 
\begin{linenomath}
\begin{equation}
OD=\sigma_0\int_{-\infty}^{\infty}{{n(x,y,z) \ dz}} =\alpha\ln\left(\frac{I_{in}}{I_{out}}\right)+\frac{I_{in}-I_{out}}{I_{sat}},
\end{equation}
\end{linenomath}
where $I_{in}$ and $I_{out}$ are, respectively, the probe intensities before and after passing through the atomic cloud.
In the limit of low saturation ($I\ll I_{sat}$) and small optical depth ($Rn\ll1$), this expression reduces to its logarithmic term, corresponding to the standard Lambert--Beer law~\citep{Lambert1760, Beer1852}. 

Experimentally, the probe intensity is inferred from the counts $C(i,j)$ recorded on each camera pixel $(x_i,y_j)$. 
These counts are proportional to the local intensity $I(x_{i},x_{j})$ and can be expressed as~\citep{Wolswijk2022a}:
\begin{linenomath}
\begin{equation}
C(i,j)=\chi_{sat}\frac{I(x_{i},x_{j})}{I_{sat}}\tau_p,
\end{equation}
\end{linenomath}
where the local intensity is assumed to be time-independent, $\tau_p$ is the probe pulse duration, and $\chi_{sat}$ encapsulates the characteristics of the imaging system and the camera sensor:
\begin{linenomath}
\begin{equation}
\chi_{sat}=\eta G T \left(\frac{L_{pix}}{M}\right)^2\frac{I_{sat}}{\hbar\omega}.
\end{equation}
\end{linenomath}
Here, $\eta$ is the quantum efficiency of the detector, $G$ the camera gain, $T$ the transmission of the imaging optics, $L_{pix}$ the physical pixel size, and $M$ the magnification of the imaging system. 

In a typical absorption imaging protocol, three images are recorded. 
The first image is taken with the atoms present and illuminated by the probe light, after being tuned into resonance, producing pixel counts $C_{at}$. 
A second image is then acquired under identical illumination conditions but without atoms, yielding pixel counts $C_{in}$, which correspond to the integrated probe intensity over the pulse duration. 
Finally, a third image is taken without atoms and without probe light, in order to measure the background contribution $C_{bg}$, arising from stray light and detector dark counts.
The reference image without atoms can be obtained either by detuning the atoms far from resonance—making the medium transparent to the probe beam—or by physically removing the atoms from the field of view, for example by releasing the trapping potential. 
For each pixel, the optical density (OD) is then related to the measured counts by
\begin{linenomath}
\begin{equation}
\label{eq:OD}
OD=\alpha\ln\left(\frac{C_{in}-C_{bg}}{C_{at}-C_{bg}}\right)+\frac{C_{in}-C_{bg}}{\chi_{sat}\tau_{p}}
\end{equation}
\end{linenomath}

Resonant absorption imaging is a relatively straightforward technique to implement and is one of the most widely used detection methods in ultracold atom experiments. However, in order to extract accurate measurements from absorption images, one has to be aware of several effects that depend on the atomic density and on the probing conditions.
To minimize structured noise in the final OD image, illumination and camera exposure conditions must be identical across the three frames.
Recently, single-shot imaging of quantum gases has been demonstrated, where only one probe exposure is required and the initial probe intensity image is reconstructed using neural networks trained on reference images collected without atoms~\citep{Ness2020}.

The calibration parameters $\alpha$ and $\chi_{sat}$, in eq.(\ref{eq:OD}), must be experimentally determined: one can follow different approaches~\citep{Horikoshi2017, Reinaudi2007}, which yield consistent results when systematic errors are properly addressed~\citep{Mordini2020a}.
At high atomic densities, multiple scattering of probe photons can occur, leading to deviations from the simple Beer--Lambert law description and to an underestimation of the true atom number~\citep{Veyron2022, Chomaz2012, Corman2017}. 
In such regimes, photons scattered by one atom may be reabsorbed by others, effectively reducing the scattering cross section, which leads to an underestimation of the actual optical density and to a distortion in the inferred spatial distribution of the cloud. 
The probe parameters themselves must also be carefully chosen~\citep{Horikoshi2017}: saturation effects, power broadening, and probe detuning can all significantly alter the effective scattering cross-section and thus the measured signal. 
Therefore, accurate modeling of the atom--light interaction is essential, and experimental protocols often need to be adapted to the specific density and optical depth of the sample.

\paragraph{Time of flight imaging}
\label{sec:det:ensemble:abs:TOF}

Performing absorption imaging directly on an atomic cloud \emph{in-situ}, i.e. keeping the particles confined within the trapping potential, is often challenging because of the high optical density. A widely employed extension of absorption imaging is \textit{time-of-flight} (TOF) imaging, where the trapping potential is suddenly switched off and the atomic cloud is allowed to freely expand for a time $t_{TOF}$, typically of the order of a few tens of ms, before being imaged.
This expansion reduces the optical density, allowing, for instance, to accurately measure the number of particles in the cloud. This procedure, moreover, for sufficiently long expansion times, maps the initial momentum distribution $n_p(\mathbf{p};0)$ of the atomic cloud onto a density distribution in coordinate space $n(\mathbf{r};t_{TOF})$:
\begin{linenomath}
\begin{equation}
n(\mathbf{r};t_{TOF})\simeq n_p(m\mathbf{r}/t_{TOF};0),
\end{equation}
\end{linenomath}
which holds in the so-called far-field regime, where the size of the cloud after expansion is much larger than the initial size of the trapped cloud. For a gas initially confined in a harmonic trap with trapping frequencies $\omega_i$, this condition is satisfied when $t_{TOF\gg1/\omega_i}$.

Under these conditions, the temperature of a thermal cloud can be directly extracted from the size of the expanded atomic distribution, after a sufficiently long expansion time ~\citep{Inguscio2013, Yavin2002}:
\begin{linenomath}
\begin{equation}
\sigma = t_{TOF}\sqrt{k_BT/m}.
\end{equation}
\end{linenomath}

TOF imaging was essential in the first demonstrations of Bose--Einstein condensation~\citep{Anderson1995, Davis1995}. The formation of a BEC within a cloud of bosonic atoms, cooled below the critical temperature $T_C$, is detected by the appearance of a pronounced peak at the center of the expanded cloud, indicating a macroscopic occupation of the lowest momentum state. Additionally, the density distribution changes shape upon BEC formation, becoming bimodal: while the thermal cloud density above $T_C$ is well-described by a Gaussian function, the BEC exhibits a different distribution. In the Thomas-Fermi limit, the BEC's density profile takes the shape of the trapping potential (e.g., an inverted parabola in the case of a harmonic trap). While the expansion of the thermal part is isotropic, the TOF expansion of a BEC initially confined in an elongated trap is anisotropic, being faster in the direction of stronger initial confinement. This is a consequence of the uncertainty principle $\Delta p_i\Delta x_i\sim \hbar$, holding for the BEC wavefunction: for the direction $x_i$ with stronger initial confinement, the momentum spread is larger. In the presence of repulsive interactions between particles, the effect is amplified, since the particles are pushed further apart in the direction in which they were initially more confined~\citep{Inguscio2013}.

Today, TOF remains a standard detection technique for both bosonic and fermionic quantum gases. It is also particularly powerful in combination with optical lattices, where interference patterns that emerge after expansion provide insights into long-range coherence and quantum phase transitions~\citep{Greiner2002}.

\subsubsection{Partial transfer absorption imaging}
\label{sec:det:ensemble:ptai}

Resonant absorption imaging (RAI), while being one of the most widely used techniques to detect ultracold atoms, suffers, however, from several limitations. When imaging samples with very high optical density, only a small fraction of probe photons is transmitted through the cloud, leading to poor signal-to-noise ratios and image saturation~\citep{Ketterle1999}. Increasing the probe intensity can partially mitigate this problem~\citep{Reinaudi2007}, but even then, only a limited dynamic range of atomic densities can be accurately measured. Moreover, the resonant nature of the probe makes RAI intrinsically destructive, allowing only a single image to be captured per atomic sample.

\emph{Partial Transfer Absorption Imaging} (PTAI) was developed to overcome these limitations while retaining the simplicity of implementation characteristic of RAI. In PTAI, only a fraction of the atoms is coherently transferred to an auxiliary internal state, via a radio-frequency or an optical transition (see Sec.~\ref{sec:manip:int:Rabi}). These transferred atoms can then be imaged via an optical cycling transition from that level to another electronically excited state, while the remaining atoms - left in the ground state and far off-resonant from the probe light -  are largely unaffected.  

This method is minimally destructive: only the transferred atoms are removed from the cloud, while the rest of the sample experiences negligible perturbation~\citep{Seroka2019, Mordini2020}. A detailed description of the technique and a comparison with other minimally or non-destructive imaging methods, such as phase-contrast imaging, can be found in~\citep{Ramanathan2012} and Section~\ref{sec:det:ensemble:phasecontr}.  

By extracting a sufficiently small fraction of atoms, the effective optical density can be tuned to a regime where the extracted atomic distribution can be imaged with high signal-to-noise ratio, avoiding saturation. If the extracted fraction is spatially uniform across the cloud, the resulting image provides an accurate measurement of the original density distribution, simply rescaled by the fraction of extracted atoms, a parameter that can be experimentally controlled and calibrated, This enables \textit{in situ} imaging of high-density atomic systems, such as Bose–Einstein condensates or quantum droplets~\citep{Semeghini2018}, where densities $n>10^{15}\text{cm}^{-3}$ correspond to optical densities $\text{OD}\gg100$, which are difficult to probe with standard RAI. 

For atomic clouds with densities spanning several orders of magnitude — such as partially condensed bosonic clouds — different extraction fractions are needed to probe regions of different density. A smaller fraction is used to image the dense BEC core, while a larger fraction ensures sufficient signal-to-noise in the dilute thermal tails. Leveraging the minimal destructiveness of PTAI, multiple images of the same ensemble can be acquired with progressively increasing extraction fractions. These images can then be combined to reconstruct the full density profile, in analogy to the High-Dynamic-Range (HDR) technique used in photography~\citep{Mordini2020}.  

This approach has enabled precise measurements of Bose gas density profiles and the determination of their equation of state~\citep{Mordini2020}. It also allows real-time monitoring of system dynamics without the necessity of repeating the full experimental sequence - an advantage for experiments with long cycle times or non-deterministically reproducible phenomena. PTAI has been successfully applied, for instance, to track vortex dynamics in Bose-Einstein condensates~\citep{Freilich2010, Serafini2015}, perform \textit{in situ} imaging of superfluid flow in annular traps~\citep{Ramanathan2012}, and measure trapping frequencies within a single experimental cycle~\citep{Seroka2019}.

\subsubsection{Fluorescence imaging}
\label{sec:det:ensemble:fluo}

Another imaging technique that employs resonant light is \emph{fluorescence imaging}. In contrast to absorption imaging, in which the detected signal comes from the attenuation of the probe beam, fluorescence imaging relies on photons emitted via spontaneous emission. When the atoms are illuminated with near-resonant light, their electronic transitions are repeatedly excited and de-excited through cycles of photon absorption and emission. The spontaneously emitted photons are radiated in all directions, and a fraction of them is collected by a lens (typically with a high numerical aperture) and focused onto the camera sensor to form an image. Fig.~\ref{fig:Detect}~(b) shows a sketch of a fluorescence imaging setup, in which the atomic cloud is imaged with magnification $M = \text{f}_2 / \text{f}_1$.

In practice, each pixel $(i,j)$ of the camera records a number of counts  proportional to the number of atoms $N(i,j)$ within its field of view, to the photon scattering rate of Eq.~\ref{eq:scattering_rate}, and to the exposure time or probe pulse duration. For a quantitative measurement, the overall collection efficiency of the imaging system also plays a crucial role and depends on the solid angle of collection, the transmission of the optics in the setup, and the quantum efficiency of the detector. Because the detected signal scales linearly with the atom number, fluorescence imaging is particularly robust for precise atom counting~\citep{Bakr2009, Sherson2010, Schlosser2001}.

The main advantage of fluorescence imaging is its remarkably high sensitivity, which enables a wide range of applications. Providing single-atom detection capability, this technique has become a powerful tool for probing optical tweezers~\citep{Schlosser2001, Endres2016} and achieving single-site resolution in optical lattices~\citep{Bakr2009, Sherson2010, Cheuk2015, Parsons2015, Haller2015}. However, the technique is inherently destructive, since each scattered photon imparts a recoil kick to the atom, leading to heating and atom loss after multiple scattering events.

\subsubsection{Phase-contrast imaging}
\label{sec:det:ensemble:phasecontr}

For RAI, the radiative pressure exerted by the resonant probe beam, heats and destroys the atomic sample.
An attractive alternative is provided by the so-called \emph{dispersive imaging techniques}~\citep{Andrews1996}, which are non-destructive due to the far-off-resonant character of the probing light. Dispersive methods are also advantageous for imaging systems with high optical density (such as BECs), where resonant probe light would be almost completely absorbed, leading to image saturation. 
The required condition $|\Delta| \gg \Gamma$ allows to neglect absorption, so that dispersive effects on the probe light dominate. In practice, however, the detuning is chosen as a compromise: larger detunings reduce destructiveness but also lower the signal-to-noise ratio. An important advantage of dispersive imaging is that it can be implemented in a nearly quantum non-demolition (QND) manner, meaning that the atom number or density are preserved during the measurement, allowing the same atomic cloud to be imaged repeatedly. Further details about QND are given in Section~\ref{sec:det:QND}.

One of the most widely used dispersive techniques is \emph{phase-contrast imaging} (PhCI)~\citep{Andrews1997}, originally demonstrated for imaging BECs. In PhCI, information about the atomic cloud is contained in the part of the incident imaging field $\mathbf{E}_{in}$ that is diffracted by the atoms, denoted $\mathbf{E}_d$. The total transmitted electric field is therefore  
\begin{equation}
\mathbf{E}_t = \mathbf{E}_{in} + \mathbf{E}_d = \mathbf{E}_{in} + \mathbf{E}_{in} (e^{i\phi_{at}} - 1),    
\end{equation} 
where the diffracted component carries the phase shift induced by the atoms. This spatially dependent complex phase shift can be expressed as $\phi_{at} = \phi + ib/2$, where $\phi$ is the dispersive phase shift and $b$ the non-resonant optical density. For large detunings of the imaging beam, $b (\Delta) \approx 0$. In that case, a direct measurement of $\phi_{at}$, provides the optical density at resonance (OD)~\citep{Frometa2025}:
\begin{equation}
    \phi_{at} = \phi (\Delta) = - (\text{OD}) \frac{\frac{\Delta}{\Gamma}}{1 + 4 \frac{\Delta^2}{\Gamma^2}} \ .
\end{equation}

A schematic representation of PhCI is shown in Fig.~\ref{fig:Detect}~(c), where the paths of the probe and diffracted beams can be followed up to image formation at the camera. A phase plate is placed at the focal plane of the first lens, featuring a phase spot that affects only the non-diffracted component of the probe beam, thereby introducing a controlled phase shift~\citep{Meppelink2010}. As a result, the non-diffracted light accumulates a phase $\phi_P$, and the output light field becomes $\mathbf{E}_{out} = \mathbf{E}_{in} e^{i\phi_P} + \mathbf{E}_{in} (e^{i\phi_{at}} - 1)$. The resulting intensity captured by the camera, expressed in terms of the probe beam intensity $I_{in}$ and the relevant phase shifts, is
\begin{equation}
    I_{out} = I_{in} \left[ 3 - 2\cos{(\phi_P)} + 2\cos{(\phi_P - \phi_{at})} - 2\cos{\phi_{at}} \right].
\end{equation}
From this expression, it follows that $I_{out}$ varies periodically with $\phi_{at}$ for a fixed $\phi_P$. When $\phi_{at}$ lies outside the interval $(\phi_P - \pi)/2 \leq \phi_{at} \leq (\phi_P + \pi)/2$, phase jumps appear in the recorded image, which require phase-unwrapping algorithms. The optimal value of $\phi_P$ depends on the specific application. A particularly convenient value is $\phi_P = \pi/3$, since in this case the minimum of $I_{out}$ remains strictly zero, providing improved image contrast and an approximately linear response for small $\phi_{at}$~\citep{Frometa2025}.

A simplified variant of PhCI consists of blocking the non-diffracted portion of the transmitted light. In this case, $\boldsymbol{E}_{out} = \boldsymbol{E}_i \left(e^{i\phi_{at}} - 1\right)$, and the intensity captured by the camera reduces to
\begin{equation}
    I_{out} = 2 I_{in} \left[ 1 - \cos{\phi_{at}} \right].
\end{equation}
This version is known as \emph{dark-ground imaging}, implemented by placing a dark spot instead of a phase spot in the Fourier plane~\citep{Andrews1996}. Because no probe reference beam reaches the camera, normalization of the intensity profile is not possible, which represents a limitation for quantitative measurements of the absolute phase shift. In practice, dark-ground imaging is therefore often employed for qualitative visualization of structured atomic clouds.

\subsubsection{Polarization-contrast imaging}
\label{sec:det:ensemble:polcontr}

Another dispersive technique is \emph{polarization contrast imaging} (PCI), also known as \emph{Faraday imaging}, which detects the local birefringence of the atomic cloud. This birefringence induces a rotation of the polarization of the off-resonant probe beam as it passes through the ensemble. The resulting rotation angle, the so-called Faraday angle $\theta_F(x,y)$, encodes information about both the optical density $OD$ and the average spin projection $\langle F_z \rangle$ along the quantization axis defined by the magnetic field direction.
In particular, $\theta_F$ is proportional to the product $\text{OD} \langle F_z \rangle$, up to proportionality constants determined by the Clebsch--Gordan factors of the atomic transition and the vector polarizability of the ground state~\citep{Geremia2006}. 
The ability to measure the collective spin projection makes PCI especially powerful for studying spin domains and magnetization dynamics~\citep{Higbie2005,Kubasik2009,Sewell2012}.

A schematic representation of PCI is shown in Fig.~\ref{fig:Detect}~(c) and is similar to that of PhCI, with the difference that instead of a phase spot at the Fourier plane of $L_1$, a linear polarizer is placed before the camera. The polarizer is rotated by a small angle $\theta$ with respect to the incident polarization axis, causing interference between the transmitted probe beam and the light component rotated by the Faraday effect.

\noindent Here, we consider an incident probe beam linearly polarized along the $x$ axis with intensity $I_{in}$. 
After passing through the atomic cloud, the polarization of the probe beam rotates by the Faraday angle, and the electric field can be expressed as 
\begin{equation}
\boldsymbol{E}_t = E_{in} \left[ \cos(\theta_F)\,\hat{\boldsymbol{x}} + \sin(\theta_F)\,\hat{\boldsymbol{y}} \right].    
\end{equation}
With the use of a linear polarizer, oriented at an angle $\theta$ with respect to $\hat{\boldsymbol{x}}$, the transmitted field is projected onto 
\begin{equation}   
\boldsymbol{E}_a = E_{in} \left[\cos(\theta_F)\cos(\theta) + \sin(\theta_F)\sin(\theta)\right]\hat{\boldsymbol{a}}.
\end{equation}
The intensity recorded by the camera is then given by~\citep{Bradley1997}:
\begin{equation}
    I_{out}(x,y) = I_{in} \cos^2 \left[\theta - \theta_F(x,y) \right].
\end{equation}
For small rotation angles ($\theta_F \ll 1$),
\begin{equation}
    I_{out}(x,y) \approx I_{in}\!\left[\cos^2\theta - 2 \theta_F(x,y)\sin\theta\cos\theta \right],
\end{equation}
which is linear in $\theta_F$ for a fixed polarizer orientation. The choice $\theta=\pi/4$ maximizes the signal sensitivity.

An interesting variation is dual-port PCI, in which the light transmitted by the atoms is sent to a polarizing beam splitter oriented at $45^\circ$ with respect to the polarization of the imaging beam. The two outputs of the beam splitter are detected by separate cameras and subtracted. The differential intensity is related to $\theta_F$ by~\citep{Kaminski2012}:
\begin{equation}
    \Delta I_{out}(x,y) = I_{in} \sin 2\theta_F(x,y) .
\end{equation}
The subtraction cancels common-mode intensity fluctuations of the probe beam, leading to significant noise reduction. This feature makes dual-port PCI particularly suitable for quantum metrology applications~\citep{Appel2009}.

\subsection{State-selective detection}
\label{sec:det:statesel}

When probing ultracold samples, it is often necessary not only to accurately probe the distribution of cold particles in space, but also to determine their internal quantum state, distinguishing between different hyperfine or Zeeman states. This capability is crucial, for instance, in contexts ranging from studies of spinor condensates and quantum simulation experiments where multiple spin components are involved. In many cases, for instance in quantum computation experiments, it is essential to have the capability to detect only atoms that are in a given quantum state, while leaving atoms in other states unperturbed. 
To this end, several state-selective detection methods have been developed and optimized over the years. 

\subsubsection{Stern-Gerlach separation}
\label{sec:det:statesel:stern-gerlach}

One of the first and most widely used state-selective detection techniques is based on the magnetic Stern--Gerlach~\citep{Gerlach1922} effect. In this approach, after switching off the trapping potential, a magnetic field gradient is applied during the time-of-flight evolution of the atomic cloud. Atoms experience a force which depends on their magnetic moment, so atoms in different spin states are deflected along distinct trajectories. With a sufficiently long time-of-flight, the different spin components of the atomic cloud become spatially separated and can then be imaged using, for example, absorption imaging~\citep{Ketterle1999}. 
This method provides a simple and robust means to resolve populations in different magnetic sublevels, without requiring spectroscopic resolution of the internal states. It was instrumental in the first observations of spin domains and dynamics in spinor BECs~\citep{Stenger1998, Ketterle1999}, and has been widely used to probe population imbalances, magnetization and spin textures in Fermi gases~\citep{Zwierlein2006, Ketterle2008}. 
For accurate detection, a few requirements must be met: the time-of-flight must be sufficiently long to achieve complete separation of the different spin components; the magnetic field must be precisely controlled, with an appropriate choice of magnetic gradient amplitude and timing in the imaging sequence, to avoid heating or distortion of the expanding cloud. Residual magnetic gradients and noise can reduce accuracy. Because the method involves releasing the trapping potential and imaging after expansion, it is inherently destructive. 
Despite these limitations, the Stern--Gerlach technique remains one of the most popular approaches for state-selective detection and is routinely applied in many quantum gases experiments~\citep{Semeghini2018, Chomaz2022, Soave2023, Rogora2024}.

\paragraph{Optical Stern-Gerlach separation}
\label{sec:det:statesel:stern-gerlach:optical}

In addition to the conventional Stern-Gerlach separation with magnetic field gradients, state-dependent forces can also be realized optically, using light. The so-called \emph{optical Stern--Gerlach effect}~\citep{Sleator1992} employs spatially inhomogeneous light fields that produce position-dependent AC Stark shifts (see Sec~\ref{sec:trap:odt:fort}). Atoms in different internal states then experience distinct optical dipole forces due to the spin-state dependence of their polarizability. This results in a differential deflection of the components in the presence of a light-intensity gradient, leading to a spatial separation after time-of-flight,  analogous to the magnetic Stern-Gerlach separation. This technique is particularly useful for distinguishing between different nuclear spin states that are only weakly sensitive to magnetic field variations, such as in strontium or ytterbium~\citep{Stellmer2011, Fallani2023}.

\subsubsection{Optical pumping and shelving techniques}
\label{sec:det:statesel:optpump}

Optical pumping and shelving provide a complementary approach to state-selective detection. The basic idea is to transfer atoms into a particular internal state, often a metastable or "dark" state, so that it can be distinguished from the others by the presence or absence of a fluorescence or absorption signal~\citep{Kuhr2003}. This method has become the standard for state readout in trapped-ion systems, where high-fidelity discrimination between different internal states is essential. A classic example is the implementation of a robust two-ion geometric phase gate, where the performance of the gate was validated through state-selective fluorescence detection~\citep{Leibfried2003}.  
Analogous techniques are widely used with neutral atoms as well, especially in optical lattices and optical tweezer arrays where high-fidelity readout of individual atoms is required~\citep{Bakr2009, Sherson2010, Kaufman2021}. Optical pumping techniques are also useful for all-optical state-selective detection of Rydberg atoms~\citep{Karlewski2015, Guenter2012, Guenter2013, Gavryusev2016, Gavryusev2016a, FerreiraCao2020, Rui2020}.

A practical implementation is to use resonant light that scatters only from a chosen hyperfine or Zeeman state, while simultaneously optically pumping atoms in other states into a dark level that remains undetected. In this way, fluorescence or absorption imaging becomes selective to the desired state. This method is particularly effective when combined with high-numerical-aperture imaging systems, such as those used in quantum gas microscopes, where single-site resolution can be achieved~\citep{Bakr2009, Sherson2010, Parsons2015}. State-selective resonant imaging enables direct visualization of spatial spin distributions, although it is typically destructive and the achievable selectivity is limited by optical pumping processes during detection, which may cause population leakage into undesired states. Nonetheless, by careful choice of probe transitions and repumping schemes, detection fidelities exceeding 99\% have been demonstrated in neutral-atom quantum computing platforms~\citep{Bernien2017, Endres2016}.

\subsubsection{Non-destructive state-selective imaging techniques}
\label{sec:det:statesel:optpump:nondistr}

In contexts where preserving the atomic sample is important - for example, for repeated interrogation, feedback control, or maintaining long coherence times - imaging techniques that only minimally perturb both the internal and motional states of the atoms are highly desirable. To probe different internal states, one can use dispersive imaging methods that exploit the real part of the atomic susceptibility, such as phase-contrast imaging (Sec~\ref{sec:det:ensemble:phasecontr}) and polarization contrast and Faraday imaging (Sec~\ref{sec:det:ensemble:polcontr}). 
 These techniques can provide information on internal (spin or hyperfine) state populations by exploiting the state-dependent optical susceptibility and by selecting a probe frequency such that the states of interest yield distinguishable signals. The use of far-detuned probe light minimizes spontaneous scattering, thereby reducing heating and loss of atoms. 

 Several implementations have been demonstrated. Phase-sensitive diffraction contrast imaging (DCI) uses optical phase shifts to infer state populations in multi-level systems without significant atom loss~\cite{Sheludko2008}. Faraday imaging, which measures state-dependent polarization rotation, provides minimally destructive, spatially resolved information on spin populations. In particular, balanced polarization-contrast configurations suppress imaging artifacts in optically dense samples, and repeated imaging of the same atomic cloud has been achieved over thousands of cycles with negligible perturbation~\cite{Kaminski2012, Gajdacz2013}.  

Even higher sensitivity can be obtained by coupling the atoms to an optical cavity (see Sec.~\ref{sec:det:QND:cavity}). State-dependent dispersive shifts of the cavity resonance can then be detected with very low probe power, enabling non-destructive state readout with single-atom sensitivity. Such methods have been employed to monitor spin polarization in both ultracold and thermal ensembles~\cite{Brennecke2007, HernandezRuiz2024}.  

Beyond detection, nondestructive imaging techniques have become essential tools for quantum feedback control and quantum-enhanced metrology. Real-time dispersive probing in optical cavities, for instance, has enabled the creation of spin-squeezed states (see Sec.~\ref{sec:det:QND:squeezing}) that surpass the standard quantum limit in precision measurements~\cite{Esteve2008, SchleierSmith2010}. 
These developments will be discussed in greater detail in the following sections.

\subsection{Correlation measurements}
\label{sec:det:corrmeas}


Correlation measurements have played a central role in the study of ultracold atomic systems, even before the advent of single-particle-resolved imaging. These techniques probe the quantum statistical properties of atoms through fluctuations in density and momentum distributions, providing access to many-body correlations beyond mean-field observables~\citep{Greiner2002a, Will2010, Lewenstein1996, Zhou2019}. A central example is the use of noise correlations in time-of-flight experiments, inspired by the Hanbury Brown--Twiss (HBT) technique originally developed in quantum optics~\citep{HanburyBrown1956, Foelling2005}. In such measurements, the density-density correlations of expanding atomic clouds reveal quantum statistical effects: bunching for bosons and antibunching for fermions, directly reflecting the underlying symmetry of the many-body wavefunction~\citep{Jeltes2007}.

Noise correlation techniques have been instrumental in exploring strongly correlated phases in optical lattices. For instance, the emergence of coherence peaks in bosonic Mott insulators was detected using second-order correlations, providing a direct signature of localized particle distributions~\citep{Foelling2005}. Similarly, the observation of fermionic antibunching in degenerate Fermi gases provided one of the clearest demonstrations of Pauli exclusion in momentum space~\citep{Rom2006}. These experiments established correlation measurements as a sensitive probe of quantum statistics and interaction-driven phase transitions.

Beyond simple bosonic and fermionic systems, correlation techniques have been extended to mixtures and higher-order correlations. In Bose–Fermi mixtures, cross-correlations revealed interaction-induced modifications of quantum noise~\citep{Ospelkaus2006}, while higher-order correlation functions have been used to probe non-Gaussian quantum states and few-body physics. Such approaches have also been applied to investigate quench dynamics, thermalization, and many-body localization, where fluctuations encode information about entanglement growth and coherence properties.

A further refinement has been the use of HBT interferometry in trapped cold atoms, where spatial and temporal correlations reveal coherence lengths and the onset of long-range order~\citep{Schellekens2005}. This method has also been employed to study pair correlations in strongly interacting gases and molecular systems, providing insight into pairing mechanisms and short-range order. Recent work has extended correlation analysis to momentum-resolved Bragg scattering and Fourier-transform spectroscopy, enabling momentum-space access to dynamical structure factors without requiring single-atom detection.

The current state of the art combines correlation techniques with advanced detection schemes such as absorption imaging, fluorescence, and time-of-flight methods that achieve high signal-to-noise ratios. While quantum gas microscopes provide single-particle resolution, correlation measurements without such resolution remain crucial due to their applicability to large ensembles and dynamical systems where single-site imaging is not feasible. Looking forward, correlation measurements are expected to play an important role in exploring non-equilibrium quantum dynamics, detecting exotic phases such as topological order, and benchmarking quantum simulators by providing robust statistical signatures of many-body states.

\subsection{Quantum Non-demolition Measurements}
\label{sec:det:QND}

When we perform a measurement on a physical system, the Heisenberg uncertainty principle gives us a fundamental limit on the uncertainty product of any pair of non-commuting observables
\begin{equation}
	\Delta \mathcal{O}_1 \Delta \mathcal{O}_2 \ge \frac{1}{2} |\braket{[\mathcal{O}_1,\mathcal{O}_2]}|
	\label{Eq:uncertainty_princ}
\end{equation} 
where $\Delta \mathcal{O}_i$ is the uncertainty of the observable $\mathcal{O}_i$ and $\braket{[\mathcal{O}_1,\mathcal{O}_2]}$ is the mean value of their commutator. For example, in the case of position and momentum we have $\Delta x\Delta p\ge \hbar/2$.
When one observable is measured precisely (i.e., with a very small uncertainty), this induces a large uncertainty on the other observable, as dictated by Eq.~\ref{Eq:uncertainty_princ}.
Although this does not directly reduce the precision of the measured observable, large fluctuations in the conjugate variable may couple back to the first one when the measurement is repeated on the system.
This effect, known in literature as \emph{measurement back action}, prevents recovery of the same measurement outcome in subsequent interrogations~\citep{Braginsky1995}.
In the 1970s Braginsky and others~\citep{Braginsky1980} introduced the concept of a \emph{Quantum Non-Demolition} (QND) measurement, where measurement back action is avoided by careful design of the experimental procedure. In such measurements, the back-action noise is confined to non-relevant observables and does not affect the quantity of interest.

The procedure relies on extracting information about an observable of a system $\mathcal{S}$ using a meter $\mathcal{M}$. The system $\mathcal{S}$ and the meter $\mathcal{M}$ become entangled through an interaction Hamiltonian $\mathcal{H}_{\mathrm{int}}$, such that each carries information about the other, in addition to its own properties.

QND measurements were first proposed in the context of gravitational-wave detection~\citep{Abadie2011} and were first demonstrated experimentally in quantum optics~\citep{Grangier1998}.
Since then, the technique has found broad applicability, and it has been used to measure optical variables~\citep{Grangier1998}, opto-mechanical systems~\citep{Heidmann1997} and atomic spin systems~\citep{Vasilakis2011}. Furthermore, QND measurements have been successfully implemented in applications such as quantum memories and proposed for quantum computing architechtures~\citep{Ralph2006, Wang2017a}.
In the context of spin ensembles, QND measurements have been applied in atomic clocks and magnetometers~\citep{Leroux2010a, Budker2007}. 
They are also widely used to generate non-classical states of matter, leveraging the entanglement established between the system and the meter during the QND interaction~\citep{Julsgaard2001}.

\subsubsection{QND measurements on atomic ensembles}
\label{sec:det:QND:intro}
QND measurements can be performed on atomic systems by monitoring appropriate changes in an incident light field. 
These measurements are very important since they allow one to monitor atomic systems continuously without too much atom loss or induced decoherence. 
In addition, QND measurements are a very powerful tool, as we will see, that allows to prepare interesting collective states such as the so-called spin squeezed states which are useful in quantum sensing and metrology~\citep{Ma2011, Pezze2018}.

A possible strategy to perform QND measurements on atoms is to detect the dispersive phase shift of the light transmitted through an atomic cloud. In the dispersive regime $|\Delta|\gg \Gamma$ and for a Gaussian beam with $N$ two-level atoms at its focus, this phase shift is given by $\phi\approx -N\eta_{\rm fs}\Gamma/(2\Delta)$~\citep{Tanji-Suzuki2011}, where $\Delta$ is the detuning of the incident field from atomic resonance, $\Gamma$ is the transition linewidth and $\eta_{\rm fs} = 6/(k^2 w^2)$ is the free-space single-atom cooperativity, where $k$ is the wavenumber and $w$ the Gaussian beam waist~\citep{Kogelnik1966}. 

The induced phase shift, ultimately due to the atomic index of refraction, can interestingly be made state-dependent. 
An example is provided by a three-level system with two lower levels $\ket{\downarrow},\ket{\uparrow}$ (arising for example from the hyperfine structure) that are coupled to an excited state $\ket{e}$ through optical transitions. 
By appropriately tuning a probe laser between the two optical transitions it is possible to make this phase shift proportional to the population imbalance $J_z$ between the states $\ket{\uparrow}$ and $\ket{\downarrow}$.
As a result, a phase shift estimation produces a measurement of the population imbalance between the two states. 
This phase shift can be monitored e.g. through a Mach-Zehnder interferometer where the atoms are located in one arm. Since, under certain conditions, the imbalance is a conserved quantity, this kind of measurement can be made of the QND type.

While the measurement precision with a fixed population imbalance is fundamentally set by photon shot noise, spontaneous emission adds decoherence through elastic Rayleigh scattering or noise and decoherence through inelastic Raman scattering~\citep{Uys2010}. Interestingly, the ultimate measurement precision because of this is set by the cooperativity $\eta_{\rm fs}$.

\begin{figure}[htb!]
  \centering
  \includegraphics[width=0.45\textwidth]{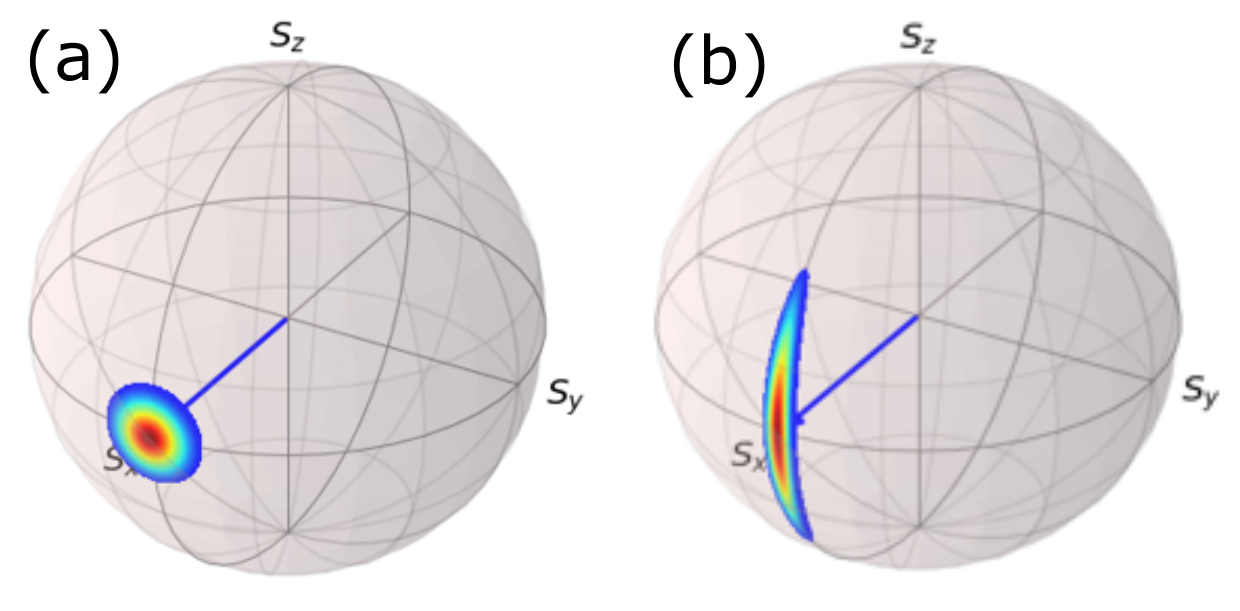}
  \caption{(\textbf{a}) Coherent and (\textbf{b}) spin squeezed state comparison for a spin-$N/2$ system, depicted on Bloch spheres. The colored areas denote the uncertainty of the Bloch vector. The Bloch sphere provides a geometric representation of the quantum state of a two-level system, where the Bloch vector indicates the mean state population. The squeezed state shows reduced uncertainty in one quadrature and increased uncertainty in the other.
  }
  \label{fig:squeezing}
\end{figure}

Experiments involving continuous QND measurements along these ideas are often related to the conditional preparation of quantum entangled spin squeezed states~\citep{Kitagawa1993}. 
In fact, when the QND measurement precision overcomes the atom shot noise $(\Delta J_z)_{\rm shot} = \sqrt{N}/2$ for a coherent spin state, the measurement projects the collective state to the one that corresponds to the measurement outcome. The variance of the prepared state is then essentially given by the measurement imprecision plus the unavoidable effects of spontaneous emission. In the context of metrological gain in atom interferometers, squeezing is often quantified through the Wineland parameter $\xi^2 = (\Delta\varphi)^2/(\Delta\varphi)^2_{\rm SQL}$ i.e. the ratio of the interferometric phase sensitivity with a squeezed state and of the phase sensitivity at the so-called Standard Quantum Limit, $(\Delta\varphi)_{\rm SQL} = N^{-1/2}$, which corresponds to an uncorrelated coherent spin state~\citep{Wineland1994,Ma2011}, see Fig.~\ref{fig:squeezing}.  An experiment using a Mach-Zehnder interferometer as considered here is reported in~\citep{Appel2009} and attained a spin noise reduction of 3.4 dB in variance. Other experiments in a similar spirit were performed in~\citep{Kuzmich1998,Kuzmich2000,Takano2009}, where the QND measurement is obtained by monitoring the atom-induced Faraday rotation of the light polarization, an effect which can also be described in terms of the dispersive phase shift.

\subsubsection{Cavity-enhanced QND detection methods}
\label{sec:det:QND:cavity}

The precision in measurements based on atom-light interaction is essentially set by the cooperativity $\eta_{\rm fs}$ which can only be increased by reducing the probe beam size $w$ and this might mean that less atoms interact with the probe or that diffraction effects start to play a role. A very effective approach in this sense involves surrounding the atomic ensemble with an optical resonator which essentially enhances the cooperativity parameter by the average number of round trips $\mathcal{F}/\pi$, where $\mathcal{F}$ is the cavity finesse. 
Considering for simplicity a two-mirror Fabry-Pérot cavity and an atom at the antinode of the corresponding mode standing wave, the single-atom cooperativity, also known as the Purcell factor~\citep{Purcell1946,Motsch2010}, becomes $\eta = 24\mathcal{F}/(\pi k^2 w^2)$. Since today's optical cavities can reach finesse values exceeding $10^5$, an enhancement e.g. of the dispersive phase shift by the same order of magnitude is possible. The phase shift leads to an intracavity optical path length variation, which translates to a shift of the cavity resonance frequency, in units of half the cavity linewidth $\kappa/2$, given by $\frac{\delta\omega_c}{\kappa/2} = \frac{N\eta\Gamma}{2\Delta}$. Any strategy that can determine this frequency shift e.g. through the change in the cavity transmission power or through phase-sensitive homodyne detection, provides a measurement of the number of atoms $N$. As discussed above, the system can also be configured to detect the population imbalance $J_z$ which is relevant for continuous monitoring and spin squeezing for quantum metrology applications~\citep{Schleier-Smith2010}. Other interesting effects occur when the cavity resonance frequency is tuned so it coincides with the atomic resonance frequency. If the collective cooperativity $N\eta\gg 1$, this produces a mixing between the cavity and the atomic modes which is manifest as a splitting of the cavity transmission profile by $2g\sqrt{N}$, where $2g$ is the single-photon Rabi frequency i.e. the Rabi frequency corresponding to the electric field amplitude of a single-photon. This effect is known as the vacuum Rabi splitting~\citep{Kimble1998, Agarwal1984, Boca2004} and can be equivalently used to measure non-destructively the number of atoms.


These features have been used in several experiments that involve single atoms interacting with an optical resonator, through cavity-enhanced fluorescence~\citep{Bochmann2010, Terraciano2009}, dispersive shifts~\citep{Puppe2007}, fiber-based optical cavities~\citep{Gehr2010, Heine2009, Trupke2007} and cavities integrated with atom chips~\citep{Teper2006}. 

The ultimate performance of cavity-based detection methods has been analyzed in~\citep{Hope2004,Hope2005,Poldy2008}.

\subsubsection{Spin squeezing-enhanced sensing}
\label{sec:det:QND:squeezing}

Squeezing can also be generated by the One-Axis Twisting technique, with a Hamiltonian $\propto J_z^2$, which induces a rotation on the Bloch sphere proportional to $J_z$, thereby creating a state with reduced quantum uncertainty along an oblique direction~\citep{Kitagawa1993}. This state preparation method differs from QND-induced squeezing in that it is deterministic, as it does not depend on the outcome of a measurement. This feature makes the method less sensitive to detection noise. Using these techniques, 5.6 and 6.6 dB of metrologically relevant spin squeezing have been achieved with~\isotope[87]{Rb} and~\isotope[171]{Yb}~\citep{Schleier-Smith2010b, Leroux2010, Braverman2019}.

The largest amounts of squeezing have been generated through cavity-enhanced QND measurements. In 2016 experiments, 20.1 and 17.7 dB of spin squeezing were achieved in~\isotope[87]{Rb} atoms prepared in the clock states~\citep{Hosten2016, Cox2016}. The frequency shift of a cavity resonance was measured using a homodyne detection system. 

All these works generally assume that detection noise must be kept below the SQL to exploit squeezed states. However, in 2016, following Ref.~\citep{Davis2016}, it was demonstrated that squeezed states can yield an advantage even with a detection noise floor  above the SQL, by employing an intermediate quantum phase magnification step~\citep{Hosten2016a}.

Extending squeezing techniques to atomic clocks and interferometers with optical transitions is particularly promising due to the superior phase stability and time-keeping precision of optical transitions. A major milestone in this direction was the observation in 2020 of atomic squeezing in a~\isotope[171]{Yb} optical lattice clock using a high-finesse cavity~\citep{pedrozo2020}. In that experiment, they reached a 5.7 dB gain in stability, although the optical clock did not operate below the SQL, since it was limited by local oscillator phase noise.

In 2021, the successor experiment to Ref.~\citep{Cox2016} involved injecting a squeezed state into a Mach–Zehnder interferometer via the One-Axis Twisting technique, instead of the QND measurement~\citep{Greve2022}. In 2022, a follow-up to the 2016 experiment of Ref.~\citep{Hosten2016} was performed, entangling a network of four atomic clocks with a shared QND measurement, obtaining a proof-of-principle metrological enhancement of 11.6 dB~\citep{Malia2022}.

More recently, in 2023, a state-of-the-art~\isotope[87]{Sr} optical clock implemented a QND measurement, enhancing the performance by 1.9 dB at the $10^{-17}$ level~\citep{Robinson2024}. However, neither this measurement was under the SQL. They anticipate that improved control of atomic motion, larger atom number and higher single-atom cooperativity could yield stronger spin squeezing.

All demonstrated entangled interferometers and clocks to date have employed Fabry–Pérot cavities. A ring bow-tie cavity configuration would represent a significant upgrade for these experiments, as it provides homogeneous atom–light interactions during QND measurements by allowing the probe light to pass through the atoms in only one direction. Therefore, to realize a fully entanglement-enhanced Mach-Zehnder interferometer, capable of actually measuring gravity and not only doing proof-of-principle demonstrations, a ring bow-tie configuration would be advantageous. Following Ref.~\citep{Salvi2018}, such a configuration is currently being developed with strontium atoms.

Interestingly, in 2025 a gravimeter based on BECs of $N = 6 \times 10^3$~\isotope[87]{Rb} atoms demonstrated a sensitivity of -1.7 dB beyond the SQL by employing squeezed states~\citep{cassens2025}. In this case, squeezing was not generated with a cavity but through spin-changing collisions.

\subsubsection{QND Magnetometry}
\label{sec:det:QND:magnetometry}

One of the most prominent applications of quantum nondemolition (QND) measurements is optical magnetometry, where the ability to measure collective spin observables without destroying their coherence enables unprecedented sensitivity to weak magnetic fields. A particularly powerful mechanism for QND-based magnetometry is provided by the Faraday effect, which couples the atomic spin projection (that is the system) along the light propagation direction to the polarization of an optical probe (that is the meter) in a dispersive and minimally destructive way. Faraday rotation occurs when a linearly polarized light beam passes through an atomic medium subject to a magnetic field aligned with the propagation direction. In this condition, light can be described as a superposition of two circularly polarized components ($\sigma^+$ and $\sigma^-$). In an isotropic medium in the absence of a magnetic field, both polarizations interact identically with the atoms, which results in equal refractive indices and in no net rotation. However, the presence of a magnetic field makes the system anisotropic. Atomic energy levels, degenerate in the absence of a field, are separated according to the Zeeman effect: sublevels with opposite projections of the magnetic moment along the field experience opposite energy shifts. The two circular polarizations of light thus couple to distinct transitions: the $\sigma^+$ component drives transitions with $\Delta m = +1$, while the $\sigma^-$ component drives those with $\Delta m = -1$. The associated resonance frequencies are no longer the same and, as the laser is detuned from resonance, the refractive indices for the two polarizations are different. When the transmitted components recombine, the result is a rotation of the polarization axis~\citep{Faraday1846, Budker2013, Barron2004, Piller1972}.

Faraday rotation is widely applied in optical magnetometry, enabling the detection of extremely weak magnetic fields down to femtotesla sensitivity~\citep{Budker2013, Alvarez2022}. Furthermore, the quantum nondemolition character of Faraday probing has been exploited to perform minimally invasive measurements of collective atomic spins~\citep{Swar2021, Kristensen2017, Kristensen2019}. This approach allows for repeated interrogation of an ensemble, providing enhanced sensitivity through spin squeezing and entanglement generation. In quantum information science, similar techniques enable optical measurements sensitive to the state of single atoms or photons, allowing nondestructive readout protocols and hybrid light–matter interfaces~\citep{Atatre2007, Takei2010}.

Important experimental advances have included demonstrations of spin-squeezed states in large ensembles probed by Faraday rotation, with direct applications to quantum-enhanced atomic clocks and magnetometers~\citep{Leroux2010}. Experiments have also realized stroboscopic backaction-evading protocols, where temporal modulation of the probe field further suppresses decoherence~\citep{Vasilakis2011}. These results establish Faraday-based QND detection as a leading method in quantum metrology, allowing one to overcome classical noise limits and enhance measurement precision beyond the standard quantum limit. Unlike cavity-enhanced methods for atomic clocks shown in sec.~\ref{sec:det:QND:cavity}, it was recently shown that a high cavity finesse is not advantageous for magnetometry. There is instead an optimum finesse that maximizes the signal-to-noise ratio~\citep{Mazzinghi2021}. 

Today, state-of-the-art QND magnetometry combines optimized probe geometries, advanced detection techniques, and entanglement-assisted strategies, achieving record-breaking sensitivities relevant for applications from biomagnetic sensing to tests of fundamental physics. Future directions will likely focus on integrating QND magnetometry with quantum networks, exploiting cavity enhancement for stronger light–matter coupling, and extending protocols to exotic systems such as Rydberg ensembles and solid-state spin defects. These avenues highlight the dual role of Faraday rotation-based QND measurements as both a workhorse of quantum sensing and a cornerstone of scalable quantum technologies.

\subsection{High-resolution detection}
\label{sec:det:highres}


High-resolution and high-sensitivity detection of ultracold atoms is a cornerstone of modern experiments, enabling direct access to microscopic occupation, spin, and correlation information with single-particle sensitivity. Several complementary technologies have achieved this goal: quantum gas microscopes (QGMs) that provide site-resolved fluorescence imaging in optical lattices, high-NA imaging of atoms trapped in optical tweezers, and cavity-enhanced detection schemes that combine single-particle sensitivity with nondestructive readout and enhanced measurement bandwidth~\citep{Bakr2009, Sherson2010, Kaufman2012, Thompson2013}. 

Typically, quantum gas microscopes achieve single-site resolution by combining an optical lattice with high-numerical-aperture (NA) objective lenses and fluorescence imaging while maintaining atoms tightly confined in deep lattice wells~\citep{Bakr2009, Sherson2010}. Key technical ingredients include (i) collection optics with NA $\gtrsim 0.5$ to gather sufficient fluorescence photons, (ii) deep optical potentials to prevent loss during imaging, and (iii) in-situ cooling during imaging (e.g. Raman- or polarization-gradient based schemes) to suppress recoil heating and keep atoms localized on single sites. These capabilities enabled direct observation of Mott shells, single-site resolved spin and density correlations, and real-space studies of quantum magnetism and out-of-equilibrium dynamics in both bosonic and fermionic Hubbard systems~\citep{Bakr2009, Sherson2010, Cheuk2015, Parsons2015, Mazurenko2017}. 

Optical tweezers provide an alternative route to single-particle-resolved detection that is particularly flexible for programmable geometries. In tweezer platforms, tightly focused beams and high-NA optics allow efficient fluorescence collection from single trapped atoms. Recent advances combine ground-state cooling in tight tweezers, high-fidelity state readout, and rapid imaging sequences that preserve internal-state coherence~\citep{Kaufman2012, Kaufman2015, Endres2016}. The deterministic assembly and imaging of defect-free arrays further leverage these detection methods: fluorescence readout of tweezers is used both for initial occupancy determination and for mid-sequence diagnostics during rearrangement protocols~\citep{Endres2016, Lis2023}. Tweezer-based imaging techniques often rely on sub-Doppler cooling, Raman sideband cooling during imaging, or shelving schemes to maximize fluorescence counts while minimizing loss and heating~\citep{Kaufman2012}.

Optical cavities enable high-sensitivity detection through the strong atom–light coupling afforded by the resonant enhancement of circulating photons. Cavity-based detection ranges from single-atom, single-photon-level readout to nondestructive dispersive measurements of atomic ensembles~\citep{Hood2000, Colombe2007, Thompson2013}. Advantages of cavity detection include (i) high photon collection efficiency independent of the external imaging NA, (ii) the possibility of dispersive, minimally destructive measurements of internal states or collective populations, and (iii) integration with quantum-network and cavity-QED schemes. Cavity readout has been used for single-atom detection, real-time monitoring of quantum jumps, and cavity-enhanced measurements of collective observables with high bandwidth~\citep{Hood2000, Thompson2013}.

A particularly important class of techniques leverages "correlation measurements at the single-particle level". When single atoms are imaged with site resolution, one can directly compute spatial correlation functions $g^{(1)}(r)$ and $g^{(2)}(r)$, measure entanglement witnesses, extract string-order parameters, and reconstruct full counting statistics. Quantum gas microscopes enabled the first direct observations of antiferromagnetic correlations in the Fermi–Hubbard model~\citep{Mazurenko2017} and the detection of string patterns associated with doped holes~\citep{Chiu2019}. Single-site readout also permits the measurement of higher-order correlations, nonlocal order, and the tomography of few-body states formed in tweezers~\citep{Mazurenko2017, Chiu2019, Kaufman2015}.

Important results and benchmarks across these platforms include:
\begin{itemize}
  \item \textbf{Quantum gas microscopes:} site-resolved detection of Mott insulating shells, single-atom resolved study of spin and density correlations, and direct measurements of quantum dynamics and thermalization~\citep{Bakr2009, Sherson2010, Parsons2015, Mazurenko2017}.
  \item \textbf{Tweezers:} ground-state cooling and high-fidelity fluorescence detection of single atoms in tight traps, deterministic assembly of defect-free arrays with mid-sequence imaging, and high-fidelity state readout suitable for quantum gates~\citep{Kaufman2012, Endres2016, Kaufman2015}.
  \item \textbf{Cavities:} nondestructive, high-bandwidth readout of collective and single-atom observables, real-time monitoring of quantum jumps, and integration with quantum-network architectures~\citep{Hood2000, Colombe2007, Thompson2013}.
\end{itemize}

Each approach has characteristic strengths and limitations. Quantum gas microscopes excel at large-scale, lattice-based studies with full spatial context but require specialized high-NA imaging systems and typically destructively scatter many photons during imaging. Tweezers are highly reconfigurable and ideal for programmable small- to intermediate-sized arrays, with straightforward single-atom readout, but face challenges in scaling optical access for very large arrays while maintaining uniform imaging fidelity. Cavity-based detection offers nondestructive, efficient readout and seamless integration with photonic interfaces, but requires careful engineering of cavity parameters and control of collective back-action when used for many-body systems.

The current state of the art combines these detection modalities in complementary ways: quantum gas microscopes routinely achieve single-site, single-atom fidelity sufficient to study quantum magnetism and nonequilibrium dynamics; tweezer arrays now permit repeated high-fidelity readout during complex assembly and gate sequences; and cavity systems provide nondestructive monitoring and interfaces for quantum networking~\citep{Bakr2009, Endres2016, Thompson2013}. Future directions include (i) faster, lower-backaction imaging methods (e.g., correlation-assisted or quantum-limited dispersive readout), (ii) integration of high-NA optics with large-scale tweezer platforms, (iii) hybrid systems combining site-resolved imaging with cavity-enhanced readout for multiplexed sensing, and (iv) the use of single-particle-resolved detection in error-corrected neutral-atom quantum processors and distributed quantum networks. Overall, high-resolution, high-sensitivity detection remains a rapidly advancing frontier that directly enables precision studies of many-body physics and the scaling of quantum technologies.

\begin{table}[ht!] 
\centering
\caption{Overview of particle and quantum state detection techniques and their applications.}
\label{tab:Detection}
\small
\begin{tabularx}{\textwidth}{X X X X}
\toprule
\textbf{Methods} & \textbf{Benefits} & \textbf{Limitations \& Requirements} & \textbf{Main applications} \\
\midrule

\textbf{Ensemble detection methods} & & & \\
\multirow[t]{3}{*}{Absorption imaging}
& -relatively easy to implement & -Resonant interaction & Ensemble imaging~\citep{Ketterle1999}\\
& -direct measurement of optical density & -limited dynamic range & TOF imaging~\citep{Ketterle1999} \\

\multirow[t]{3}{*}{Fluorescence imaging}
& -Single atom detection & -Strongly destructive & Quantum gas microscopes~\citep{Bakr2009, Sherson2010} \\
& -Site resolved microscopy & -Limited scattered photons & Optical tweezers~\citep{Schlosser2001, Endres2016} \\
& -Very high sensitivity & -Requires low-noise detection &  \\

\multirow[t]{3}{*}{Partial transfer abs. imaging}
& -Minimally destructive & -Relies on resonant interaction & Real-time dynamics~\citep{Ramanathan2012}\\
& -Tunable probed fraction & -Low signal for small probed fraction & Dense atomic clouds~\citep{Mordini2020}  \\
& -High dynamic range & -Requires well-controlled fields & \\

\multirow[t]{3}{*}{Phase-contrast imaging}
& -Non-destructive & -Precise phase plate alignment & Condensate profiles~\citep{Meppelink2010} \\
& -Works for high OD & -Requires phase unwrapping & Real-time dynamics~\citep{Andrews1997} \\

\multirow[t]{3}{*}{Polarization contrast imaging}
& -Non-destructive & -Critical stable polarization & Imaging magnetization~\citep{Higbie2005}\\
& -Spin sensitive &  & Quantum metrology~\citep{Appel2009} \\
& -Dual-port cancels noise &  & \\

\multirow[t]{3}{*}{Cavity-enhanced atom detection}
& -Higher cooperativity & -High-finesse optical cavity & Spin squeezing~\citep{Hosten2016, Cox2016}\\
& -Higher dispersive phase shift & -Technical difficulties &  Fiber-based cavities~\citep{Gehr2010,Trupke2007}\\
& -Enhanced fluorescence & -Frequency noise &  Cavities with atom chips~\citep{Teper2006}\\
\midrule

\textbf{State-selective detection} & & & \\
\multirow[t]{3}{*}{Stern--Gerlach}
& -Spatial separation of different spin states & -Magnetic field gradient  & Spin-state resolved imaging~\citep{Ketterle1999}\\
& -Absorption imaging can be used & -States sensitive to magnetic field &  \\
& -Simple to implement & Long enough time-of-flight &  \\
\multirow[t]{3}{*}{Optical pumping, shelving}
& -High fidelity state discrimination & -Requires precise optical control & Trapped ions, Rydberg and neutral atoms~\citep{Kuhr2003, Leibfried2003, Karlewski2015}\\
& -Compatible with single-atom detection & -Shelving state must be long-lived & Quantum computation and quantum gas microscopy~\citep{Bakr2009, Sherson2010} \\
& -Non-destructive for other states (or for selected state)& -Sensitive to off-resonant scattering &  \\


\bottomrule
\end{tabularx}
\end{table}


\section{Conclusions}

The precise and robust methods for cooling, trapping, controlling and detecting  neutral atoms and molecules that have been covered in this tutorial review are the foundation that is enabling ultracold quantum gases and atomic physics to stand at the forefront of quantum science and technology. 

Looking ahead, the convergence of control techniques, robust detection, and large-scale engineering can further extend the frontiers of the field. On the fundamental science side, ultracold atoms will continue to provide unique insights into strongly interacting quantum matter, many-body phenomena, topological phases, and nonequilibrium dynamics. Concurrently, these tools and their future refinement are becoming the foundation of quantum-enhanced technologies for computation, communication, and metrology. This dual role, as both a fundamental and applied platform, ensures that this area of work will inspire and challenge many generations of scientists, engineers and entrepreneurs, with a strong outlook for generating significant impact on society.
\vspace{6pt} 

\authorcontributions{Conceptualization: Vladislav Gavryusev, Giulia Del Pace, Louise Wolswijk, Paolo Vezio. Writing, both original draft preparation and review and editing, and visualization: Louise Wolswijk, Luca Cavicchioli, Giuseppe Vinelli, Mauro Chiarotti, Ludovica Donati, Marcia Frometa Fernandez, Diego Hern\'andez Rajkov, Christian Mancini, Paolo Vezio, Tianwei Zhou, Giulia Del Pace, Chiara Mazzinghi, Nicolò Antolini, Leonardo Salvi and Vladislav Gavryusev. Supervision and project administration: Vladislav Gavryusev. All authors have read and agreed to the published version of the manuscript.}

\funding{This project has received funding from Consiglio Nazionale delle Ricerche (CNR) PNRR MUR project IR0000016 I-PHOQS ’Integrated Infrastructure Initiative in Photonic and Quantum Sciences’, QuantERA ERA-NET Cofund in Quantum Technologies project SQUEIS (Squeezing enhanced inertial sensing), from the Italian Ministry of Education and Research (MUR) PRIN 2022 Project ‘Quantum Sensing and Precision Measurements with Nonclassical States', and, in the context of the National Recovery and Resilience Plan and Next Generation EU, from project PE0000023-NQSTI, from M4C2 investment 1.2 project MicroSpinEnergy (V.G.), from M4C2 investment 1.3 project nr. CN\_00000013 (L.W.), from M4C2 investment 1.3 Extended Partenariat PE00000001 - "RESTART" (P.V.), from Space It Up project funded by the Italian Space Agency, ASI, and the Ministry of University and Research, MUR, under contract n. 2024-5-E.0 - CUP n. I53D24000060005 (G.V.), from ASI and CNR under the Joint Project “Laboratori congiunti ASI-CNR nel settore delle Quantum Technologies (QASINO)” (Accordo Attuativo n. 2023-47-HH.0), from  INFN sez. di Napoli with the project QGSKY (C.M.), from the Horizon Europe program (Grant ID 101080164, UVQuanT) (M.C.) and from the Horizon Europe program HORIZON-CL4-2022-QUANTUM-02-SGA via project 101113690 (PASQuanS2.1) (M.F.F and D.H.R).}

\dataavailability{No new data were created or analyzed in this study. Data sharing is not applicable to this article.}

\acknowledgments{We acknowledge insightful discussions with Gabriele Rosi, Stefano Finelli and Alessio Ciamei.}

\conflictsofinterest{The authors declare no conflicts of interest. The funders had no role in the design of the study; in the collection, analyses, or interpretation of data; in the writing of the manuscript; or in the decision to publish the results.}

\isPreprints{}{
\begin{adjustwidth}{-\extralength}{0cm}
} 

~\reftitle{References}

\bibliography{ReferencesVladGavr, PaperRef_AI, QuantumMetrology, LiLab, nico, imaging, optical_lattice, evaporative_cooling, rabi-Chiarotti} 

%

\PublishersNote{}
\isPreprints{}{
\end{adjustwidth}
} 
\end{document}